\documentclass[sn-basic, Namedate, iicol, pdflatex]{sn-jnl}

\usepackage{graphicx}%
\usepackage{multirow}%
\usepackage{amsmath,amssymb,amsfonts}%
\usepackage{amsthm}%
\usepackage{mathrsfs}%
\usepackage[title]{appendix}%
\usepackage{xcolor}%
\usepackage{textcomp}%
\usepackage{manyfoot}%
\usepackage{booktabs}%
\usepackage{algorithm}%
\usepackage{algorithmicx}%
\usepackage{algpseudocode}%
\usepackage{listings}%

\theoremstyle{thmstyleone}%
\theoremstyle{thmstyletwo}%

\theoremstyle{thmstylethree}%

\begin{document}

\title[Making waves in massive star asteroseismology]{\bf Making waves in massive star asteroseismology}


\author[1,2]{\fnm{Dominic M.} \sur{Bowman}}\email{dominic.bowman@newcastle.ac.uk}

\affil[1]{\orgdiv{School of Mathematics, Statistics and Physics}, \orgname{Newcastle University}, \orgaddress{\street{Herschel Building}, \city{Newcastle upon Tyne}, \postcode{NE1 7RU}, \country{United Kingdom}}}

\affil[2]{\orgdiv{Institute of Astronomy}, \orgname{KU Leuven}, \orgaddress{\street{Celestijnenlaan 200D}, \city{Leuven}, \postcode{B-3001}, \country{Belgium}}}


\abstract{Massive stars play a major role not only in stellar evolution but also galactic evolution theory. This is because of their dynamical interaction with binary companions, and because their strong winds and explosive deaths as supernovae provide chemical, radiative and kinematic feedback to their environments. Yet this feedback strongly depends on the physics of the supernova progenitor star. It is only in recent decades that asteroseismology -- the study of stellar pulsations -- has developed the necessary tools to a high level of sophistication to become a prime method at the forefront of astronomical research for constraining the physical processes at work within stellar interiors. For example, precise and accurate asteroseismic constraints on interior rotation, magnetic field strength and geometry, mixing and angular momentum transport processes of massive stars are becoming increasingly available across a wide range of masses. Moreover, ongoing large-scale time-series photometric surveys with space telescopes have revealed a large diversity in the variability of massive stars, including widespread coherent pulsations across a large range in mass and age, and the discovery of ubiquitous stochastic low-frequency (SLF) variability in their light curves. In this invited review, I discuss the progress made in understanding the physical processes at work within massive star interiors thanks to modern asteroseismic techniques, and conclude with a future outlook.}




\maketitle


\section{Introduction}
\label{section: intro}

Massive stars play a pivotal role in the chemical, radiative and kinetic evolution of their host galaxies \citep{Maeder_rotation_BOOK, Langer2012}. This is because massive stars, typically defined as having birth masses larger than $M_{\rm ini} \gtrsim 8$~M$_{\odot}$, end their lives as a violent supernova explosion before forming a neutron star or black hole. Hence massive stars are an important driver of feedback to their host galaxy \citep{Bromm2009b, deRossi2010d}. In addition, during their lives massive stars have strong radiatively driven winds (see e.g. \citealt{Puls2020a, Sundqvist2019, Bjorklund2021a, Bjorklund2023a} and references therein) that contribute to interstellar feedback and new generations of star formation. Another important aspect of massive stars is that they are commonly found in binary or higher-order multiple systems (e.g. \citealt{Vanbeveren1998a, Sana2012b}), with dynamical interaction and mass transfer between component stars of a multiple system having a significant impact on a star's evolution (e.g. \citealt{deMink2013, Schneider_F_2015a, Marchant2023a*}). This makes massive stars, in general, complex but fruitful laboratories for testing a variety of physics.

It is clear that understanding massive star evolution is critical for a broad range of topics within stellar astrophysics. However, in addition to the physics associated to stellar winds and binarity, all evolution models of massive stars currently suffer from large theoretical uncertainties associated with the interior physical processes of rotation and mixing \citep{Langer2012, Ekstrom2012a, Martins2013c}. In particular, the amount and shape of mixing at the boundary of convective and radiative regions, commonly referred to as convective boundary mixing (CBM), is currently controlled by uncalibrated parameterisations in evolution models \citep{Bowman2020c, Kaiser2020a, Scott2021a, Johnston2021b, Anders_E_2023a}. Such uncertainties are common to all massive stars regardless of their metallicity, rotation rate or presence of a binary companion, since all massive stars have a hydrogen-burning convective core during the main sequence. Because of the unknown amount and shape of CBM at the boundary of convective cores, the uncertainties in core masses and ages of massive stars can easily reach 100\% \citep{Bowman2020c, Johnston2021b}. Furthermore, the rotation rate of a massive star influences the amount of interior mixing and interstellar feedback (see e.g. \citealt{Hirschi2005c, Ekstrom2012a, Chieffi2013, Georgy2013c}), but also the efficiency of angular momentum transport \citep{Aerts2019b}.

Today, evolution models include numerical prescriptions for interior rotation, mixing and angular momentum transport physics based on important theoretical work (see e.g. \citealt{Zahn1991, Talon1997b}). However, testing the validity and applicability of such prescriptions using observational constraints has some catching up to do. A powerful method for such a task is asteroseismology \citep{ASTERO_BOOK, Kurtz2022a}, which is the method of quantitative comparison of observed pulsation mode frequencies to a grid of evolution models and their theoretically predicted pulsation mode frequencies to determine the best-fitting model \citep{Aerts2021a}. Asteroseismology has been highly successful when applied to low-mass stars, such as red giants \citep{Chaplin2013c, Hekker2017a, Garcia_R_2019}. In particularly, asteroseismology has concretely demonstrated large discrepancies in the observed and theoretically predicted radial rotation profiles of low-mass stars, thus calling for improved theory and numerical prescriptions for angular momentum transport across stellar evolution \citep{Aerts2019b}. On the other hand, asteroseismology of massive stars is lagging behind low-mass stars predominantly because such stars are intrinsically rare in the Universe, but also because high-precision, long-duration time-series data for a large sample of stars was not available until recently \citep{Bowman2020c}. It is primarily thanks to the ongoing NASA TESS mission \citep{Ricker2015} that the situation has changed, with the TESS mission providing high-precision light curves of thousands of pulsating massive stars across the entire sky.

In this invited review, I discuss recent advances in our understanding of massive star interiors thanks to asteroseismology. As this review is part of the Springer Nature 2023 Astronomy Prize Awardees Collection, I pay particular attention to work that I have personally been involved with. Indeed, as mentioned previously, it is thanks to the ongoing NASA TESS mission that has revolutionised the observational data-driven paradigm shift of asteroseismology to massive stars. The review by \citet{Bowman2020c} discussed the constraints on interior rotation and mixing based on massive star asteroseismology determined up until the launch of the TESS mission. In this complementary review, I focus on the advances made using TESS data and the new questions and puzzles that have been revealed since circa 2020.


\section{Diverse variability in massive stars revealed by space photometry}
\label{section: diverse}

It was the launch of space telescopes in recent decades, such as the MOST \citep{Walker2003}, CoRoT \citep{Auvergne2009}, Kepler \citep{Koch2010, Borucki2010}, BRITE \citep{Weiss2014}, and TESS \citep{Ricker2015} missions, that opened pandora's box for variability studies of massive stars. This is because the typical variability time scales, such as from rotational modulation, pulsations or binarity, can be of the order of a few days to several hours for massive stars making them difficult to probe using ground-based telescopes. Herculean efforts that assembled many nights of observations of individual massive stars with some of the world's largest ground-based telescopes have yielded significant asteroseismic results (e.g. \citealt{Aerts2003d, Handler2004b}), but only for a handful of massive stars. For an in-depth discussion and examples of asteroseismology of massive stars using ground-based telescopes, I refer the reader to the review by \citet{Bowman2020c}. 

The MOST mission performed valuable initial studies of massive star variability, including the characterisation of pulsations \citep{Walker2005a, Walker2005b, Saio2006b}, rotation \citep{Rami2014}, and rotation period spin-down caused by large-scale magnetic fields \citep{Townsend2013a}. However, it was space photometry assembled by CoRoT that demonstrated the feasibility of massive star asteroseismology on a large scale, although the CoRoT mission was modest in terms of sample size by today's standards. One of the important contributions from the CoRoT mission was the first detection of a gravity-mode period spacing in a massive star \citep{Degroote2010a}. Later, the nominal Kepler mission data indirectly included a few massive stars (see e.g. \citealt{Aerts2017a, Pope2019b}) but generally avoided them to focus instead on low-mass stars with potential exoplanets. As an example, Kepler data allowed pulsations to be characterised in high-mass eclipsing binaries with far better photometric precision than possible from the ground \citep{Tkachenko2012b, Tkachenko2016}. In the revised K2 mission \citep{Howell2014}, a few hundred massive stars were included \citep{Bowman2019b}, but K2 light curves were maximally 80~d in length and of lesser quality than nominal Kepler mission light curves. On the other hand, the BRITE-constellation of cube-satellites, launched over a decade ago and continues to assemble long-term light curves of selected bright massive stars. For a review of the BRITE mission, which includes key examples of major results, and an ensemble analysis of its first data release, I refer the reader to \citet{Weiss2021a} and \citet{Zwintz2023a*}, respectively.

In this review, I focus on work motivated primarily by newly available light curves from the ongoing TESS mission \citep{Ricker2015}, which has a fundamentally different observational strategy compared to previous time-series photometry space missions. The CoRoT and Kepler missions focussed on assembling light curves within fixed fields in the sky, pre-chosen to contain targets with high scientific potential. Whereas the TESS mission is operating as a fully community-driven survey in which essentially all stars brighter than approximately $V \lesssim 14$~mag in the sky have a light curve spanning at least 27~d with a cadence of 30~min in the nominal mission. As of the extended and second extended TESS missions, this cadence was significantly improved to be 10~min and 200~sec, respectively, in the so-called full-frame images (FFIs; \citealt{Jenkins2016b}). 

The community are able to propose a limited number of high-priority targets for even shorter cadences of 2~min and even 20~sec in the form of TESS Guest Investigator (GI) proposals\footnote{\url{https://heasarc.gsfc.nasa.gov/docs/tess/approved-programs.html}}. Another difference and advantage is that all TESS data products are made immediately publicly available with no proprietary period. Such an approach has facilitated many serendipitous discoveries and has the major advantage of providing a large sample of massive stars across the whole sky, albeit at the expense of providing typically shorter continuous light curves compared to CoRoT and Kepler.


\section{Discovery and exploitation of stochastic low-frequency variability in massive stars}
\label{section: SLF}

A major step forward in our understanding of massive stars is in the ongoing work to comprehend stochastic low-frequency (SLF) variability in their light curves. The first discussion of SLF variability, although it was simply referred to as `astrophysical red noise', in time-series photometry was based on the only three O~stars observed by CoRoT \citep{Blomme2011b}. However, the conclusions were speculative as to the physical mechanism giving rise to such a phenomenon. 

Later, the presence of SLF variability at the surface of these three massive O stars was postulated to be connected to convectively driven gravity waves by \citet{Aerts2015c} based on a qualitative comparison of the observed frequency spectra of the same three O~stars observed by CoRoT to theoretical frequency spectra of gravity waves driven by core convection in 2D hydrodynamical simulations \citep{Rogers2013b, Rogers2015}. At the time, these were the only numerical simulations with the necessary global spatial scale, resolution and input physics that permitted a physical comparison to the inferred gravity wave spectra of observed stars. It was this hypothesis that sparked much of the subsequent research into finding and characterising the signatures of gravity waves in many more massive stars using both photometry and spectroscopy.


	\subsection{Constraints from time series photometry}
	\label{subsection: photometry}
	
\begin{figure}
	\includegraphics[width=0.99\columnwidth]{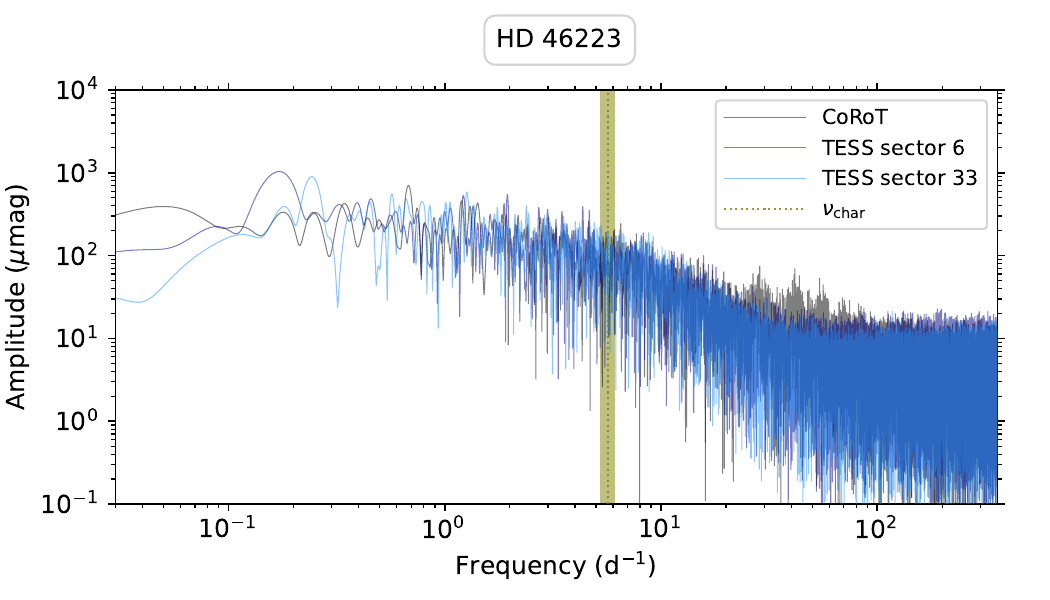}
	\includegraphics[width=0.99\columnwidth]{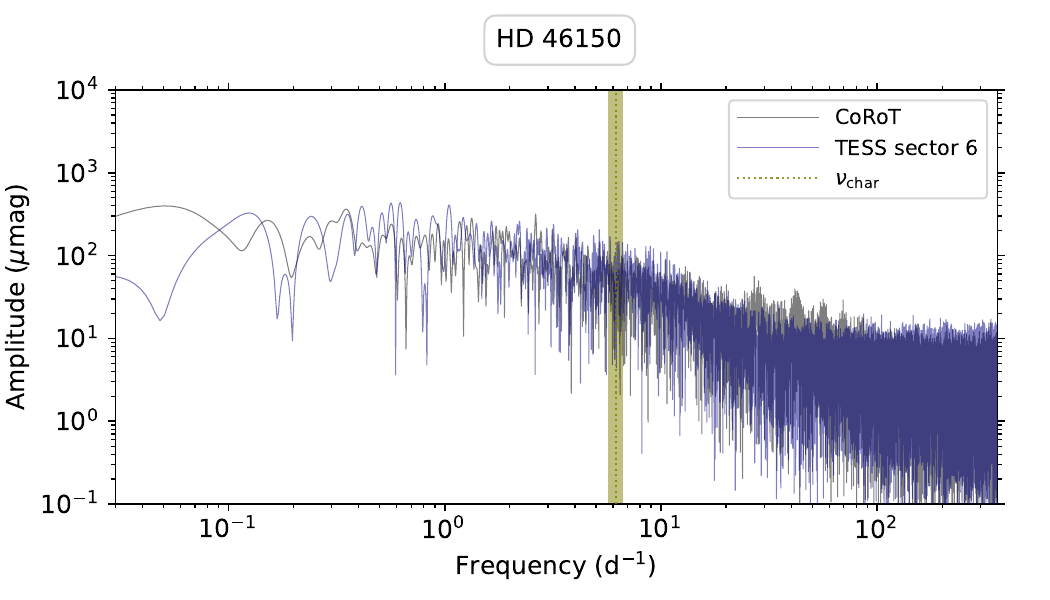}
	\caption{Example of SLF variability profiles in the frequency spectra of two O stars, HD\,46223 and HD\,46150, observed by CoRoT and later re-observed by the TESS mission in 2-min cadence in its sectors 6 and 33 (if available). These two examples demonstrate the broad frequency range over which SLF variability is significant (i.e. periods of tens of days to minutes). The reported $\nu_{\rm char}$ values and their 2$\sigma$ confidence intervals from \citet{Bowman2022b} are also shown as the yellow shaded regions.}
	\label{figure: SLF}
	\end{figure}

	The first large-scale search for SLF variability using dozens of early-type stars observed by CoRoT demonstrated that a common morphology exists in the frequency spectra of such stars \citep{Bowman2019a}. The shape of the SLF variability in the amplitude spectra of early-type stars is well described by a semi-Lorentzian function:

	\begin{equation}
	\alpha \left( \nu \right) = \frac{ \alpha_{0} } { 1 + \left( \frac{\nu}{\nu_{\rm char}} \right)^{\gamma}} + C_{\rm w} ~ ,	
	\label{equation: red noise}
	\end{equation}
	
	\noindent where $\alpha_{0}$ represents the amplitude at a frequency of zero, $\gamma$ is the logarithmic amplitude gradient, $\nu_{\rm char}$ is the characteristic frequency, and $C_{\rm w}$ is a frequency-independent (i.e. white) noise term \citep{Bowman2019a}. The characteristic frequency, which represents the inverse of the typical timescale of the variability, is of order several d$^{-1}$ for massive dwarf stars \citep{Bowman2019a}. Moreover, this was the first time that a quantitative approach to characterising SLF variability in observations was used, with results shown to be consistent with the theoretical frequency spectra of gravity waves from 2D hydrodynamical simulations \citep{Rogers2013b, Rogers2015} and predictions of standing wave gravity eigenfrequencies of such stars \citep{Bowman2019a}. Examples of the SLF variability morphology in the frequency spectra of two O stars, HD\,46223 and HD\,46150, observed by CoRoT and TESS are shown in Fig.~\ref{figure: SLF}, which demonstrate the broad frequency range of SLF variability spanning from minutes to several days and includes the reported $\nu_{\rm char}$ values from \citet{Bowman2022b}.
			
	With a larger sample of massive stars available from the K2 and TESS missions, it was later shown that SLF variability is essentially ubiquitous in massive stars when sufficiently high-quality light curves are available \citep{Bowman2019b}. A large sample of over 160 massive stars observed by K2 and TESS, which included metal-poor extra-galactic and metal-rich galactic stars across the entire sky, again demonstrated that all stars had a similar morphology in their frequency spectra. Two important conclusions from this large sample are: (i) the presence and morphology of SLF variability and specifically the inferred $\nu_{\rm char}$ parameter is seemingly insensitive to metallicity; and (ii) the amplitude and characteristic frequency, $\nu_{\rm char}$, of SLF variability depends on the brightness and luminosity of the star \citep{Bowman2019b}. 

\begin{figure}
	\includegraphics[width=0.99\columnwidth]{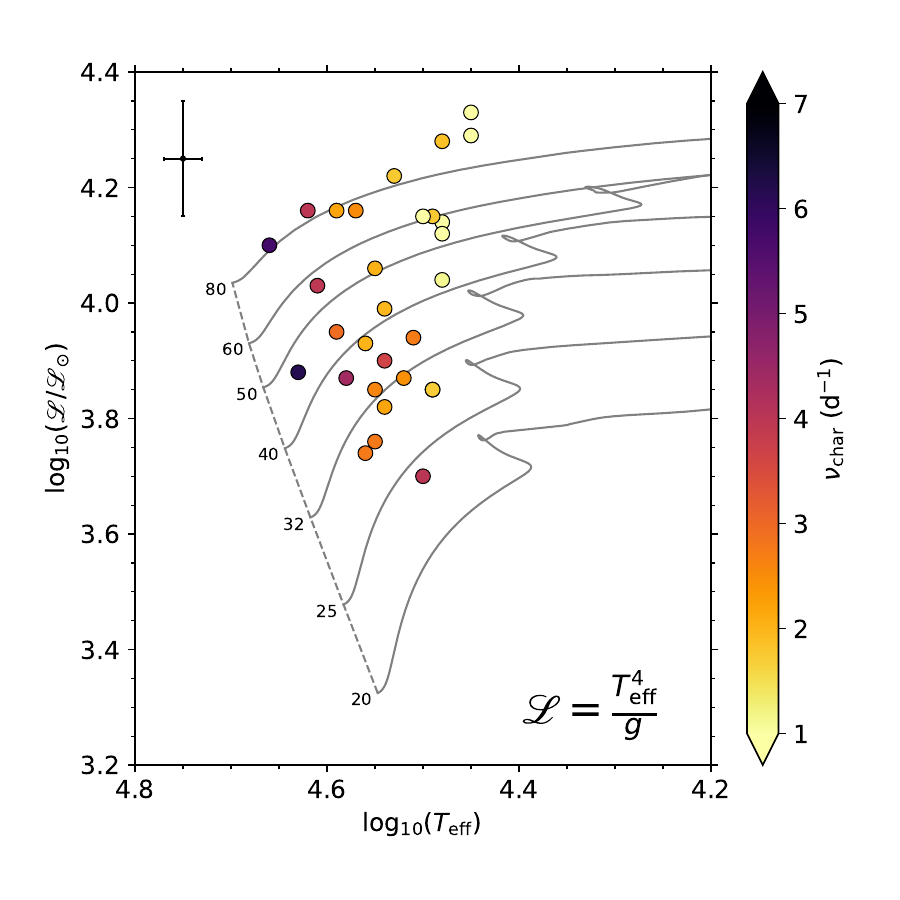}
	\caption{Relationship between the measured characteristic frequency, $\nu_{\rm char}$, of SLF variability and location in the spectroscopic HR~diagram for 30 galactic O stars \citep{Bowman2020b, Bowman2022b}, demonstrating that as stars evolve the value of $\nu_{\rm char}$ decreases. Evolutionary tracks starting at the ZAMS are shown as solid grey lines with initial masses given in units of M$_{\odot}$ \citep{Burssens2020a}. A typical spectroscopic error bar for the sample is shown in the top-left corner.}
	\label{figure: HRD}
	\end{figure}

	Using a combination of TESS light curves and high-resolution spectroscopy, including from the HERMES spectrograph mounted on the 1.2-m KU Leuven telescope on La Palma \citep{Raskin2011}, it was demonstrated that a strong correlation exists between the morphology of observed SLF variability and the location of a star in the Hertzsprung--Russell (HR) diagram \citep{Bowman2020b}. More specifically, this means that the morphology of SLF variability directly probes the mass and age of a massive star \citep{Bowman2020b}. Such a discovery has opened up a new sub-branch of asteroseismology in which constraints on the fundamental properties of massive stars can be achieved without the prerequisite of identifying the spherical geometry of individual pulsation mode frequencies. A demonstration of the change in the measured $\nu_{\rm char}$ parameter as stars evolve during the main sequence based on the results of \citet{Bowman2020b} is shown in Fig.~\ref{figure: HRD}. The conclusion from this transition is that the physical mechanism causing SLF variability tends to produce longer periods for older stars, but the full frequency range of SLF variability remains broad in all cases.
	
	More recently, the correlation between SLF variability morphology and location of massive stars in the HR~diagram was confirmed using the novel light-curve fitting method of Gaussian process (GP) regression \citep{Bowman2022b}. A GP regression model describes stochastic variability in a light curve by fitting hyperparameters and defining a covariance matrix. This provides posterior distributions that represent structures in covariance matrices, which are conditioned on the input data to evaluate the optimum representation of the input time series data \citep{Rasmussen2006}. For a detailed discussion of GP regression methods and applications to astronomical time series data, I refer the reader to \citet{Aigrain2023a}. In the study by \citet{Bowman2022b}, a GP kernel of a damped simple harmonic oscillator was used within the {\sc celerite2}\footnote{\url{https://celerite2.readthedocs.io/en/latest/}} software package \citep{Foreman-Mackey2017}, which allowed $\nu_{\rm char}$ as well as the quality factor, $Q$, of the input light curves to be efficiently and accurately inferred. Interestingly, the GP regression methodology revealed that stars are more stochastic (i.e. lower $Q$ values) near the zero-age main sequence and are more quasi-periodic (i.e. higher $Q$ values) near the terminal-age main sequence \citep{Bowman2022b}. This transition further demonstrates that the physical mechanism responsible for SLF variability not only depends on age, but also that SLF variability has an age-dependent stochasticity and coherency.
		
	\subsection{Constraints from complementary spectroscopy}
	\label{subsection: spectroscopy}
	
	Complementary to photometric time series observations, spectroscopic variability also provides valuable constraints on the physical processes within massive stars. Spectral line profile variability (LPV) and the time-dependence of non-rotational spectral line profile broadening mechanisms, for example macroturbulence, has historically been linked to pulsations \citep{Lucy1976e, Fullerton1996, Howarth1997, Aerts2009b}. The triangular shapes of spectral lines in massive stars cannot exclusively be explained by rotational broadening \citep{Simon-Diaz2014a}, but despite the necessity of including a macroturbulent broadening term, the physics explaining this `nuisance' term in spectroscopy has remained illusive. 
	
	In a detailed study of macroturbulence in galactic massive stars, it was found that early- to late-B main-sequence stars have a diverse range in macroturbulent broadening values, whereas main-sequence O and B supergiants typically have macroturbulence rather than rotation as the dominant broadening mechanism in their spectral lines \citep{Simon-Diaz2017a, deBurgos2023a, deBurgos2023b*}. Large amounts of macroturbulence are also necessary to fit the spectra lines of supergiants in the Large Magellanic Cloud (LMC) galaxy \citep{Serebriakova2023a}, demonstrating its importance across different metallicity regimes. In order to explain such large values of macroturbulence, which can exceed 100~km\,s$^{-1}$ for O stars, non-radial gravity-mode pulsations have been shown to provide the necessary predominantly tangential velocity field near the stellar surface \citep{Aerts2009b}. Such pulsations can be excited by an opacity heat-engine mechanism and/or turbulent pressure fluctuations within massive star envelopes \citep{Aerts2009b, Grassitelli2015a}. Alternatively, the dynamics of turbulent massive star envelopes can give rise to large velocity fields \citep{Schultz2022a, Jiang_Y_2023a}, but the interplay of these non-mutually exclusive mechanisms is yet to be investigated.
	
	In addition to coherent gravity-mode pulsations described above, it is also plausible for an ensemble of stochastic gravity waves to produce the necessary predominantly tangential velocity field at the stellar surface to explain LPV and macroturbulence. This has been demonstrated using numerical simulations \citep{Aerts2015c}. In support of this conclusion, the amplitude of SLF variability in time-series photometry was also found to strongly correlate with the amount of macroturbulence in a large sample of massive stars \citep{Bowman2020b}. Therefore, the physics that causes SLF variability in photometry is likely also responsible for macroturbulence in spectroscopy and connected to pulsations.


	\subsection{Plausible physical mechanisms}
	\label{subsection: physics}
	
	Inspired by the new observational discovery that SLF variability is seemingly ubiquitous across a wide range of stellar mass and age, there have been several theoretical interpretations offered in the literature each with their own advantages and disadvantages. As mentioned previously, gravity waves excited by turbulent core convection predicted by 2D and 3D hydrodynamical simulations produce a frequency spectrum with similar morphology to that of SLF variability in time-series observations \citep{Rogers2013b, Rogers2015, Rogers2017c, Edelmann2019a, Ratnasingam2019a, Ratnasingam2020a, Ratnasingam2023a, Horst2020a, Varghese2023a, Vanon2023a, Herwig2023a, Thompson_W_2023a*}.
	
	Of course, all hydrodynamical studies have specific numerical setups and assumptions in an attempt to most accurately reproduce reality. It is an advisable and valuable exercise to compare different numerical methods, especially if different results are obtained from codes attempting to solve the same physical problem (see e.g. \citealt{Andrassy2022a, Lecoanet2023a}). The differences in some setups and apparent disagreement in numerical predictions as outputs of specific simulations, such as the frequency spectrum of core-excited gravity waves, has led to the physical mechanism behind SLF variability being hotly debated in the literature. For example, some hydrodynamical simulations with different numerical setups conclude that gravity waves excited by turbulent core convection do not reach the stellar surface with observable amplitudes because of radiative damping \citep{Lecoanet2019a, Lecoanet2021a, LeSaux2023a, Anders_E_2023b}. However, almost all simulations are performed using zero or very small rotation rates compared to observed stars, with rotation being both an important restoring force but also a strong contributory factor for setting the (non-linear) amplitudes of waves within stellar interiors. In their 3D hydrodynamical simulations, \citet{Anders_E_2023b} predict around an order of magnitude larger wave amplitudes for moderately rotating simulations compared to slowly rotating simulations, with predicted wave amplitudes that approach the measured amplitude of SLF variability in observed rotating stars. Therefore, for an accurate comparison of numerical predictions from simulations to measured parameters of observed stars, realistic rotation rates and large-scale spherical simulations are necessary before concluding if the cause of SLF variability is (or is not) gravity waves excited by turbulent core convection.
	
	On the other hand, other studies hypothesise that if the physical mechanism of SLF variability is presumably not gravity waves excited by core convection, then it may be instead gravity waves and/or turbulence induced from sub-surface convection zones \citep{Cantiello2009a, Cantiello2021b}. Some hydrodynamical simulations of subsurface convection zones in massive stars also show them to be a plausible mechanism for SLF variability and macroturbulence in massive stars \citep{Schultz2022a}. However, it is not clear whether all massive stars have well-defined subsurface convection zones, as they are strongly dependent on a star's mass, rotation profile, age, opacity profile and metallicity. For example, main-sequence stars with masses between approximately 8 and 20 M$_{\odot}$ at metallicities similar to the Small Magellanic Cloud (SMC) galaxy do not have significant subsurface convection zones according to the Rayleigh criterion for convection \citep{Jermyn2022a}. Yet stars expected to not have subsurface convection zones show similar SLF variability in time series photometry to those that do \citep{Bowman2024a**}.
	
	For completeness, the modulation of a star's photosphere and atmosphere by an optically thick wind has also been shown to be a plausible mechanism for SLF variability in time-series photometry of post-main sequence massive stars \citep{Krticka2018e, Krticka2021b}. However, such a mechanism is likely negligible for the optically thin winds of main-sequence massive stars. This is especially true for late-O and early-B main-sequence stars which do not have a significant radiatively driven wind.

	It is still an open question of which physical mechanisms that produce SLF variability (and macroturbulence) dominate in particular parameter regimes of the HR diagram. This is because the properties of (sub-surface) convection, stellar winds, but also the excitation, propagation and damping of gravity waves depend on stellar structure properties, such as mass, age, rotation rate and metallicity. From the observations, it is clear that any mechanism aiming to explain SLF variability in main-sequence massive stars must explain a broad range in mass (i.e. $3 \lesssim M \lesssim 100$~M$_{\odot}$), ages that span from the zero-age main sequence (ZAMS) to the terminal-age main sequence (TAMS) and beyond, rotation rates between zero and approaching critical, and metal-poor (i.e. $Z \leq 0.002$) and metal-rich (i.e. $Z \geq 0.014$) stars. More specifically, any mechanism must additionally explain the broad range in frequency of SLF variability and the correlation between its morphology and mass and age, and apparent insensitivity to metallicity \citep{Bowman2019a, Bowman2019b, Bowman2020b, Bowman2022b, Bowman2024a**}.


\section{Physics of stellar interiors derived from forward asteroseismic modelling}
\label{section: modelling}

The TESS mission data have and continue to provide a vast treasure trove for constraining the interior physical processes of stars using asteroseismology. Recent reviews focussing on low-to-intermediate-mass stars are available \citep{Aerts2021a, Kurtz2022a}, but in this work I focus on massive stars in particular. As mentioned previously, massive stars with coherent pulsations studied from the ground were limited to relatively high-amplitude $\beta$~Cephei stars, which have masses larger than approximately 6~M$_{\odot}$ and low-radial order pressure and gravity modes with periods of order several hours \citep{Stankov2005}. A handful of asteroseismic studies of $\beta$~Cep stars using ground-based data were able to constrain the interior rotation profiles and to a lesser extent the interior mixing properties (e.g. \citealt{Aerts2003d, Dupret2004b}). Such constraints remain extremely valuable to this day for improving our understanding of angular momentum transport within stellar interiors (e.g. \citealt{Salmon2022b}). 

The constraints on interior rotation and mixing of massive stars from asteroseismology prior to the TESS mission were summarised by \citet{Bowman2020c}. In this section, I provide an update by highlighting major advances in the field of massive star asteroseismology since circa 2020 and were hence not included in the review by \citet{Bowman2020c}. It goes without saying that many of the discussed studies would not have been possible without open-source and excellently documented software tools, such as the {\sc MESA}\footnote{\url{https://docs.mesastar.org/en/}} evolution code \citep{Paxton2011, Paxton2013, Paxton2015, Paxton2018, Paxton2019, Jermyn2023a} and the {\sc GYRE}\footnote{\url{https://gyre.readthedocs.io/en/}} pulsation code \citep{Townsend2013b, Townsend2018a, Townsend2020a, Goldstein_J_2020a, Sun_M_2023b}.

\begin{figure}
\includegraphics[width=0.99\columnwidth]{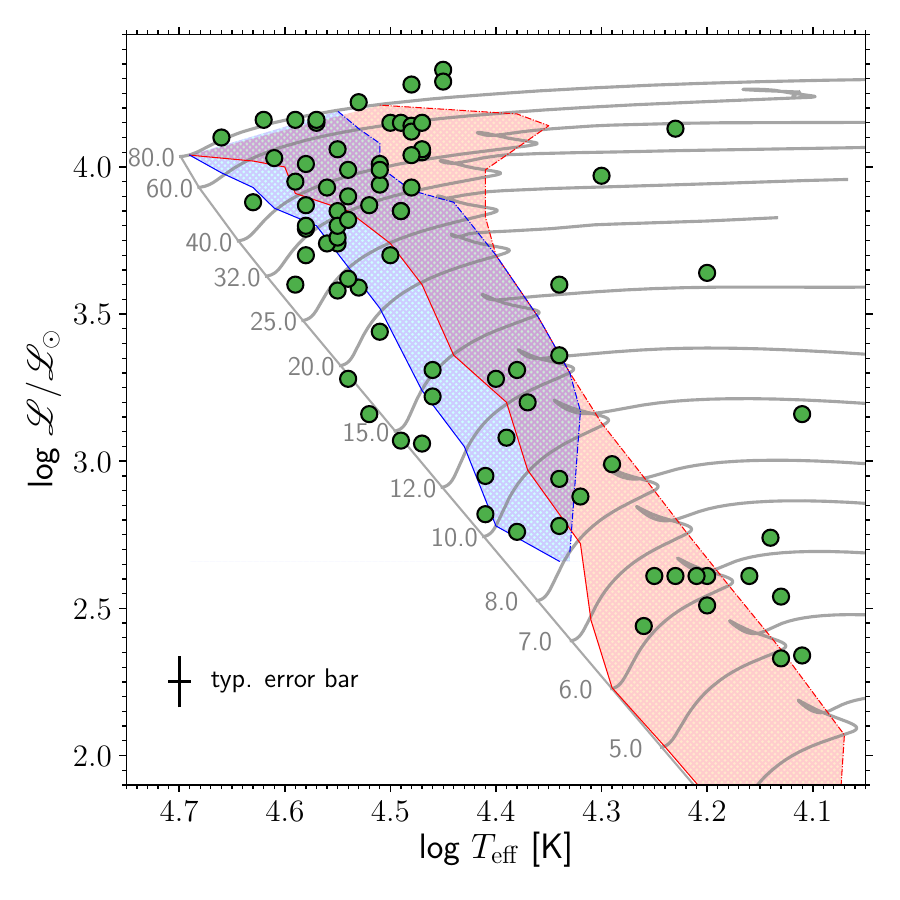}
\caption{The spectroscopic HR diagram of pulsating massive stars identified by \citet{Burssens2020a} in TESS sectors 1-13 that have high-resolution spectroscopy are shown as green circles. The ordinate axis shows spectroscopic luminosity such that $\mathscr{L} := T_{\rm eff}^{4} / g$ \citep{Langer2014a}. The blue and red hatched areas are the theoretical instability regions of pressure and gravity modes, respectively, calculated for non-rotating main-sequence stars, and non-rotating evolutionary tracks with solar metallicity (in units of $M_{\odot}$) are given as grey lines from \citet{Burssens2020a}. A typical error bar for a star's location is also shown in the bottom-left corner. Figure adapted from \citet{Bowman2020c}, his figure 4.}
\label{figure: TESS HRD}
\end{figure}

Early ensemble papers using the first few sectors of TESS data demonstrated that there is a diverse range of variability mechanisms at work in massive stars, which includes coherent pulsations but also binarity and rotational modulation \citep{Bowman2019b, Pedersen2019a, David-Uraz2019b, Southworth2022a}. Moreover, space photometry has demonstrated that $\beta$~Cep stars are relatively common among massive stars and have a large range in pulsation periods and amplitudes (e.g. \citealt{Burssens2019a, Burssens2020a}). With the addition of high-resolution spectroscopy, the pulsations seen in photometry can be placed in the context of a star's location in the HR~diagram, but this also allows some initial constraints on a star's mass and evolutionary stage \citep{Burssens2020a, Gebruers2022a, Serebriakova2023a}. All of the newly discovered pulsating massive stars within cycle 1 (i.e. TESS sectors 1-13) identified by \citet{Burssens2020a} and for which high-resolution optical spectroscopy was available are plotted in the spectroscopic HR~diagram in Fig.~\ref{figure: TESS HRD}, which demonstrates the broad range in mass and age that pulsating massive stars occupy.

The first-light ensemble papers of massive star variability as seen by TESS have demonstrated two important conclusions: (i) the variability fraction is extremely high ($> 95\%$) in photometry of massive stars; and (ii) pulsating stars are often found outside of the theoretical pulsation instability regions, but also vice versa. There are two likely explanations for how non-pulsating stars can be found inside theoretical instability regions, and vice versa. The first is our incomplete understanding of stellar opacities and how accurate pulsation excitation calculations are based on such opacity tables \citep{Seaton1994, Iglesias1996}. The excitation of coherent pulsations in massive stars is expected because of the opacity ($\kappa$) heat-engine mechanism operating at the iron bump in opacity at approximately 200\,000~K \citep{Dziembowski1993e, Dziembowski1993f}. However, the modification of opacity table data in stellar models can drastically alter the parameter space in which coherent pulsations are (not) expected in the HR~diagram \citep{Daszy2009b, Moravveji2016a, Paxton2015, Walczak2015, Walczak2019a, Szewczuk2018a}. On the other hand, the opacity within a star is indeed lower for low-metallicity stars such as those in the SMC and LMC galaxies. This has historically been the explanation of why a dearth of coherent pulsators are found outside of the metal-rich disk of our Milky Way galaxy (see e.g. \citealt{Salmon2012}).

The second important physical influence on the pulsational instability regions of massive stars is rotation. The importance of rotation for pulsations in massive stars is perhaps best illustrated using the sub-group of Be stars (e.g. \citealt{Baade1988c, Huat2009c, Kurtz2015b, Neiner2020b}). These make up around 20\% of B dwarfs and are defined by having emission lines in spectroscopy, although this is known to be a transient phenomenon. They are also generally rapidly rotating stars that seem to lack binary companions making them thought to be the products of binary evolution (see e.g. \citealt{Abt1984b, Bodensteiner2020b}). However, Be stars in binary or multiple systems do exist (see e.g. \citealt{Rivinius2013c}), so their origin remains unclear. Despite their interesting properties for binary evolution theory (e.g. \citealt{Abdul-Masih2020b, Shenar2020b, Bodensteiner2020c, Marchant2023a*}), which is not the subject of this review, the rapid rotation of Be stars has been found to correlate with the incidence of non-radial pulsation using time-series photometry \citep{Naze2020c, Labadie-Bartz2021a, Labadie-Bartz2022a}. Most importantly, rapid rotation breaks the spherical symmetry of a star and spherical geometry of pulsation modes. Such an impact is significant enough to modify the location of instability regions of pulsating massive stars in the HR~diagram \citep{Townsend2005b, Townsend2005e, Szewczuk2017a}. Moreover, the high incidence of non-linear pulsations in Be stars (e.g. \citealt{Kurtz2015b, Neiner2020b}) also demonstrates the importance of including rotation in future hydrodynamical simulations studying wave generation, propagation and damping.


	\subsection{Interior rotation and mixing from asteroseismology}
	\label{subsection: interiors}
	
	Generally speaking, forward asteroseismic modelling of massive stars requires sophisticated modelling techniques, with large grids of evolutionary models containing different prescriptions of physical processes to be constrained, and complex statistical treatment of random and systematic uncertainties (e.g. \citealt{Daszy2010a, Daszy2013b, Aerts2019a, Handler2019a, Salmon2022a}). This is because the theoretical uncertainties -- i.e. correlations and degeneracies among theoretical parameters to be constrained -- are considerably larger than the observed pulsation mode frequencies in the era of space photometry \citep{Bowman2021c}.
	
	In principle, forward asteroseismic modelling in a linear framework makes use of a set of observed pulsation mode frequencies with identified mode geometry, $n$, $\ell$, and $m$, in terms of spherical harmonics. These are then compared using a statistical likelihood, or merit function, to predicted pulsation mode frequencies with the same spherical harmonic geometry from a grid of stellar evolution models. Such a theoretical grid contains various prescriptions for physical processes, such as mixing, which are controlled by free parameters. A minimal set of free parameters to be determined in this way includes a star's mass ($M$), initial hydrogen and metal mass fractions ($X$, $Z$), age ($t$) or, for example, the central hydrogen mass fraction, $X_{\rm c}$ by proxy, and parameterisations for interior mixing profiles such as CBM and radiative envelope mixing, $f_{\rm CBM}(r)$ and $D_{\rm env}(r)$, respectively. The model grid is calculated to cover the required parameter space in the HR~diagram inferred from spectroscopic estimates of effective temperature, $T_{\rm eff}$, and luminosity, $L$, or surface gravity by proxy, $\log\,g$. Moreover, the determination of the interior radial rotation profile, $\Omega(r)$, a priori using a mostly model-independent methodology, such as from rotational splitting within pulsation mode multiplets and/or gravity-mode period spacing patterns, is a major advantage for any subsequent forward asteroseismic modelling (see e.g. \citealt{Aerts2018b, Burssens2023a}).
	
	Historically, a simplistic $\chi^2$ merit function has been used to ascertain the best fitting model in forward asteroseismic modelling of early-type stars (e.g. \citealt{Moravveji2015b, Moravveji2016b}). However, such an approach neglects to include the correlation structure of the theoretical model grid, with a more suitable, robust and superior statistical merit function for forward asteroseismic modelling being the Mahalanobis Distance \citep{Aerts2018b}. Comparisons of using the Mahalanobis Distance and $\chi^2$ as merit functions in forward asteroseismic modelling of early-type stars demonstrated how the former is a significant improvement for probing the posterior parameter distribution \citep{Michielsen2021a}. This is particularly true for forward asteroseismic modelling using gravity modes that are extremely sensitive to the mass and size of the convective core in main-sequence early-type stars \citep{Michielsen2021a}. Building on this framework, detailed forward asteroseismic modelling that robustly handles random and systematic observational and theoretical uncertainties is now possible and facilitates high-precision constraints for pulsation massive stars \citep{Bowman2021c}. In some cases, a 1\% precision on the core mass has been achieved \citep{Michielsen2021a, Bowman2021c}. Nevertheless, asteroseismology across a wide range of masses and ages has concretely shown that core masses are underestimated in current evolution models \citep{Johnston2021b}, meaning that our understanding of CBM is incomplete \citep{Anders_E_2023a}.
	
	\begin{figure}
	\centering
	\includegraphics[width=0.99\columnwidth]{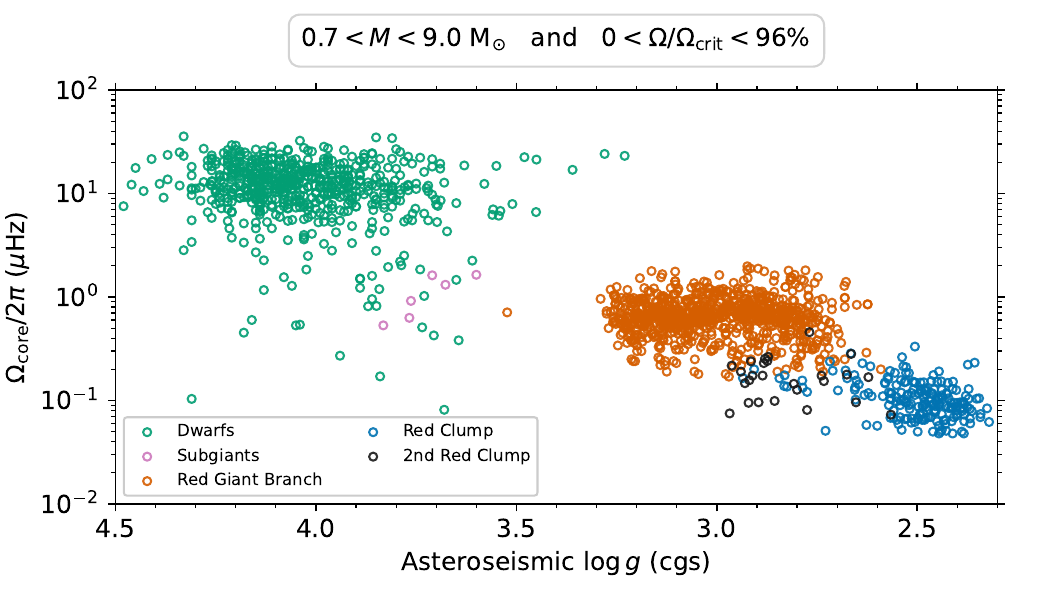}
	\includegraphics[width=0.99\columnwidth]{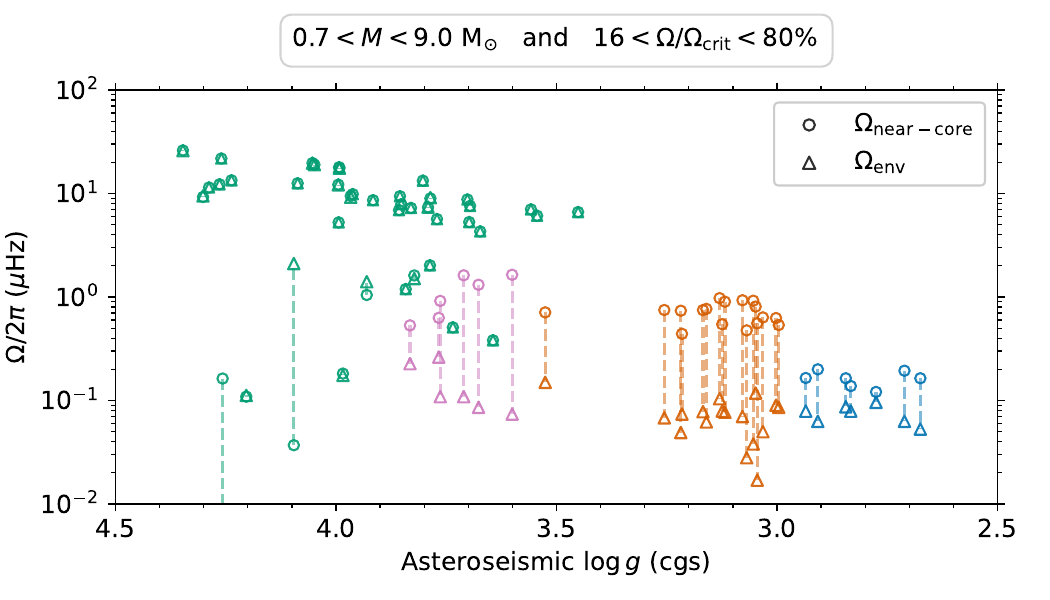}
	\includegraphics[width=0.99\columnwidth]{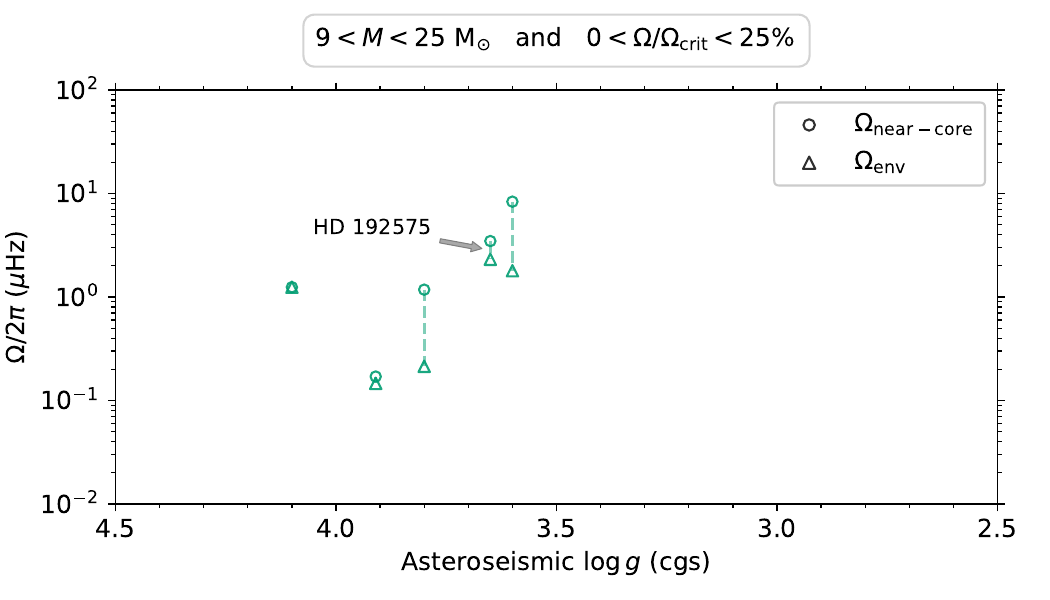}
	\caption{Top Panel: low- and intermediate-mass stars with asteroseismic estimates of near-core rotation rates ($\Omega_{\rm near-core}$) versus asteroseismic $\log\,g$ as a proxy for age. Middle Panel: sub-sample of stars at different evolutionary stages with an asteroseismic estimate of envelope rotation ($\Omega_{\rm env}$) as a function of asteroseismic $\log\,g$. Uncertainties in $\log\,g$ range between 0.01 and 0.30 dex, and uncertainties in rotation rates are smaller than the symbol size. Bottom Panel: same as middle panel but exclusively for massive stars, of which there are very few known in the literature. These figures were produced using asteroseismic near-core rotation measurements for 1210 low- and intermediate-mass stars from the review by \citet{Aerts2019b}, 26 SPB stars from \citet{Pedersen2021a}, 611 $\gamma$~Dor stars from \citet{Li_G_2020a}, the few available massive stars from the review by \citet{Bowman2020c}, and the new $\beta$~Cep star HD\,192575 \citep{Burssens2023a}. The colours of all symbols are consistent across all panels and denote the evolutionary stage of the stars. Note that individual error bars for the massive stars are not plotted for clarity because they vary considerably within the subsample.}
	\label{figure: interior rotation}
	\end{figure}
	
	One of the biggest success stories of asteroseismology has been the determination of interior rotation rates of stars across all stages of stellar evolution. In cases of stars with rotationally split pulsation multiplets, demonstrating non-rigid radial rotation rates can be achieved using a largely model-independent method \citep{Aerts2003d, Kurtz2014, Saio2015b}. However the exact rotation profile is model dependent \citep{Salmon2022b, Burssens2023a}. Given that asteroseismology is an ever advancing field of stellar astrophysics, reviews of interior rotation rates become out of date rather soon after publication. Nevertheless, the literature rotation rates for 1210 dwarfs and giant stars \citep{Aerts2019b, Aerts2021a}, and additionally 26 slowly pulsating B (SPB) stars \citep{Pedersen2021a}, 611 $\gamma$~Dor stars \citep{Li_G_2020a}, the few available massive stars from the review by \citet{Bowman2020c}, and the new $\beta$~Cep star HD\,192575 \citep{Burssens2023a} are summarised in Fig.~\ref{figure: interior rotation}. 
	
	In the top panel of Fig.~\ref{figure: interior rotation}, the near-core rotation rates for low- and intermediate-mass (i.e. $0.7 \leq M \leq 9$~M$_{\odot}$) stars are shown as a function of their so-called asteroseismic $\log\,g$, which is calculated from the asteroseismic masses and radii from forward asteroseismic modelling, and acts as a proxy for a star's age or evolutionary stage. In the middle panel, the stars that also have an envelope rotation rate measured from asteroseismology or rotational modulation are shown as triangle symbols in addition to their corresponding circle symbols denoting the near-core rotation rate. The sample in Fig.~\ref{figure: interior rotation} demonstrates the difference in differential rotation as a function of evolutionary stage. Most interesting is that core-hydrogen burning main-sequence stars and core-helium burning red clump stars have the smallest amount of differential rotation on average. This implies that the presence of a convective core has a direct impact on the efficiency of angular momentum transport within stars across their evolution \citep{Aerts2019b, Aerts2021a}. 
	
	In the bottom panel of Fig.~\ref{figure: interior rotation}, the massive stars (i.e. $M \geq 9$~M$_{\odot}$) with asteroseismic near-core and envelope rotation rates have been purposefully separated to demonstrate that not many exist in the literature. It is important to emphasise that, with the exception of HD~192575, there is a large amount of uncertainty for the rotation rates for the sub-sample of massive stars. This is because the other massive stars reported in the literature typically have only an upper or lower limit on the core and envelope rotation rates, which in some cases was derived from interpreting only a single or a few fiducial structure models (see e.g. \citealt{Dupret2004b}). The exception is HD~192575 for which full error propagation was performed to derive its rotation profile. In the next section, I focus on the interesting and novel case of the $\beta$~Cep star HD~192575, for which its rotation and mixing profiles have been measured to high precision thanks to excellent TESS mission data \citep{Burssens2023a}.


	\subsection{The case of HD~192575}
	\label{subsection: HD192575}
	
	A valuable new addition to the sparsely populated bottom panel of Fig.~\ref{figure: interior rotation} is the $\beta$~Cep star HD~192575 \citep{Burssens2023a}. This massive star was first discovered to pulsate in many low-radial pressure and gravity modes thanks to TESS cycle 2 data. Even though it is a relatively bright star, it was not subject to an in-depth photometric study prior to the TESS mission. A quantitative spectroscopic analysis using multi-epoch high-resolution {\sc HERMES} spectroscopy \citep{Raskin2011} showed no evidence for a binary companion, and allowed high-precision atmospheric parameters of $T_{\rm eff} = 23\,900 \pm 900$~K, $\log\,g = 3.65 \pm 0.15$, $v\,\sin\,i = 27^{+6}_{-8}$~km\,s$^{-1}$, and solar metallicity to be determined. Moreover, based on the Gaia DR3 parallax \citep{Gaia2016a}, a luminosity of $L/L_{\odot} =  4.30 \pm 0.07$ was derived. These complementary constraints were used to hugely delimit the parameter space within the HR~diagram for calculating structure models and corresponding pulsation mode frequencies with the {\sc MESA} and {\sc GYRE} codes \citep{Burssens2023a}.
	
\begin{figure}
	\centering
	\includegraphics[width=0.99\columnwidth]{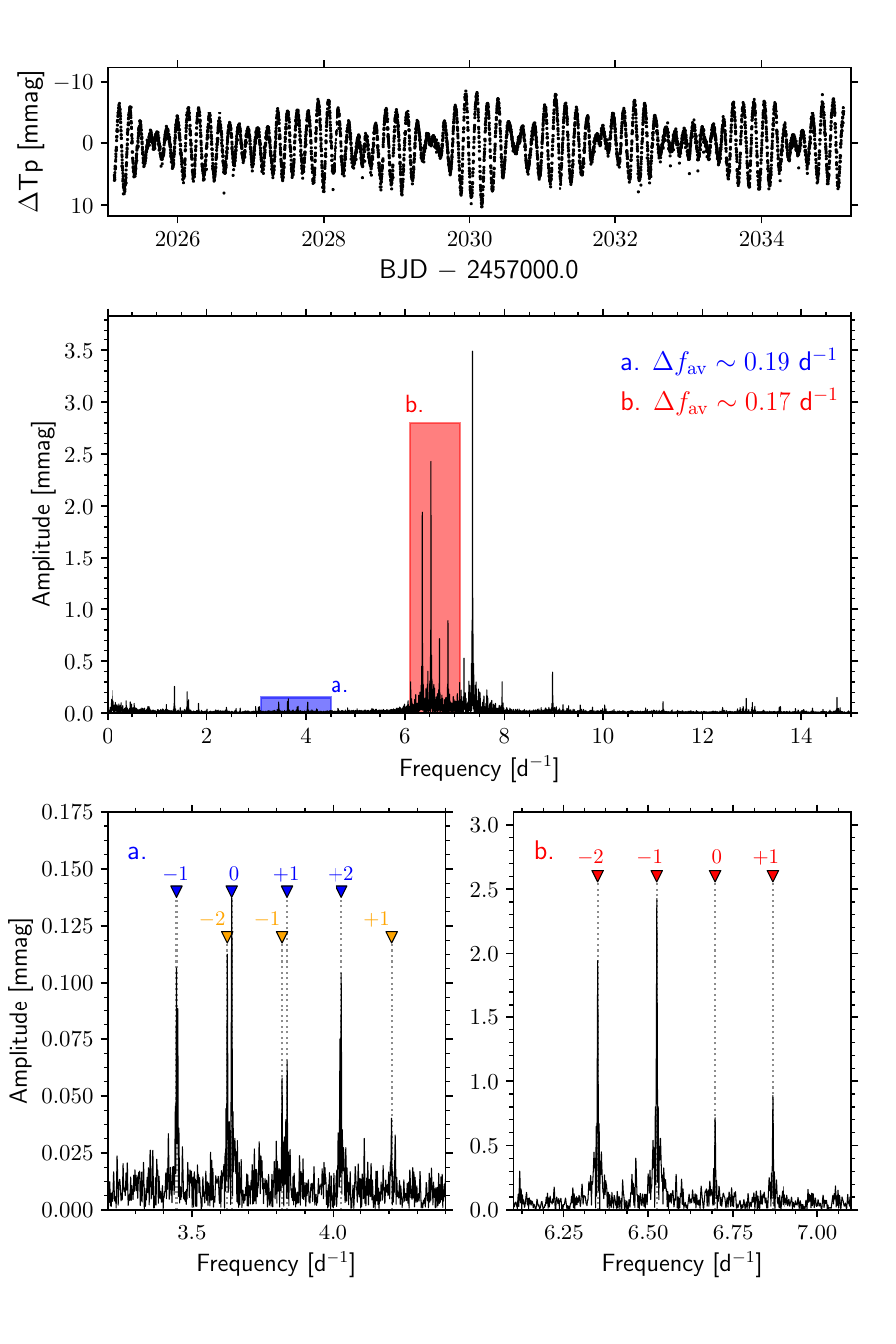}
	\caption{Top panel: 10-d section of the cycle 2 TESS light curve for the $\beta$~Cep star HD~192575. Middle panel: frequency spectrum of the full 1-year cycle 2 TESS light curve, with the average rotational splitting of the utilised rotational multiplets labelled and colour-coded. Bottom panels: zoom-in of the same rotational multiplets and the corresponding azimuthal order, $m$, of each component pulsation mode as assigned by \citet{Burssens2023a} in the best-fitting solution from the forward asteroseismic modelling. Figure adapted from \citet{Burssens2023a}, their figure 1.}
	\label{figure: HD192575}
	\end{figure}

	Frequency analysis of the 1-yr cycle 2 TESS light curve revealed several rotational multiplets, which are pulsation modes of the same radial order and angular degree but different azimuthal order, and have $2\ell + 1$ $m$-components that are separated by the rotation rate within each mode-specific pulsation cavity. An illustration of the rotational multiplets in HD~192575 from a frequency analysis of the TESS cycle 2 data is shown in Fig.~\ref{figure: HD192575}. The differences in the frequency spacings of different rotational multiplets allowed \citet{Burssens2023a} to extract a core-to-surface radial rotation profile and conclusively exclude rigid rotation. Based on different assumed prescriptions for the shape of the shear layer between the core and surface, the amount of differential rotation was found to differ, such that core was inferred to be rotating between 1.4 and 6.3 times faster than the envelope \citep{Burssens2023a}. This is the first time that such an analysis was performed. However, the most likely scenario is that the chemical composition gradient left behind by the receding convective core during the main sequence is the true shear layer, which yielded a core-to-surface rotation ratio of $1.49^{+0.56}_{-0.33}$ \citep{Burssens2023a}. This result has been subsequently shown to be consistent with 2D stellar structure models using the ESTER code \citep{EspinosaLara2011, EspinosaLara2013, Rieutord2016c} by \citet{Mombarg2023b}. Therefore, the asteroseismic modelling of HD~192575 serves as an extremely useful anchor point for testing angular momentum transport within the massive star regime (see e.g. \citealt{Salmon2022b, Mombarg2023b}).
	
	In addition to its interior rotation profile, \citet{Burssens2023a} performed forward asteroseismic modelling to constrain its mass, radius, age, and core mass to unprecedentedly high precision for a massive star. Ultimately the fractional uncertainties for mass, age and core mass were better than about 15\%, which is remarkable given the complexity of the physics that was included in the input grid of evolution models. The derived parameters from only 1~yr of TESS data and their $2\sigma$ confidence intervals obtained through forward asteroseismic modelling were a mass of $M = 12.0 \pm 1.5$~M$_{\odot}$, a radius of $R = 9.1^{+0.8}_{-1.7}$~R$_{\odot}$, an age of $t = 17.0^{+4.7}_{-5.4}$~Myr, and a core mass of $M_{\rm cc} = 2.9^{+0.5}_{-0.8}$~M$_{\odot}$ \citep{Burssens2023a}.
	
	An important detail is that the work of \citet{Burssens2023a} represents the first forward asteroseismic modelling study of a massive star to incorporate sources of observational and theoretical uncertainties, and implement using the Mahalanobis distance as a statistical merit function, thus appropriately handling the stellar model grid parameter correlations and degeneracies. As discussed briefly by \citet{Aerts2023c*}, additional TESS data of HD~192575 during cycle 4 have recently become available, which when combined with the previous cycle 2 data provide improved frequency resolution and precision for forward asteroseismic modelling. With these higher quality data a more complete assignment of $m$ values for components of the $\ell =2$ rotational multiplets is possible (cf. Fig.~\ref{figure: HD192575}). With these updated $m$ values, a slightly more evolved star is found from the forward asteroseismic modelling \citep{Vanlaer2024a**}, but the derived mass, age, radius and core mass remain within the $2\sigma$ confidence intervals reported by \citet{Burssens2023a}. Interestingly, there are very slight asymmetries detectable in the $\ell=2$ multiplets, which can also be exploited to constrain the interior magnetic field and geometry within the deep interior of HD~192575 \citep{Vanlaer2024a**}.


	\subsection{Magnetoasteroseismology}
	\label{subsection: magneto}
	
	All of the asteroseismic modelling discussed in this review up until this point has been based on the use of non-magnetic stellar evolution models. Of course, magnetism is an important aspect of stellar structure and evolution since it can drastically change the interior rotation, mixing and angular momentum transport \citep{Maeder2005b, Keszthelyi2019, Keszthelyi2020a, Keszthelyi2021a}. However, until recently all constraints on the strength and geometry of magnetic fields for massive stars were limited to the stellar surface. Even though the detection of rotational modulation in a massive star's light curve suggests the presence of a magnetic field, this remains an indirect and limited detection method (e.g. \citealt{Buysschaert2018b, David-Uraz2019b, David-Uraz2021b}). Using a more targeted and suitable strategy, large campaigns using spectropolarimetry have demonstrated that approximately 10\% of massive dwarf stars have strong large-scale magnetic fields at their surfaces \citep{Wade2016a, Grunhut2017, Shultz2019d}. Such fields are presumed to extend much deeper than the stellar photosphere, and thought to be created during star formation \citep{Mestel1999a, Neiner2015d}, or formed during binary mergers \citep{Schneider_F_2019a, Schneider_F_2020a}.
	
	There has been a great deal of theoretical work studying the impact of strong magnetic fields for gravity-mode pulsators. For example, \citet{Prat2019a} and \citet{Prat2020a} investigated the topology and obliquity of a strong magnetic field in the deep interior of a fiducial intermediate-mass main-sequence star, and specifically how it changes the morphology of gravity-mode period spacing patterns to take on a ``saw-tooth'' pattern. Later, \citet{VanBeeck2020a} applied the same theoretical framework to various structure models for a range of masses and ages on the main-sequence and demonstrated that more evolved stars have the greater potential for detecting deep magnetic fields using pulsations. However, including a strong large-scale interior magnetic field using an a posteriori perturbative approach has limitations when comparing theoretical predictions to observations.
	
	One particular magnetic star that has received much attention in the literature is the B3.5\,V star HD~43317 which has a predominantly dipolar field of strength $B_{\rm p} = 1312 \pm 332$~G detected and characterised using spectropolarimetry \citep{Briquet2013, Buysschaert2017b}. HD~43317 is also a known pulsator with dozens of high-frequency gravity mode frequencies detected in its 150-d CoRoT light curve \citep{Papics2012a, Buysschaert2018c}. Interestingly, unlike other magnetic early-type dwarfs, HD~43317 is a relatively rapid rotator with a rotation period of $0.897673 \pm 0.000004$~d \citep{Buysschaert2017b}. Also, it has no signatures of being part of a binary system making HD~43317 a single, rapidly rotating, pulsating magnetic star. Although these characteristics do not make it unique, they do make it rare among the wider population of B dwarfs. 
	
	Thanks to space photometry and dedicated spectropolarimetry campaigns to charactertise the geometry of magnetic fields, there is a growing number of magnetic pulsating stars \citep{Shibahashi2000b, Briquet2012, Handler2012a, Aerts2014a, Wade2020a}. However, only two confirmed pulsating magnetic early-type stars have undergone forward asteroseismic modelling: the $\beta$~Cep star V2052~Ophiuchi \citep{Briquet2012} and the SPB star HD~43317 \citep{Buysschaert2018c}. In both cases, a negligible amount of CBM was required to best fit the observed pulsation mode frequencies. This led to the tentative conclusion, although it is based on a very small sample size, that the presence of a magnetic field in the near core region is inhibiting mixing at the interface of the convective core and the radiative envelope \citep{Briquet2012, Buysschaert2018c}. Indeed, a magnetic field in the near-core region could be supported by either a global magnetic field as detected at the surface that extends much deeper within a star and/or a convective core dynamo with magnetic field lines that penetrate outside the convective core (see e.g. \citealt{Augustson2016a, Augustson2019a, Augustson2020c}).
	
	The study of pulsating magnetic stars naturally leads to defining the field of magneto-asteroseismology, with an important proof-of-concept of its application being the recent analysis of HD~43317 by \citet{Lecoanet2022a}. In this work, the best-fitting non-magnetic structure models of the pulsating magnetic star HD~43317 from the forward asteroseismic modelling by \citet{Buysschaert2018c} were used as input into 3D rotating magnetohydrodynamical calculations with the {\sc Dedalus}\footnote{\url{https://dedalus-project.org}} code \citep{Burns2020}. Some assumptions inevitably had to be made, which included: (i) the utilised structure models based on non-magnetic evolution models were accurate; (ii) the measured rotation period at the surface from spectropolarimetry was used to assume a rigid interior rotation profile; and (iii) the measured surface magnetic field strength was used as an anchor point to assume a purely dipolar magnetic field geometry within the star's interior. Given all these assumptions, the 3D magnetohydrodynamical calculations allowed wave-wave interactions to be studied, and specifically the critical magnetic field strength required to suppress a standing gravity wave (i.e. gravity mode) and convert it into an Alfven wave. Based on a comparison of the observed pulsation modes identified by \citet{Buysschaert2018c} in the CoRoT light curve of HD~43317, \citet{Lecoanet2022a} were able to place an upper limit of the magnetic field strength of approximately $5 \times 10^5$~G in the near-core region. Such a value is consistent with the expected magnetic field strength in a star with a convective core dynamo supported by a large-scale global magnetic field of fossil origin \citep{Augustson2016a}. Therefore, not only pulsation frequencies but also their observed amplitudes provide invaluable insight of the physics of stellar interiors and facilitate magneto-asteroseismology.


\section{Upcoming instrumentation and projects}
\label{section: next gen}

It is not only TESS that is currently driving advances in our understanding of massive star interiors. For example, the European Space Agency (ESA) Gaia mission \citep{Gaia2016a} is providing astrometry and light curves of a vast number of variable stars, including potentially hundreds of thousands of massive stars \citep{Gaia2019a, Gaia2023e}. The all-sky Gaia survey provides a truly huge data set for identifying new pulsating massive stars to follow-up. There remains much work to do to fully exploit the natural synergies of Gaia with asteroseismology (e.g. \citealt{Aerts2023a}).

Spectroscopic campaigns unprecedented in terms of size and scope for studying massive stars are also currently underway, which are necessary to supplement forward asteroseismic modelling. One such example is the XShootU collaboration\footnote{\url{https://massivestars.org/xshootu/}}, which is assembling ground-based European Southern Observatory (ESO) XShooter optical and near-infrared spectroscopy and NASA Hubble Space telescope (HST) UV spectroscopy to provide the largest sample of massive stars at low metallicity to date \citep{Vink2023a}. Large programs with ESO instrumentation targeting massive stars in the galaxy and beyond are also assembling extremely useful spectroscopy for placing stars in the HR~diagram and extracting their abundances, rotational and macroturbulent broadening velocities \citep{Gebruers2022a, Serebriakova2023a}, with the number of targets expected to grow much larger in the coming years. An example of a large sample of high-resolution optical spectra of galactic massive stars is the {\sc IACOB}\footnote{\url{http://research.iac.es/proyecto/iacob}} project, which continues to act as a robust data set for studying massive stars (e.g. \citealt{deBurgos2023a, deBurgos2023b*}).


	\subsection{The CubeSpec space mission}
	\label{subsection: cubespec}
	
\begin{figure}
	\centering
	\includegraphics[width=0.99\columnwidth]{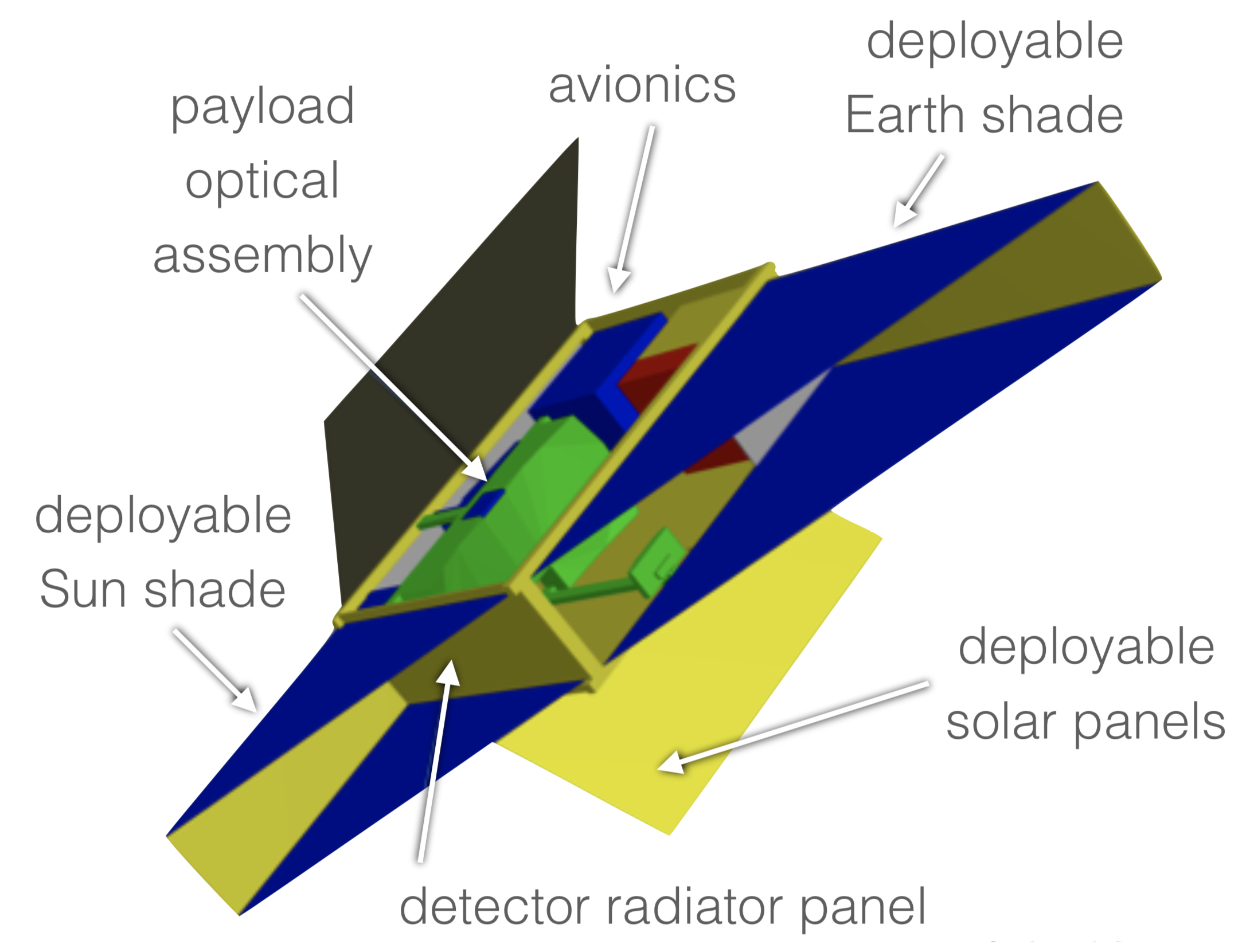}
	\caption{Schematic of the platform layout of the CubeSpec mission with important components labelled.}
	\label{figure: cubespec}
	\end{figure}

	A particularly exciting project specifically tailored to empower massive star asteroseismology in the coming decade is the CubeSpec space mission \citep{Bowman2022a}, which is being designed and built by KU Leuven, Belgium\footnote{\url{https://fys.kuleuven.be/ster/instruments/cubespec}}. The CubeSpec mission is an in-orbit demonstration of a 6-unit cubesat within the European Space Agency's GSTP technology programme, and is funded by the Belgian federal science policy office (BELSPO). A schematic of the platform layout of CubeSpec is shown in Fig.~\ref{figure: cubespec}. The primary engineering goal of the CubeSpec mission is to validate as a proof-of-concept the feasibility of high spectral resolution spectroscopy from a cubesat \citep{Raskin2018}. 
	
	On the other hand, the primary science objective of CubeSpec is an in-orbit demonstration of massive star asteroseismology using long-term, high-cadence, and high-resolution time-series optical spectroscopy. These data will facilitate the study of spectral line profile variability (LPV) caused by pulsations, such that the geometry of pulsation modes in terms of spherical harmonics can be identified \citep{Bowman2022a}. Such a method is highly complementary to the methods of pulsation mode identification based on time-series photometry, and had huge success using ground-based telescopes in the past (see e.g. \citealt{Aerts1992b, Fullerton1996, Telting1998c, Uytterhoeven2004b, Uytterhoeven2005a}). However, ground-based telescopes are inevitably hampered by daily aliasing and multi-site coordination to achieve a maximal duty cycle is expensive, which is a negligible issue for the space-based CubeSpec mission. 
	
	The original design concept of CubeSpec is discussed in \citet{Raskin2018}. Updated engineering and scientific specifications as well as a discussion of expected performance and identification of an initial taregt list of pulsating massive stars are provided by \citet{Bowman2022a}. The prospect of combining exquisite time-series spectroscopy from CubeSpec with concomitant time-series photometry from TESS is truly a novel endeavour and will lead to exciting new results for massive stars.
	
	\subsection{PLATO}
	\label{subsection: plato}

	In the not too distant future, the ESA PLATO mission \citep{Rauer2014} will facilitate major advancements across a range of topics in astrophysics. Selected as ESA's M3 mission in the 2015-2023 Cosmic Vision program and an expected launch date at the end of 2026, PLATO has the primary science goal to detect and accurately characterise large numbers of exoplanets using the transit method in time-series photometry. To achieve this goal, the PLATO platform consists of 26 white-light cameras each with a field of view of 1037 deg$^2$, which partially overlap to create a combined field of view of 2132 deg$^2$ \citep{Rauer2014}. Such a large area to be observed with high photometric precision means that light curves spanning years will be assembled for hundreds of thousands of pulsating stars. Unlike the Kepler mission, PLATO is not purposefully avoiding massive stars, nor is it avoiding (very) bright stars allowing it to maximise the synergy with ground-based spectrographs, which are typically brightness-limited for the purposes of radial velocity follow-up.
	
	The location of PLATO's long pointing fields of view on sky will be decided no later than two years before launch. Nevertheless, with such a large area many more massive stars will be included compared to previous missions, such as CoRoT, Kepler and TESS. Moreover, the duration of PLATO's high-precision light curves in its long pointing fields of view will far exceed the 150-d length of a CoRoT campaign and an individual 28-d TESS sector, and be comparable to the 4-yr length of the Kepler mission. Asteroseismic results based on Kepler mission data of early-type stars have demonstrated that years-long light curves are necessary to resolve individual pulsation mode frequencies of coherent pulsators \citep{Bowman2016a, Bowman2021c}. In this regard, PLATO data will have a huge advantage compared to the individual 28-d TESS sectors for many pulsating massive stars in the sky.
	
	With a deluge of high-quality light curves for so many massive stars already available from TESS and expected soon from PLATO, which include not only pulsators but also eclipsing binaries and stars with rotational modulation (etc), the asteroseismic community are already moving from a star-by-star approach to more ensemble-driven analysis techniques. Even the first task of classification of different types of variability in light curves is a computationally expensive and time-consuming task to perform manually. Hence machine-learning techniques are becoming increasingly popular because of their efficiency and overall high success rates (see e.g. \citealt{Armstrong2016b, Audenaert2021a, Audenaert2022a, Barbara2022a}). Similar efforts to maximise the asteroseismic output of massive stars from PLATO are being prepared as part of an international asteroseismic consortium\footnote{\url{https://tasoc.dk}} and will be needed once the data begin to arrive.


\section{Concluding remarks}
\label{section: conclusion}

In this invited review, I have discussed the exciting results that have been made possible because of space-based time-series photometry missions, paying particular attention to the new results since the launch of the NASA TESS mission \citep{Ricker2015}. In this way, the current review is complementary to the pre-TESS review of massive star asteroseismology by \citet{Bowman2020c}. The all-sky approach taken by the TESS mission has demonstrated that diverse variability caused by pulsations, binarity, rotation and magnetic fields are widespread across the upper part of the HR~diagram (see e.g. \citealt{Burssens2020a}), which has opened up many new avenues for stellar astrophysics research.

An important new discovery continuing to fuel debate in the literature is the near-ubiquitous detection of SLF variability in massive star light curves \citep{Bowman2019b, Bowman2020b}. Compelling mechanisms proposed in the literature include gravity waves excited by turbulent core convection \citep{Rogers2013b, Rogers2015, Rogers2017c, Edelmann2019a, Ratnasingam2019a, Ratnasingam2020a, Ratnasingam2023a, Horst2020a, Varghese2023a, Herwig2023a, Thompson_W_2023a*}, the dynamics of turbulent massive star envelopes \citep{Schultz2022a, Jiang_Y_2023a}, and optically thick line-driven winds in evolved massive stars \citep{Krticka2018e, Krticka2021b}. The plausibility of some mechanisms is debated since they depend on the specific numerical setup of 2D or 3D hydrodynamical simulations \citep{Lecoanet2019a, Lecoanet2021a, LeSaux2023a, Anders_E_2023b, Herwig2023a, Thompson_W_2023a*}. However, it is critical to emphasise that these mechanisms are not mutually exclusive in explaining SLF variability, and that there is growing body of observational evidence for how the morphology of SLF variability in both the time and frequency domains depends on a star's mass, age, rotation rate and metallicity \citep{Bowman2020b, Bowman2022b}.

With high-precision time-series photometry from space missions, asteroseismology has been able to place tight constraints on physical processes that are currently controlled by free parameters in state-of-the-art evolution models. For example, precise masses, radii, ages and interior rotation profiles are available for thousands of low-mass stars \citep{Aerts2021a}. However, insight of massive stars is lacking behind in numbers because of their intrinsic rarity but also because of their complexity, which includes their rapid rotation and possible presence of a companion star to name a few. The first forward asteroseismic modelling that appropriately takes sources of observational and theoretical uncertainties into account was recently performed for the $\beta$~Cep star HD~192575 by \citet{Burssens2023a}. This work yielded fractional uncertainties of less than 15\% for mass, age and core mass of a supernovae progenitor \citep{Burssens2023a}, which is unprecedented and thanks to the combination of TESS mission photometry, Gaia astrometry and high-resolution spectroscopy. Moreover, a rigid interior rotation profile is conclusively ruled out for HD~192575, which has already proven an invaluable result for improving 2D stellar evolution models and prescriptions for angular momentum transport \citep{Mombarg2023b}.

In addition to constraints on interior rotation and mixing, the emergence of magneto-asteroseismology has allowed the strength and geometry of interior magnetic fields to be probed. For example, \citet{Lecoanet2022a} infer a magnetic field strength of $5 \times 10^5$~G in the near-core region of the rapidly rotating SPB star HD~43317 \citep{Buysschaert2018c}. With longer and longer light curves becoming available for pulsating massive stars and larger samples of stars targeted with spectropolarimetry, additional magneto-asteroseismic studies are expected soon.

In the near future, building on the successes of CoRoT, Kepler and now TESS, the field of asteroseismology is moving towards a more holistic approach to understanding the physics of stellar interiors rather than focusing on a niche part of it for individual stars. These efforts will be maximised by ESA's upcoming PLATO mission providing high-precision light curves for potentially tens of thousands of pulsating massive stars \citep{Rauer2014}. By including complementary observables within a forward asteroseismic modelling framework, such as parallaxes to derive distances from Gaia \citep{Gaia2016a}, magnetic field strengths from spectropolarimetry, and constraints on the spherical harmonic geometry of additional pulsation modes from spectral LPV with the CubeSpec mission \citep{Bowman2022a}, asteroseismology is expected to reach new heights in the decades to come. This is ultimately the goal of the ongoing SYMPHONY\footnote{\url{https://research.ncl.ac.uk/symphony/}} project (P.I. Bowman), which has a holistic approach to studying different aspects of massive stars with asteroseismology as its main tool.


\backmatter

\bmhead{Acknowledgments}

DMB gratefully acknowledges the financial and logistical support in the form of a senior postdoctoral fellowship from the Research Foundation Flanders (FWO; grant number [1286521N]), the Engineering and Physical Sciences Research Council (EPSRC) of UK Research and Innovation (UKRI) in the form of a Frontier Research grant under the UK government’s ERC Horizon Europe funding guarantee (SYMPHONY; grant number [EP/Y031059/1]), and a Royal Society University Research Fellowship (grant number: URF{\textbackslash}R1{\textbackslash}231631). The author thanks the CubeSpec team and the European Space Agency (ESA) and the Belgian Federal Science Policy Office (BELSPO) for their support in the framework of the CubeSat IOD GSTP Programme. DMB is appreciative to the awards committee of the Royal Astronomical Society (RAS) for awarding him the 2023 George Darwin Lectureship, which was the main motivation to write this review as part of the Springer Nature 2023 Astronomy Prize Awardees Collection. Finally, DMB is ever grateful to the continued support of his colleagues, in particular Prof. Donald Kurtz and Prof. Conny Aerts for their inspirational motivation, the rewarding experience of being PhD advisor to Dr. Siemen Burssens, and the academic hosts of his research: Newcastle University, UK and KU Leuven, Belgium.


\bmhead{Software availability}

This review has made use of the {\sc python} library for publication quality graphics ({\sc matplotlib}; \citealt{Matplotlib_2007}), {\sc seaborn} \citep{Seaborn_2021}, and {\sc numpy} \citep{Numpy_2020}.

\bmhead{Data availability}

The TESS data presented in this paper were obtained from the Mikulski Archive for Space Telescopes (MAST) at the Space Telescope Science Institute (STScI), which is operated by the Association of Universities for Research in Astronomy, Inc., under NASA contract NAS5-26555. Support to MAST for these data is provided by the NASA Office of Space Science via grant NAG5-7584 and by other grants and contracts. Funding for the TESS mission was provided by the NASA Explorer Program. The CoRoT space mission was developed and operated by the French space agency CNES, with participation of ESA's RSSD and Science Programmes, Austria, Belgium, Brazil, Germany, and Spain.

For the purpose of open access, a Creative Commons Attribution (CC BY) licence has been applied to this author accepted version of this manuscript and stored publicly on the \url{https://arxiv.org} repository.


\bibliography{/Users/dominic/Documents/RESEARCH/Bibliography/master_bib.bib}

\begin{thebibliography}{218}
\providecommand{\natexlab}[1]{#1}
\providecommand{\url}[1]{{#1}}
\providecommand{\urlprefix}{URL }
\providecommand{\doi}[1]{\url{https://doi.org/#1}}
\providecommand{\eprint}[2][]{\url{#2}}
 \bibcommenthead

\bibitem[{{Abdul-Masih} et~al(2020){Abdul-Masih}, {Banyard}, {Bodensteiner},
  {Bordier}, {Bowman}, {Dsilva}, {Fabry}, {Hawcroft}, {Mahy}, {Marchant},
  {Raskin}, {Reggiani}, {Shenar}, {Tkachenko}, {Van Winckel}, {Vermeylen}, and
  {Sana}}]{Abdul-Masih2020b}
{Abdul-Masih} M, {Banyard} G, {Bodensteiner} J, et~al (2020) {On the signature
  of a 70-solar-mass black hole in LB-1}. \nat 580(7805):E11--E15.
  \doi{10.1038/s41586-020-2216-x},
  {\href{https://arxiv.org/abs/1912.04092}{{arXiv:1912.04092}}} {[astro-ph.SR]}

\bibitem[{{Abt} and {Cardona}(1984)}]{Abt1984b}
{Abt} HA, {Cardona} O (1984) {Be stars in binaries}. \apj 285:190--194.
  \doi{10.1086/162490}

\bibitem[{{Aerts}(2021)}]{Aerts2021a}
{Aerts} C (2021) {Probing the interior physics of stars through
  asteroseismology}. Reviews of Modern Physics 93(1):015001.
  \doi{10.1103/RevModPhys.93.015001}

\bibitem[{{Aerts} and {Rogers}(2015)}]{Aerts2015c}
{Aerts} C, {Rogers} TM (2015) {Observational Signatures of Convectively Driven
  Waves in Massive Stars}. \apjl 806:L33. \doi{10.1088/2041-8205/806/2/L33},
  {\href{https://arxiv.org/abs/1505.06648}{{arXiv:1505.06648}}} {[astro-ph.SR]}

\bibitem[{{Aerts} and {Tkachenko}(2023)}]{Aerts2023c*}
{Aerts} C, {Tkachenko} A (2023) {Asteroseismic Modelling of Fast Rotators and
  its Opportunities for Astrophysics}. arXiv e-prints arXiv:2311.08453.
  \doi{10.48550/arXiv.2311.08453},
  {\href{https://arxiv.org/abs/2311.08453}{{arXiv:2311.08453}}} {[astro-ph.SR]}

\bibitem[{{Aerts} et~al(1992){Aerts}, {de Pauw}, and {Waelkens}}]{Aerts1992b}
{Aerts} C, {de Pauw} M, {Waelkens} C (1992) {Mode identification of pulsating
  stars from line profile variations with the moment method. an example : the
  beta Cephei star delta Ceti.} \aap 266:294--306

\bibitem[{{Aerts} et~al(2003){Aerts}, {Thoul}, {Daszy{\'n}ska}, {Scuflaire},
  {Waelkens}, {Dupret}, {Niemczura}, and {Noels}}]{Aerts2003d}
{Aerts} C, {Thoul} A, {Daszy{\'n}ska} J, et~al (2003) {Asteroseismology of HD
  129929: Core Overshooting and Nonrigid Rotation}. Science 300:1926--1928.
  \doi{10.1126/science.1084993}

\bibitem[{{Aerts} et~al(2009){Aerts}, {Puls}, {Godart}, and
  {Dupret}}]{Aerts2009b}
{Aerts} C, {Puls} J, {Godart} M, et~al (2009) {Collective pulsational velocity
  broadening due to gravity modes as a physical explanation for macroturbulence
  in hot massive stars}. \aap 508:409--419. \doi{10.1051/0004-6361/200810471},
  {\href{https://arxiv.org/abs/0909.3585}{{arXiv:0909.3585}}} {[astro-ph.SR]}

\bibitem[{{Aerts} et~al(2010){Aerts}, {Christensen-Dalsgaard}, and
  {Kurtz}}]{ASTERO_BOOK}
{Aerts} C, {Christensen-Dalsgaard} J, {Kurtz} DW (2010) Asteroseismology.
  Springer

\bibitem[{{Aerts} et~al(2014){Aerts}, {Molenberghs}, {Kenward}, and
  {Neiner}}]{Aerts2014a}
{Aerts} C, {Molenberghs} G, {Kenward} MG, et~al (2014) {The Surface Nitrogen
  Abundance of a Massive Star in Relation to its Oscillations, Rotation, and
  Magnetic Field}. \apj 781:88. \doi{10.1088/0004-637X/781/2/88},
  {\href{https://arxiv.org/abs/1312.4144}{{arXiv:1312.4144}}} {[astro-ph.SR]}

\bibitem[{{Aerts} et~al(2017){Aerts}, {S{\'{\i}}mon-D{\'{\i}}az}, {Bloemen},
  {Debosscher}, {P{\'a}pics}, {Bryson}, {Still}, {Moravveji}, {Williamson},
  {Grundahl}, {Fredslund Andersen}, {Antoci}, {Pall{\'e}},
  {Christensen-Dalsgaard}, and {Rogers}}]{Aerts2017a}
{Aerts} C, {S{\'{\i}}mon-D{\'{\i}}az} S, {Bloemen} S, et~al (2017) {Kepler
  sheds new and unprecedented light on the variability of a blue supergiant:
  Gravity waves in the O9.5Iab star HD 188209}. \aap 602:A32.
  \doi{10.1051/0004-6361/201730571},
  {\href{https://arxiv.org/abs/1703.01514}{{arXiv:1703.01514}}} {[astro-ph.SR]}

\bibitem[{{Aerts} et~al(2018){Aerts}, {Molenberghs}, {Michielsen}, {Pedersen},
  {Bj{\"o}rklund}, {Johnston}, {Mombarg}, {Bowman}, {Buysschaert},
  {P{\'a}pics}, {Sekaran}, {Sundqvist}, {Tkachenko}, {Truyaert}, {Van Reeth},
  and {Vermeyen}}]{Aerts2018b}
{Aerts} C, {Molenberghs} G, {Michielsen} M, et~al (2018) {Forward Asteroseismic
  Modeling of Stars with a Convective Core from Gravity-mode Oscillations:
  Parameter Estimation and Stellar Model Selection}. \apjs 237:15.
  \doi{10.3847/1538-4365/aaccfb},
  {\href{https://arxiv.org/abs/1806.06869}{{arXiv:1806.06869}}} {[astro-ph.SR]}

\bibitem[{{Aerts} et~al(2019{\natexlab{a}}){Aerts}, {Mathis}, and
  {Rogers}}]{Aerts2019b}
{Aerts} C, {Mathis} S, {Rogers} TM (2019{\natexlab{a}}) {Angular Momentum
  Transport in Stellar Interiors}. \araa 57:35--78.
  \doi{10.1146/annurev-astro-091918-104359},
  {\href{https://arxiv.org/abs/1809.07779}{{arXiv:1809.07779}}} {[astro-ph.SR]}

\bibitem[{{Aerts} et~al(2019{\natexlab{b}}){Aerts}, {Pedersen}, {Vermeyen},
  {Hendriks}, {Johnston}, {Tkachenko}, {P{\'a}pics}, {Debosscher}, {Briquet},
  {Thoul}, {Rainer}, and {Poretti}}]{Aerts2019a}
{Aerts} C, {Pedersen} MG, {Vermeyen} E, et~al (2019{\natexlab{b}}) {Combined
  asteroseismology, spectroscopy, and astrometry of the CoRoT B2V target HD
  170580}. \aap 624:A75. \doi{10.1051/0004-6361/201834762},
  {\href{https://arxiv.org/abs/1902.04093}{{arXiv:1902.04093}}} {[astro-ph.SR]}

\bibitem[{{Aerts} et~al(2023){Aerts}, {Molenberghs}, and {De
  Ridder}}]{Aerts2023a}
{Aerts} C, {Molenberghs} G, {De Ridder} J (2023) {Astrophysical properties of
  15062 Gaia DR3 gravity-mode pulsators. Pulsation amplitudes, rotation, and
  spectral line broadening}. \aap 672:A183. \doi{10.1051/0004-6361/202245713},
  {\href{https://arxiv.org/abs/2302.07870}{{arXiv:2302.07870}}} {[astro-ph.SR]}

\bibitem[{{Aigrain} and {Foreman-Mackey}(2023)}]{Aigrain2023a}
{Aigrain} S, {Foreman-Mackey} D (2023) {Gaussian Process Regression for
  Astronomical Time Series}. \araa 61:329--371.
  \doi{10.1146/annurev-astro-052920-103508},
  {\href{https://arxiv.org/abs/2209.08940}{{arXiv:2209.08940}}} {[astro-ph.IM]}

\bibitem[{{Anders} and {Pedersen}(2023)}]{Anders_E_2023a}
{Anders} EH, {Pedersen} MG (2023) {Convective Boundary Mixing in Main-Sequence
  Stars: Theory and Empirical Constraints}. Galaxies 11(2):56.
  \doi{10.3390/galaxies11020056},
  {\href{https://arxiv.org/abs/2303.12099}{{arXiv:2303.12099}}} {[astro-ph.SR]}

\bibitem[{{Anders} et~al(2023){Anders}, {Lecoanet}, {Cantiello}, {Burns},
  {Hyatt}, {Kaufman}, {Townsend}, {Brown}, {Vasil}, {Oishi}, and
  {Jermyn}}]{Anders_E_2023b}
{Anders} EH, {Lecoanet} D, {Cantiello} M, et~al (2023) {The photometric
  variability of massive stars due to gravity waves excited by core
  convection}. Nature Astronomy 7:1228--1234. \doi{10.1038/s41550-023-02040-7},
  {\href{https://arxiv.org/abs/2306.08023}{{arXiv:2306.08023}}} {[astro-ph.SR]}

\bibitem[{{Andrassy} et~al(2022){Andrassy}, {Higl}, {Mao}, {Moc{\'a}k},
  {Vlaykov}, {Arnett}, {Baraffe}, {Campbell}, {Constantino}, {Edelmann},
  {Goffrey}, {Guillet}, {Herwig}, {Hirschi}, {Horst}, {Leidi}, {Meakin},
  {Pratt}, {Rizzuti}, {R{\"o}pke}, and {Woodward}}]{Andrassy2022a}
{Andrassy} R, {Higl} J, {Mao} H, et~al (2022) {Dynamics in a stellar convective
  layer and at its boundary: Comparison of five 3D hydrodynamics codes}. \aap
  659:A193. \doi{10.1051/0004-6361/202142557},
  {\href{https://arxiv.org/abs/2111.01165}{{arXiv:2111.01165}}} {[astro-ph.SR]}

\bibitem[{{Armstrong} et~al(2016){Armstrong}, {Kirk}, {Lam}, {McCormac},
  {Osborn}, {Spake}, {Walker}, {Brown}, {Kristiansen}, {Pollacco}, {West}, and
  {Wheatley}}]{Armstrong2016b}
{Armstrong} DJ, {Kirk} J, {Lam} KWF, et~al (2016) {K2 variable catalogue - II.
  Machine learning classification of variable stars and eclipsing binaries in
  K2 fields 0-4}. \mnras 456:2260--2272. \doi{10.1093/mnras/stv2836},
  {\href{https://arxiv.org/abs/1512.01246}{{arXiv:1512.01246}}} {[astro-ph.SR]}

\bibitem[{{Audenaert} and {Tkachenko}(2022)}]{Audenaert2022a}
{Audenaert} J, {Tkachenko} A (2022) {Multiscale entropy analysis of
  astronomical time series. Discovering subclusters of hybrid pulsators}. \aap
  666:A76. \doi{10.1051/0004-6361/202243469},
  {\href{https://arxiv.org/abs/2206.13529}{{arXiv:2206.13529}}} {[astro-ph.SR]}

\bibitem[{{Audenaert} et~al(2021){Audenaert}, {Kuszlewicz}, {Handberg},
  {Tkachenko}, {Armstrong}, {Hon}, {Kgoadi}, {Lund}, {Bell}, {Bugnet},
  {Bowman}, {Johnston}, {Garc{\'\i}a}, {Stello}, {Moln{\'a}r}, {Plachy},
  {Buzasi}, {Aerts}, and {T'DA Collaboration}}]{Audenaert2021a}
{Audenaert} J, {Kuszlewicz} JS, {Handberg} R, et~al (2021) {TESS Data for
  Asteroseismology (T'DA) Stellar Variability Classification Pipeline: Setup
  and Application to the Kepler Q9 Data}. \aj 162(5):209.
  \doi{10.3847/1538-3881/ac166a},
  {\href{https://arxiv.org/abs/2107.06301}{{arXiv:2107.06301}}} {[astro-ph.SR]}

\bibitem[{{Augustson} and {Mathis}(2019)}]{Augustson2019a}
{Augustson} KC, {Mathis} S (2019) {A Model of Rotating Convection in Stellar
  and Planetary Interiors. I. Convective Penetration}. \apj 874(1):83.
  \doi{10.3847/1538-4357/ab0b3d},
  {\href{https://arxiv.org/abs/1902.10593}{{arXiv:1902.10593}}} {[astro-ph.SR]}

\bibitem[{{Augustson} et~al(2016){Augustson}, {Brun}, and
  {Toomre}}]{Augustson2016a}
{Augustson} KC, {Brun} AS, {Toomre} J (2016) {The Magnetic Furnace: Intense
  Core Dynamos in B Stars}. \apj 829(2):92. \doi{10.3847/0004-637X/829/2/92},
  {\href{https://arxiv.org/abs/1603.03659}{{arXiv:1603.03659}}} {[astro-ph.SR]}

\bibitem[{{Augustson} et~al(2020){Augustson}, {Mathis}, and
  {Astoul}}]{Augustson2020c}
{Augustson} KC, {Mathis} S, {Astoul} A (2020) {A Model of Rotating Convection
  in Stellar and Planetary Interiors. II. Gravito-inertial Wave Generation}.
  \apj 903(2):90. \doi{10.3847/1538-4357/abba1c},
  {\href{https://arxiv.org/abs/2009.10473}{{arXiv:2009.10473}}} {[astro-ph.SR]}

\bibitem[{{Auvergne} et~al(2009){Auvergne}, {Bodin}, {Boisnard}, {Buey},
  {Chaintreuil}, {Epstein}, {Jouret}, {Lam-Trong}, {Levacher}, {Magnan},
  {Perez}, {Plasson}, {Plesseria}, {Peter}, {Steller}, {Tiph{\`e}ne}, {Baglin},
  {Agogu{\'e}}, {Appourchaux}, {Barbet}, {Beaufort}, {Bellenger}, {Berlin},
  {Bernardi}, {Blouin}, {Boumier}, {Bonneau}, {Briet}, {Butler}, {Cautain},
  {Chiavassa}, {Costes}, {Cuvilho}, {Cunha-Parro}, {de Oliveira Fialho},
  {Decaudin}, {Defise}, {Djalal}, {Docclo}, {Drummond}, {Dupuis}, {Exil},
  {Faur{\'e}}, {Gaboriaud}, {Gamet}, {Gavalda}, {Grolleau}, {Gueguen},
  {Guivarc'h}, {Guterman}, {Hasiba}, {Huntzinger}, {Hustaix}, {Imbert},
  {Jeanville}, {Johlander}, {Jorda}, {Journoud}, {Karioty}, {Kerjean},
  {Lafond}, {Lapeyrere}, {Landiech}, {Larqu{\'e}}, {Laudet}, {Le Merrer},
  {Leporati}, {Leruyet}, {Levieuge}, {Llebaria}, {Martin}, {Mazy}, {Mesnager},
  {Michel}, {Moalic}, {Monjoin}, {Naudet}, {Neukirchner}, {Nguyen-Kim},
  {Ollivier}, {Orcesi}, {Ottacher}, {Oulali}, {Parisot}, {Perruchot},
  {Piacentino}, {Pinheiro da Silva}, {Platzer}, {Pontet}, {Pradines},
  {Quentin}, {Rohbeck}, {Rolland}, {Rollenhagen}, {Romagnan}, {Russ}, {Samadi},
  {Schmidt}, {Schwartz}, {Sebbag}, {Smit}, {Sunter}, {Tello}, {Toulouse},
  {Ulmer}, {Vandermarcq}, {Vergnault}, {Wallner}, {Waultier}, and
  {Zanatta}}]{Auvergne2009}
{Auvergne} M, {Bodin} P, {Boisnard} L, et~al (2009) {The CoRoT satellite in
  flight: description and performance}. \aap 506:411--424.
  \doi{10.1051/0004-6361/200810860},
  {\href{https://arxiv.org/abs/0901.2206}{{arXiv:0901.2206}}} {[astro-ph.SR]}

\bibitem[{{Baade}(1988)}]{Baade1988c}
{Baade} D (1988) {Nonradial Pulsations and the be Phenomenon}. In: {Cayrel de
  Strobel} G, {Spite} M (eds) The Impact of Very High S/N Spectroscopy on
  Stellar Physics, p 217

\bibitem[{{Barbara} et~al(2022){Barbara}, {Bedding}, {Fulcher}, {Murphy}, and
  {Van Reeth}}]{Barbara2022a}
{Barbara} NH, {Bedding} TR, {Fulcher} BD, et~al (2022) {Classifying Kepler
  light curves for 12 000 A and F stars using supervised feature-based machine
  learning}. \mnras 514(2):2793--2804. \doi{10.1093/mnras/stac1515},
  {\href{https://arxiv.org/abs/2205.03020}{{arXiv:2205.03020}}} {[astro-ph.SR]}

\bibitem[{{Bj{\"o}rklund} et~al(2021){Bj{\"o}rklund}, {Sundqvist}, {Puls}, and
  {Najarro}}]{Bjorklund2021a}
{Bj{\"o}rklund} R, {Sundqvist} JO, {Puls} J, et~al (2021) {New predictions for
  radiation-driven, steady-state mass-loss and wind-momentum from hot, massive
  stars. II. A grid of O-type stars in the Galaxy and the Magellanic Clouds}.
  \aap 648:A36. \doi{10.1051/0004-6361/202038384},
  {\href{https://arxiv.org/abs/2008.06066}{{arXiv:2008.06066}}} {[astro-ph.SR]}

\bibitem[{{Bj{\"o}rklund} et~al(2023){Bj{\"o}rklund}, {Sundqvist}, {Singh},
  {Puls}, and {Najarro}}]{Bjorklund2023a}
{Bj{\"o}rklund} R, {Sundqvist} JO, {Singh} SM, et~al (2023) {New predictions
  for radiation-driven, steady-state mass-loss and wind-momentum from hot,
  massive stars. III. Updated mass-loss rates for stellar evolution}. \aap
  676:A109. \doi{10.1051/0004-6361/202141948},
  {\href{https://arxiv.org/abs/2203.08218}{{arXiv:2203.08218}}} {[astro-ph.SR]}

\bibitem[{{Blomme} et~al(2011){Blomme}, {Mahy}, {Catala}, {Cuypers}, {Gosset},
  {Godart}, {Montalban}, {Ventura}, {Rauw}, {Morel}, {Degroote}, {Aerts},
  {Noels}, {Michel}, {Baudin}, {Baglin}, {Auvergne}, and
  {Samadi}}]{Blomme2011b}
{Blomme} R, {Mahy} L, {Catala} C, et~al (2011) {Variability in the CoRoT
  photometry of three hot O-type stars. HD 46223, HD 46150, and HD 46966}. \aap
  533:A4. \doi{10.1051/0004-6361/201116949},
  {\href{https://arxiv.org/abs/1107.0267}{{arXiv:1107.0267}}} {[astro-ph.SR]}

\bibitem[{{Bodensteiner} et~al(2020{\natexlab{a}}){Bodensteiner}, {Shenar},
  {Mahy}, {Fabry}, {Marchant}, {Abdul-Masih}, {Banyard}, {Bowman}, {Dsilva},
  {Frost}, {Hawcroft}, {Reggiani}, and {Sana}}]{Bodensteiner2020c}
{Bodensteiner} J, {Shenar} T, {Mahy} L, et~al (2020{\natexlab{a}}) {Is HR 6819
  a triple system containing a black hole?. An alternative explanation}. \aap
  641:A43. \doi{10.1051/0004-6361/202038682},
  {\href{https://arxiv.org/abs/2006.10770}{{arXiv:2006.10770}}} {[astro-ph.SR]}

\bibitem[{{Bodensteiner} et~al(2020{\natexlab{b}}){Bodensteiner}, {Shenar}, and
  {Sana}}]{Bodensteiner2020b}
{Bodensteiner} J, {Shenar} T, {Sana} H (2020{\natexlab{b}}) {Investigating the
  lack of main-sequence companions to massive Be stars}. \aap 641:A42.
  \doi{10.1051/0004-6361/202037640},
  {\href{https://arxiv.org/abs/2006.13229}{{arXiv:2006.13229}}} {[astro-ph.SR]}

\bibitem[{{Borucki} et~al(2010){Borucki}, {Koch}, {Basri}, {Batalha}, {Brown},
  {Caldwell}, {Caldwell}, {Christensen-Dalsgaard}, {Cochran}, {DeVore},
  {Dunham}, {Dupree}, {Gautier}, {Geary}, {Gilliland}, {Gould}, {Howell},
  {Jenkins}, {Kondo}, {Latham}, {Marcy}, {Meibom}, {Kjeldsen}, {Lissauer},
  {Monet}, {Morrison}, {Sasselov}, {Tarter}, {Boss}, {Brownlee}, {Owen},
  {Buzasi}, {Charbonneau}, {Doyle}, {Fortney}, {Ford}, {Holman}, {Seager},
  {Steffen}, {Welsh}, {Rowe}, {Anderson}, {Buchhave}, {Ciardi}, {Walkowicz},
  {Sherry}, {Horch}, {Isaacson}, {Everett}, {Fischer}, {Torres}, {Johnson},
  {Endl}, {MacQueen}, {Bryson}, {Dotson}, {Haas}, {Kolodziejczak}, {Van Cleve},
  {Chandrasekaran}, {Twicken}, {Quintana}, {Clarke}, {Allen}, {Li}, {Wu},
  {Tenenbaum}, {Verner}, {Bruhweiler}, {Barnes}, and {Prsa}}]{Borucki2010}
{Borucki} WJ, {Koch} D, {Basri} G, et~al (2010) {Kepler Planet-Detection
  Mission: Introduction and First Results}. Science 327:977--.
  \doi{10.1126/science.1185402}

\bibitem[{{Bowman}(2020)}]{Bowman2020c}
{Bowman} DM (2020) {Asteroseismology of high-mass stars: new insights of
  stellar interiors with space telescopes}. Frontiers in Astronomy and Space
  Sciences 7:70. \doi{10.3389/fspas.2020.578584},
  {\href{https://arxiv.org/abs/2008.11162}{{arXiv:2008.11162}}} {[astro-ph.SR]}

\bibitem[{{Bowman} and {Dorn-Wallenstein}(2022)}]{Bowman2022b}
{Bowman} DM, {Dorn-Wallenstein} TZ (2022) {Photometric detection of internal
  gravity waves in upper main-sequence stars. III. Comparison of amplitude
  spectrum fitting and Gaussian process regression using CELERITE2}. \aap
  668:A134. \doi{10.1051/0004-6361/202243545},
  {\href{https://arxiv.org/abs/2211.08347}{{arXiv:2211.08347}}} {[astro-ph.SR]}

\bibitem[{{Bowman} and {Michielsen}(2021)}]{Bowman2021c}
{Bowman} DM, {Michielsen} M (2021) {Towards a systematic treatment of
  observational uncertainties in forward asteroseismic modelling of
  gravity-mode pulsators}. \aap 656:A158. \doi{10.1051/0004-6361/202141726},
  {\href{https://arxiv.org/abs/2109.10776}{{arXiv:2109.10776}}} {[astro-ph.SR]}

\bibitem[{{Bowman} et~al(2016){Bowman}, {Kurtz}, {Breger}, {Murphy}, and
  {Holdsworth}}]{Bowman2016a}
{Bowman} DM, {Kurtz} DW, {Breger} M, et~al (2016) {Amplitude modulation in
  {$\delta$} Sct stars: statistics from an ensemble study of Kepler targets}.
  \mnras 460:1970--1989. \doi{10.1093/mnras/stw1153},
  {\href{https://arxiv.org/abs/1605.03955}{{arXiv:1605.03955}}} {[astro-ph.SR]}

\bibitem[{{Bowman} et~al(2019{\natexlab{a}}){Bowman}, {Aerts}, {Johnston},
  {Pedersen}, {Rogers}, {Edelmann}, {Sim{\'o}n-D{\'{\i}}az}, {Van Reeth},
  {Buysschaert}, {Tkachenko}, and {Triana}}]{Bowman2019a}
{Bowman} DM, {Aerts} C, {Johnston} C, et~al (2019{\natexlab{a}}) {Photometric
  detection of internal gravity waves in upper main-sequence stars. I.
  Methodology and application to CoRoT targets}. \aap 621:A135.
  \doi{10.1051/0004-6361/201833662},
  {\href{https://arxiv.org/abs/1811.08023}{{arXiv:1811.08023}}} {[astro-ph.SR]}

\bibitem[{{Bowman} et~al(2019{\natexlab{b}}){Bowman}, {Burssens}, {Pedersen},
  {Johnston}, {Aerts}, {Buysschaert}, {Michielsen}, {Tkachenko}, {Rogers},
  {Edelmann}, {Ratnasingam}, {Sim{\'o}n-D{\'\i}az}, {Castro}, {Moravveji},
  {Pope}, {White}, and {De Cat}}]{Bowman2019b}
{Bowman} DM, {Burssens} S, {Pedersen} MG, et~al (2019{\natexlab{b}})
  {Low-frequency gravity waves in blue supergiants revealed by high-precision
  space photometry}. Nature Astronomy 3:760--765.
  \doi{10.1038/s41550-019-0768-1},
  {\href{https://arxiv.org/abs/1905.02120}{{arXiv:1905.02120}}} {[astro-ph.SR]}

\bibitem[{{Bowman} et~al(2020){Bowman}, {Burssens}, {Sim{\'o}n-D{\'\i}az},
  {Edelmann}, {Rogers}, {Horst}, {R{\"o}pke}, and {Aerts}}]{Bowman2020b}
{Bowman} DM, {Burssens} S, {Sim{\'o}n-D{\'\i}az} S, et~al (2020) {Photometric
  detection of internal gravity waves in upper main-sequence stars. II.
  Combined TESS photometry and high-resolution spectroscopy}. \aap 640:A36.
  \doi{10.1051/0004-6361/202038224},
  {\href{https://arxiv.org/abs/2006.03012}{{arXiv:2006.03012}}} {[astro-ph.SR]}

\bibitem[{{Bowman} et~al(2022){Bowman}, {Vandenbussche}, {Sana}, {Tkachenko},
  {Raskin}, {Delabie}, {Vandoren}, {Royer}, {Garcia}, {Van Reeth}, and
  {CubeSpec Collaboration}}]{Bowman2022a}
{Bowman} DM, {Vandenbussche} B, {Sana} H, et~al (2022) {The CubeSpec space
  mission. I. Asteroseismology of massive stars from time-series optical
  spectroscopy: Science requirements and target list prioritisation}. \aap
  658:A96. \doi{10.1051/0004-6361/202142375},
  {\href{https://arxiv.org/abs/2111.09814}{{arXiv:2111.09814}}} {[astro-ph.SR]}

\bibitem[{{Bowman} et~al(2024){Bowman}, {Van Daele}, {Michielsen}, and {Van
  Reeth}}]{Bowman2024a**}
{Bowman} DM, {Van Daele} P, {Michielsen} M, et~al (2024) {Stochastic
  low-frequency variability in time-series photometry of massive stars is
  insensitive to metallicity}. in prep

\bibitem[{{Briquet} et~al(2012){Briquet}, {Neiner}, {Aerts}, {Morel}, {Mathis},
  {Reese}, {Lehmann}, {Costero}, {Echevarria}, {Handler}, {Kambe}, {Hirata},
  {Masuda}, {Wright}, {Yang}, {Pintado}, {Mkrtichian}, {Lee}, {Han}, {Bruch},
  {De Cat}, {Uytterhoeven}, {Lefever}, {Vanautgaerden}, {de Batz},
  {Fr{\'e}mat}, {Henrichs}, {Geers}, {Martayan}, {Hubert}, {Thizy}, and
  {Tijani}}]{Briquet2012}
{Briquet} M, {Neiner} C, {Aerts} C, et~al (2012) {Multisite spectroscopic
  seismic study of the {$\beta$} Cep star V2052 Ophiuchi: inhibition of mixing
  by its magnetic field}. \mnras 427:483--493.
  \doi{10.1111/j.1365-2966.2012.21933.x},
  {\href{https://arxiv.org/abs/1208.4250}{{arXiv:1208.4250}}} {[astro-ph.SR]}

\bibitem[{{Briquet} et~al(2013){Briquet}, {Neiner}, {Leroy}, {P{\'a}pics}, and
  {MiMeS Collaboration}}]{Briquet2013}
{Briquet} M, {Neiner} C, {Leroy} B, et~al (2013) {Discovery of a magnetic field
  in the CoRoT hybrid B-type pulsator HD 43317}. \aap 557:L16.
  \doi{10.1051/0004-6361/201321779},
  {\href{https://arxiv.org/abs/1308.4636}{{arXiv:1308.4636}}} {[astro-ph.SR]}

\bibitem[{{Bromm} et~al(2009){Bromm}, {Yoshida}, {Hernquist}, and
  {McKee}}]{Bromm2009b}
{Bromm} V, {Yoshida} N, {Hernquist} L, et~al (2009) {The formation of the first
  stars and galaxies}. \nat 459(7243):49--54. \doi{10.1038/nature07990},
  {\href{https://arxiv.org/abs/0905.0929}{{arXiv:0905.0929}}} {[astro-ph.CO]}

\bibitem[{Burns et~al(2020)Burns, Vasil, Oishi, Lecoanet, and
  Brown}]{Burns2020}
Burns KJ, Vasil GM, Oishi JS, et~al (2020) Dedalus: A flexible framework for
  numerical simulations with spectral methods. Phys Rev Research 2:023068.
  \doi{10.1103/PhysRevResearch.2.023068},
  \urlprefix\url{https://link.aps.org/doi/10.1103/PhysRevResearch.2.023068}

\bibitem[{{Burssens} et~al(2019){Burssens}, {Bowman}, {Aerts}, {Pedersen},
  {Moravveji}, and {Buysschaert}}]{Burssens2019a}
{Burssens} S, {Bowman} DM, {Aerts} C, et~al (2019) {New {\ensuremath{\beta}}
  Cep pulsators discovered with K2 space photometry}. \mnras 489(1):1304--1320.
  \doi{10.1093/mnras/stz2165},
  {\href{https://arxiv.org/abs/1908.02836}{{arXiv:1908.02836}}} {[astro-ph.SR]}

\bibitem[{{Burssens} et~al(2020){Burssens}, {Sim{\'o}n-D{\'\i}az}, {Bowman},
  {Holgado}, {Michielsen}, {de Burgos}, {Castro}, {Barb{\'a}}, and
  {Aerts}}]{Burssens2020a}
{Burssens} S, {Sim{\'o}n-D{\'\i}az} S, {Bowman} DM, et~al (2020) {Variability
  of OB stars from TESS southern Sectors 1-13 and high-resolution IACOB and OWN
  spectroscopy}. \aap 639:A81. \doi{10.1051/0004-6361/202037700},
  {\href{https://arxiv.org/abs/2005.09658}{{arXiv:2005.09658}}} {[astro-ph.SR]}

\bibitem[{{Burssens} et~al(2023){Burssens}, {Bowman}, {Michielsen},
  {Sim{\'o}n-D{\'\i}az}, {Aerts}, {Vanlaer}, {Banyard}, {Nardetto}, {Townsend},
  {Handler}, {Mombarg}, {Vanderspek}, and {Ricker}}]{Burssens2023a}
{Burssens} S, {Bowman} DM, {Michielsen} M, et~al (2023) {A calibration point
  for stellar evolution from massive star asteroseismology}. Nature Astronomy
  7:913--930. \doi{10.1038/s41550-023-01978-y},
  {\href{https://arxiv.org/abs/2306.11798}{{arXiv:2306.11798}}} {[astro-ph.SR]}

\bibitem[{{Buysschaert} et~al(2017){Buysschaert}, {Neiner}, {Briquet}, and
  {Aerts}}]{Buysschaert2017b}
{Buysschaert} B, {Neiner} C, {Briquet} M, et~al (2017) {Magnetic
  characterization of the SPB/{$\beta$} Cep hybrid pulsator HD 43317}. \aap
  605:A104. \doi{10.1051/0004-6361/201731012},
  {\href{https://arxiv.org/abs/1707.09148}{{arXiv:1707.09148}}} {[astro-ph.SR]}

\bibitem[{{Buysschaert} et~al(2018{\natexlab{a}}){Buysschaert}, {Aerts},
  {Bowman}, {Johnston}, {Van Reeth}, {Pedersen}, {Mathis}, and
  {Neiner}}]{Buysschaert2018c}
{Buysschaert} B, {Aerts} C, {Bowman} DM, et~al (2018{\natexlab{a}}) {Forward
  seismic modeling of the pulsating magnetic B-type star HD 43317}. \aap
  616:A148. \doi{10.1051/0004-6361/201832642},
  {\href{https://arxiv.org/abs/1805.00802}{{arXiv:1805.00802}}} {[astro-ph.SR]}

\bibitem[{{Buysschaert} et~al(2018{\natexlab{b}}){Buysschaert}, {Neiner},
  {Martin}, {Aerts}, {Bowman}, {Oksala}, and {Van Reeth}}]{Buysschaert2018b}
{Buysschaert} B, {Neiner} C, {Martin} AJ, et~al (2018{\natexlab{b}}) {Detection
  of magnetic fields in chemically peculiar stars observed with the K2 space
  mission}. \mnras 478:2777--2793. \doi{10.1093/mnras/sty1190},
  {\href{https://arxiv.org/abs/1805.00781}{{arXiv:1805.00781}}} {[astro-ph.SR]}

\bibitem[{{Cantiello} et~al(2009){Cantiello}, {Langer}, {Brott}, {de Koter},
  {Shore}, {Vink}, {Voegler}, {Lennon}, and {Yoon}}]{Cantiello2009a}
{Cantiello} M, {Langer} N, {Brott} I, et~al (2009) {Sub-surface convection
  zones in hot massive stars and their observable consequences}. \aap
  499:279--290. \doi{10.1051/0004-6361/200911643},
  {\href{https://arxiv.org/abs/0903.2049}{{arXiv:0903.2049}}} {[astro-ph.SR]}

\bibitem[{{Cantiello} et~al(2021){Cantiello}, {Lecoanet}, {Jermyn}, and
  {Grassitelli}}]{Cantiello2021b}
{Cantiello} M, {Lecoanet} D, {Jermyn} AS, et~al (2021) {On the Origin of
  Stochastic, Low-Frequency Photometric Variability in Massive Stars}. \apj
  915(2):112. \doi{10.3847/1538-4357/ac03b0}

\bibitem[{{Chaplin} and {Miglio}(2013)}]{Chaplin2013c}
{Chaplin} WJ, {Miglio} A (2013) {Asteroseismology of Solar-Type and Red-Giant
  Stars}. \araa 51:353--392. \doi{10.1146/annurev-astro-082812-140938},
  {\href{https://arxiv.org/abs/1303.1957}{{arXiv:1303.1957}}} {[astro-ph.SR]}

\bibitem[{{Chieffi} and {Limongi}(2013)}]{Chieffi2013}
{Chieffi} A, {Limongi} M (2013) {Pre-supernova Evolution of Rotating Solar
  Metallicity Stars in the Mass Range 13-120 M $_{\odot}$ and their Explosive
  Yields}. \apj 764(1):21. \doi{10.1088/0004-637X/764/1/21}

\bibitem[{{Daszy{\'n}ska-Daszkiewicz} and {Walczak}(2009)}]{Daszy2009b}
{Daszy{\'n}ska-Daszkiewicz} J, {Walczak} P (2009) {Constraints on opacities
  from complex asteroseismology of B-type pulsators: the {\ensuremath{\beta}}
  Cephei star {\ensuremath{\theta}} Ophiuchi}. \mnras 398(4):1961--1969.
  \doi{10.1111/j.1365-2966.2009.15229.x},
  {\href{https://arxiv.org/abs/0906.3121}{{arXiv:0906.3121}}} {[astro-ph.SR]}

\bibitem[{{Daszy{\'n}ska-Daszkiewicz} and {Walczak}(2010)}]{Daszy2010a}
{Daszy{\'n}ska-Daszkiewicz} J, {Walczak} P (2010) {Complex asteroseismology of
  the {\ensuremath{\beta}} Cep/slowly pulsating B-type pulsator
  {\ensuremath{\nu}} Eridani: constraints on opacities}. \mnras
  403(1):496--504. \doi{10.1111/j.1365-2966.2009.16141.x},
  {\href{https://arxiv.org/abs/0912.0622}{{arXiv:0912.0622}}} {[astro-ph.SR]}

\bibitem[{{Daszy{\'n}ska-Daszkiewicz} et~al(2013){Daszy{\'n}ska-Daszkiewicz},
  {Szewczuk}, and {Walczak}}]{Daszy2013b}
{Daszy{\'n}ska-Daszkiewicz} J, {Szewczuk} W, {Walczak} P (2013) {The {$\beta$}
  Cep/SPB star 12 Lacertae: extended mode identification and complex seismic
  modelling}. \mnras 431:3396--3407. \doi{10.1093/mnras/stt418},
  {\href{https://arxiv.org/abs/1304.4049}{{arXiv:1304.4049}}} {[astro-ph.SR]}

\bibitem[{{David-Uraz} et~al(2019){David-Uraz}, {Neiner}, {Sikora}, {Bowman},
  {Petit}, {Chowdhury}, {Handler}, {Pergeorelis}, {Cantiello}, and
  {Cohen}}]{David-Uraz2019b}
{David-Uraz} A, {Neiner} C, {Sikora} J, et~al (2019) {Magnetic OB[A] Stars with
  TESS: probing their Evolutionary and Rotational properties (MOBSTER) - I.
  First-light observations of known magnetic B and A stars}. \mnras
  487(1):304--317. \doi{10.1093/mnras/stz1181},
  {\href{https://arxiv.org/abs/1904.11539}{{arXiv:1904.11539}}} {[astro-ph.SR]}

\bibitem[{{David-Uraz} et~al(2021){David-Uraz}, {Shultz}, {Petit}, {Bowman},
  {Erba}, {Fine}, {Neiner}, {Pablo}, {Sikora}, {ud-Doula}, and
  {Wade}}]{David-Uraz2021b}
{David-Uraz} A, {Shultz} ME, {Petit} V, et~al (2021) {MOBSTER - IV. Detection
  of a new magnetic B-type star from follow-up spectropolarimetric observations
  of photometrically selected candidates}. \mnras 504(4):4841--4849.
  \doi{10.1093/mnras/stab899},
  {\href{https://arxiv.org/abs/2004.09698}{{arXiv:2004.09698}}} {[astro-ph.SR]}

\bibitem[{{de Burgos} et~al(2023{\natexlab{a}}){de Burgos},
  {Sim{\'o}n-D{\'\i}az}, {Urbaneja}, and {Negueruela}}]{deBurgos2023a}
{de Burgos} A, {Sim{\'o}n-D{\'\i}az} S, {Urbaneja} MA, et~al
  (2023{\natexlab{a}}) {The IACOB project. IX. Building a modern empirical
  database of Galactic O9 - B9 supergiants: Sample selection, description, and
  completeness}. \aap 674:A212. \doi{10.1051/0004-6361/202346179},
  {\href{https://arxiv.org/abs/2305.00305}{{arXiv:2305.00305}}} {[astro-ph.SR]}

\bibitem[{{de Burgos} et~al(2023{\natexlab{b}}){de Burgos},
  {Sim{\'o}n-D{\'\i}az}, {Urbaneja}, and {Puls}}]{deBurgos2023b*}
{de Burgos} A, {Sim{\'o}n-D{\'\i}az} S, {Urbaneja} MA, et~al
  (2023{\natexlab{b}}) {The IACOB project X. Large-scale quantitative
  spectroscopic analysis of Galactic blue supergiants}. arXiv e-prints
  arXiv:2312.00241. \doi{10.48550/arXiv.2312.00241},
  {\href{https://arxiv.org/abs/2312.00241}{{arXiv:2312.00241}}} {[astro-ph.SR]}

\bibitem[{{de Mink} et~al(2013){de Mink}, {Langer}, {Izzard}, {Sana}, and {de
  Koter}}]{deMink2013}
{de Mink} SE, {Langer} N, {Izzard} RG, et~al (2013) {The Rotation Rates of
  Massive Stars: The Role of Binary Interaction through Tides, Mass Transfer,
  and Mergers}. \apj 764(2):166. \doi{10.1088/0004-637X/764/2/166},
  {\href{https://arxiv.org/abs/1211.3742}{{arXiv:1211.3742}}} {[astro-ph.SR]}

\bibitem[{{de Rossi} et~al(2010){de Rossi}, {Tissera}, and
  {Pedrosa}}]{deRossi2010d}
{de Rossi} ME, {Tissera} PB, {Pedrosa} SE (2010) {Impact of supernova feedback
  on the Tully-Fisher relation}. \aap 519:A89.
  \doi{10.1051/0004-6361/201014058},
  {\href{https://arxiv.org/abs/1005.4960}{{arXiv:1005.4960}}} {[astro-ph.CO]}

\bibitem[{{Degroote} et~al(2010){Degroote}, {Aerts}, {Baglin}, {Miglio},
  {Briquet}, {Noels}, {Niemczura}, {Montalban}, {Bloemen}, {Oreiro}, {Vu{\v
  c}kovi{\'c}}, {Smolders}, {Auvergne}, {Baudin}, {Catala}, and
  {Michel}}]{Degroote2010a}
{Degroote} P, {Aerts} C, {Baglin} A, et~al (2010) {Deviations from a uniform
  period spacing of gravity modes in a massive star}. \nat 464:259--261.
  \doi{10.1038/nature08864}

\bibitem[{{Dupret} et~al(2004){Dupret}, {Thoul}, {Scuflaire},
  {Daszy{\'n}ska-Daszkiewicz}, {Aerts}, {Bourge}, {Waelkens}, and
  {Noels}}]{Dupret2004b}
{Dupret} MA, {Thoul} A, {Scuflaire} R, et~al (2004) {Asteroseismology of the
  {$\beta$} Cep star HD 129929. II. Seismic constraints on core overshooting,
  internal rotation and stellar parameters}. \aap 415:251--257.
  \doi{10.1051/0004-6361:20034143}

\bibitem[{{Dziembowski} and {Pamyatnykh}(1993)}]{Dziembowski1993e}
{Dziembowski} WA, {Pamyatnykh} AA (1993) {The opacity mechanism in B-type
  stars. I - Unstable modes in Beta Cephei star models}. \mnras 262:204--212

\bibitem[{{Dziembowski} et~al(1993){Dziembowski}, {Moskalik}, and
  {Pamyatnykh}}]{Dziembowski1993f}
{Dziembowski} WA, {Moskalik} P, {Pamyatnykh} AA (1993) {The Opacity Mechanism
  in B-Type Stars - Part Two - Excitation of High-Order G-Modes in Main
  Sequence Stars}. \mnras 265:588

\bibitem[{{Edelmann} et~al(2019){Edelmann}, {Ratnasingam}, {Pedersen},
  {Bowman}, {Prat}, and {Rogers}}]{Edelmann2019a}
{Edelmann} PVF, {Ratnasingam} RP, {Pedersen} MG, et~al (2019)
  {Three-dimensional Simulations of Massive Stars. I. Wave Generation and
  Propagation}. \apj 876:4. \doi{10.3847/1538-4357/ab12df},
  {\href{https://arxiv.org/abs/1903.09392}{{arXiv:1903.09392}}} {[astro-ph.SR]}

\bibitem[{{Ekstr{\"o}m} et~al(2012){Ekstr{\"o}m}, {Georgy}, {Eggenberger},
  {Meynet}, {Mowlavi}, {Wyttenbach}, {Granada}, {Decressin}, {Hirschi},
  {Frischknecht}, {Charbonnel}, and {Maeder}}]{Ekstrom2012a}
{Ekstr{\"o}m} S, {Georgy} C, {Eggenberger} P, et~al (2012) {Grids of stellar
  models with rotation. I. Models from 0.8 to 120 M$_{\odot}$ at solar
  metallicity (Z = 0.014)}. \aap 537:A146. \doi{10.1051/0004-6361/201117751},
  {\href{https://arxiv.org/abs/1110.5049}{{arXiv:1110.5049}}} {[astro-ph.SR]}

\bibitem[{{Espinosa Lara} and {Rieutord}(2011)}]{EspinosaLara2011}
{Espinosa Lara} F, {Rieutord} M (2011) {Gravity darkening in rotating stars}.
  \aap 533:A43. \doi{10.1051/0004-6361/201117252},
  {\href{https://arxiv.org/abs/1109.3038}{{arXiv:1109.3038}}} {[astro-ph.SR]}

\bibitem[{{Espinosa Lara} and {Rieutord}(2013)}]{EspinosaLara2013}
{Espinosa Lara} F, {Rieutord} M (2013) {Self-consistent 2D models of
  fast-rotating early-type stars}. \aap 552:A35.
  \doi{10.1051/0004-6361/201220844},
  {\href{https://arxiv.org/abs/1212.0778}{{arXiv:1212.0778}}} {[astro-ph.SR]}

\bibitem[{{Foreman-Mackey} et~al(2017){Foreman-Mackey}, {Agol}, {Ambikasaran},
  and {Angus}}]{Foreman-Mackey2017}
{Foreman-Mackey} D, {Agol} E, {Ambikasaran} S, et~al (2017) {Fast and Scalable
  Gaussian Process Modeling with Applications to Astronomical Time Series}. \aj
  154(6):220. \doi{10.3847/1538-3881/aa9332},
  {\href{https://arxiv.org/abs/1703.09710}{{arXiv:1703.09710}}} {[astro-ph.IM]}

\bibitem[{{Fullerton} et~al(1996){Fullerton}, {Gies}, and
  {Bolton}}]{Fullerton1996}
{Fullerton} AW, {Gies} DR, {Bolton} CT (1996) {Absorption Line Profile
  Variations among the O Stars. I. The Incidence of Variability}. \apjs
  103:475. \doi{10.1086/192285}

\bibitem[{{Gaia Collaboration} et~al(2016){Gaia Collaboration}, {Prusti}, {de
  Bruijne}, {Brown}, {Vallenari}, {Babusiaux}, {Bailer-Jones}, {Bastian},
  {Biermann}, {Evans}, and et~al.}]{Gaia2016a}
{Gaia Collaboration}, {Prusti} T, {de Bruijne} JHJ, et~al (2016) {The Gaia
  mission}. \aap 595:A1. \doi{10.1051/0004-6361/201629272},
  {\href{https://arxiv.org/abs/1609.04153}{{arXiv:1609.04153}}} {[astro-ph.IM]}

\bibitem[{{Gaia Collaboration} et~al(2019){Gaia Collaboration}, {Eyer},
  {Rimoldini}, {Audard}, {Anderson}, {Nienartowicz}, {Glass}, {Marchal},
  {Grenon}, {Mowlavi}, {Holl}, {Clementini}, {Aerts}, {Mazeh}, {Evans},
  {Szabados}, {Brown}, {Vallenari}, {Prusti}, {de Bruijne}, {Babusiaux},
  {Bailer-Jones}, {Biermann}, {Jansen}, {Jordi}, {Klioner}, {Lammers},
  {Lindegren}, {Luri}, {Mignard}, {Panem}, {Pourbaix}, {Randich}, {Sartoretti},
  {Siddiqui}, {Soubiran}, {van Leeuwen}, {Walton}, {Arenou}, {Bastian},
  {Cropper}, {Drimmel}, {Katz}, {Lattanzi}, {Bakker}, {Cacciari},
  {Casta{\~n}eda}, {Chaoul}, {Cheek}, {De Angeli}, {Fabricius}, {Guerra},
  {Masana}, {Messineo}, {Panuzzo}, {Portell}, {Riello}, {Seabroke}, {Tanga},
  {Th{\'e}venin}, {Gracia-Abril}, {Comoretto}, {Garcia-Reinaldos}, {Teyssier},
  {Altmann}, {Andrae}, {Bellas-Velidis}, {Benson}, {Berthier}, {Blomme},
  {Burgess}, {Busso}, {Carry}, {Cellino}, {Clotet}, {Creevey}, {Davidson}, {De
  Ridder}, {Delchambre}, {Dell'Oro}, {Ducourant},
  {Fern{\'a}ndez-Hern{\'a}ndez}, {Fouesneau}, {Fr{\'e}mat}, {Galluccio},
  {Garc{\'\i}a-Torres}, {Gonz{\'a}lez-N{\'u}{\~n}ez}, {Gonz{\'a}lez-Vidal},
  {Gosset}, {Guy}, {Halbwachs}, {Hambly}, {Harrison}, {Hern{\'a}ndez},
  {Hestroffer}, {Hodgkin}, {Hutton}, {Jasniewicz}, {Jean-Antoine-Piccolo},
  {Jordan}, {Korn}, {Krone-Martins}, {Lanzafame}, {Lebzelter}, {L{\"o}ffler},
  {Manteiga}, {Marrese}, {Mart{\'\i}n-Fleitas}, {Moitinho}, {Mora}, {Muinonen},
  {Osinde}, {Pancino}, {Pauwels}, {Petit}, {Recio-Blanco}, {Richards}, {Robin},
  {Sarro}, {Siopis}, {Smith}, {Sozzetti}, {S{\"u}veges}, {Torra}, {van Reeven},
  {Abbas}, {Abreu Aramburu}, {Accart}, {Altavilla}, {{\'A}lvarez}, {Alvarez},
  {Alves}, {Andrei}, {Anglada Varela}, {Antiche}, {Antoja}, {Arcay},
  {Astraatmadja}, {Bach}, {Baker}, {Balaguer-N{\'u}{\~n}ez}, {Balm}, {Barache},
  {Barata}, {Barbato}, {Barblan}, {Barklem}, {Barrado}, {Barros}, {Barstow},
  {Bartholom{\'e} Mu{\~n}oz}, {Bassilana}, {Becciani}, {Bellazzini},
  {Berihuete}, {Bertone}, {Bianchi}, {Bienaym{\'e}}, {Blanco-Cuaresma}, {Boch},
  {Boeche}, {Bombrun}, {Borrachero}, {Bossini}, {Bouquillon}, {Bourda},
  {Bragaglia}, {Bramante}, {Breddels}, {Bressan}, {Brouillet},
  {Br{\"u}semeister}, {Brugaletta}, {Bucciarelli}, {Burlacu}, {Busonero},
  {Butkevich}, {Buzzi}, {Caffau}, {Cancelliere}, {Cannizzaro}, {Cantat-Gaudin},
  {Carballo}, {Carlucci}, {Carrasco}, {Casamiquela}, {Castellani},
  {Castro-Ginard}, {Charlot}, {Chemin}, {Chiavassa}, {Cocozza}, {Costigan},
  {Cowell}, {Crifo}, {Crosta}, {Crowley}, {Cuypers}, {Dafonte}, {Damerdji},
  {Dapergolas}, {David}, {David}, {de Laverny}, {De Luise}, {De March}, {de
  Martino}, {de Souza}, {de Torres}, {Debosscher}, {del Pozo}, {Delbo},
  {Delgado}, {Delgado}, {Diakite}, {Diener}, {Distefano}, {Dolding},
  {Drazinos}, {Dur{\'a}n}, {Edvardsson}, {Enke}, {Eriksson}, {Esquej}, {Eynard
  Bontemps}, {Fabre}, {Fabrizio}, {Faigler}, {Falc{\~a}o}, {Farr{\`a}s Casas},
  {Federici}, {Fedorets}, {Fernique}, {Figueras}, {Filippi}, {Findeisen},
  {Fonti}, {Fraile}, {Fraser}, {Fr{\'e}zouls}, {Gai}, {Galleti}, {Garabato},
  {Garc{\'\i}a-Sedano}, {Garofalo}, {Garralda}, {Gavel}, {Gavras}, {Gerssen},
  {Geyer}, {Giacobbe}, {Gilmore}, {Girona}, {Giuffrida}, {Gomes}, {Granvik},
  {Gueguen}, {Guerrier}, {Guiraud}, {Guti{\'e}rrez-S{\'a}nchez}, {Haigron},
  {Hatzidimitriou}, {Hauser}, {Haywood}, {Heiter}, {Helmi}, {Heu}, {Hilger},
  {Hobbs}, {Hofmann}, {Holland}, {Huckle}, {Hypki}, {Icardi}, {Jan{\ss}en},
  {Jevardat de Fombelle}, {Jonker}, {Juh{\'a}sz}, {Julbe}, {Karampelas},
  {Kewley}, {Klar}, {Kochoska}, {Kohley}, {Kolenberg}, {Kontizas}, {Kontizas},
  {Koposov}, {Kordopatis}, {Kostrzewa-Rutkowska}, {Koubsky}, {Lambert},
  {Lanza}, {Lasne}, {Lavigne}, {Le Fustec}, {Le Poncin-Lafitte}, {Lebreton},
  {Leccia}, {Leclerc}, {Lecoeur-Taibi}, {Lenhardt}, {Leroux}, {Liao}, {Licata},
  {Lindstr{\o}m}, {Lister}, {Livanou}, {Lobel}, {L{\'o}pez}, {Lorenz},
  {Managau}, {Mann}, {Mantelet}, {Marchant}, {Marconi}, {Marinoni},
  {Marschalk{\'o}}, {Marshall}, {Martino}, {Marton}, {Mary}, {Massari},
  {Matijevi{\v{c}}}, {McMillan}, {Messina}, {Michalik}, {Millar}, {Molina},
  {Molinaro}, {Moln{\'a}r}, {Montegriffo}, {Mor}, {Morbidelli}, {Morel},
  {Morgenthaler}, {Morris}, {Mulone}, {Muraveva}, {Musella}, {Nelemans},
  {Nicastro}, {Noval}, {O'Mullane}, {Ord{\'e}novic}, {Ord{\'o}{\~n}ez-Blanco},
  {Osborne}, {Pagani}, {Pagano}, {Pailler}, {Palacin}, {Palaversa}, {Panahi},
  {Pawlak}, {Piersimoni}, {Pineau}, {Plachy}, {Plum}, {Poggio}, {Poujoulet},
  {Pr{\v{s}}a}, {Pulone}, {Racero}, {Ragaini}, {Rambaux}, {Ramos-Lerate},
  {Regibo}, {Reyl{\'e}}, {Riclet}, {Ripepi}, {Riva}, {Rivard}, {Rixon},
  {Roegiers}, {Roelens}, {Romero-G{\'o}mez}, {Rowell}, {Royer}, {Ruiz-Dern},
  {Sadowski}, {Sagrist{\`a} Sell{\'e}s}, {Sahlmann}, {Salgado}, {Salguero},
  {Sanna}, {Santana-Ros}, {Sarasso}, {Savietto}, {Schultheis}, {Sciacca},
  {Segol}, {Segovia}, {S{\'e}gransan}, {Shih}, {Siltala}, {Silva}, {Smart},
  {Smith}, {Solano}, {Solitro}, {Sordo}, {Soria Nieto}, {Souchay}, {Spagna},
  {Spoto}, {Stampa}, {Steele}, {Steidelm{\"u}ller}, {Stephenson}, {Stoev},
  {Suess}, {Surdej}, {Szegedi-Elek}, {Tapiador}, {Taris}, {Tauran}, {Taylor},
  {Teixeira}, {Terrett}, {Teyssandier}, {Thuillot}, {Titarenko}, {Torra
  Clotet}, {Turon}, {Ulla}, {Utrilla}, {Uzzi}, {Vaillant}, {Valentini},
  {Valette}, {van Elteren}, {Van Hemelryck}, {van Leeuwen}, {Vaschetto},
  {Vecchiato}, {Veljanoski}, {Viala}, {Vicente}, {Vogt}, {von Essen}, {Voss},
  {Votruba}, {Voutsinas}, {Walmsley}, {Weiler}, {Wertz}, {Wevers},
  {Wyrzykowski}, {Yoldas}, {{\v{Z}}erjal}, {Ziaeepour}, {Zorec}, {Zschocke},
  {Zucker}, {Zurbach}, and {Zwitter}}]{Gaia2019a}
{Gaia Collaboration}, {Eyer} L, {Rimoldini} L, et~al (2019) {Gaia Data Release
  2. Variable stars in the colour-absolute magnitude diagram}. \aap 623:A110.
  \doi{10.1051/0004-6361/201833304},
  {\href{https://arxiv.org/abs/1804.09382}{{arXiv:1804.09382}}} {[astro-ph.SR]}

\bibitem[{{Gaia Collaboration} et~al(2023){Gaia Collaboration}, {De Ridder},
  {Ripepi}, {Aerts}, {Palaversa}, {Eyer}, {Holl}, {Audard}, {Rimoldini},
  {Brown}, {Vallenari}, {Prusti}, {de Bruijne}, {Arenou}, {Babusiaux},
  {Biermann}, {Creevey}, {Ducourant}, {Evans}, {Guerra}, {Hutton}, {Jordi},
  {Klioner}, {Lammers}, {Lindegren}, {Luri}, {Mignard}, {Panem}, {Pourbaix},
  {Randich}, {Sartoretti}, {Soubiran}, {Tanga}, {Walton}, {Bailer-Jones},
  {Bastian}, {Drimmel}, {Jansen}, {Katz}, {Lattanzi}, {van Leeuwen}, {Bakker},
  {Cacciari}, {Casta{\~n}eda}, {De Angeli}, {Fabricius}, {Fouesneau},
  {Fr{\'e}mat}, {Galluccio}, {Guerrier}, {Heiter}, {Masana}, {Messineo},
  {Mowlavi}, {Nicolas}, {Nienartowicz}, {Pailler}, {Panuzzo}, {Riclet}, {Roux},
  {Seabroke}, {Sordo}, {Th{\'e}venin}, {Gracia-Abril}, {Portell}, {Teyssier},
  {Altmann}, {Andrae}, {Bellas-Velidis}, {Benson}, {Berthier}, {Blomme},
  {Burgess}, {Busonero}, {Busso}, {C{\'a}novas}, {Carry}, {Cellino}, {Cheek},
  {Clementini}, {Damerdji}, {Davidson}, {de Teodoro}, {Nu{\~n}ez Campos},
  {Delchambre}, {Dell'Oro}, {Esquej}, {Fern{\'a}ndez-Hern{\'a}ndez}, {Fraile},
  {Garabato}, {Garc{\'\i}a-Lario}, {Gosset}, {Haigron}, {Halbwachs}, {Hambly},
  {Harrison}, {Hern{\'a}ndez}, {Hestroffer}, {Hilger}, {Hodgkin}, {Jan{\ss}en},
  {Jevardat de Fombelle}, {Jordan}, {Krone-Martins}, {Lanzafame},
  {L{\"o}ffler}, {Marchal}, {Marrese}, {Moitinho}, {Muinonen}, {Osborne},
  {Pancino}, {Pauwels}, {Recio-Blanco}, {Reyl{\'e}}, {Riello}, {Roegiers},
  {Rybizki}, {Sarro}, {Siopis}, {Smith}, {Sozzetti}, {Utrilla}, {van Leeuwen},
  {Abbas}, {{\'A}brah{\'a}m}, {Abreu Aramburu}, {Aguado}, {Ajaj},
  {Aldea-Montero}, {Altavilla}, {{\'A}lvarez}, {Alves}, {Anders}, {Anderson},
  {Anglada Varela}, {Antoja}, {Baines}, {Baker}, {Balaguer-N{\'u}{\~n}ez},
  {Balbinot}, {Balog}, {Barache}, {Barbato}, {Barros}, {Barstow},
  {Bartolom{\'e}}, {Bassilana}, {Bauchet}, {Becciani}, {Bellazzini},
  {Berihuete}, {Bernet}, {Bertone}, {Bianchi}, {Binnenfeld}, {Blanco-Cuaresma},
  {Boch}, {Bombrun}, {Bossini}, {Bouquillon}, {Bragaglia}, {Bramante},
  {Breedt}, {Bressan}, {Brouillet}, {Brugaletta}, {Bucciarelli}, {Burlacu},
  {Butkevich}, {Buzzi}, {Caffau}, {Cancelliere}, {Cantat-Gaudin}, {Carballo},
  {Carlucci}, {Carnerero}, {Carrasco}, {Casamiquela}, {Castellani},
  {Castro-Ginard}, {Chaoul}, {Charlot}, {Chemin}, {Chiaramida}, {Chiavassa},
  {Chornay}, {Comoretto}, {Contursi}, {Cooper}, {Cornez}, {Cowell}, {Crifo},
  {Cropper}, {Crosta}, {Crowley}, {Dafonte}, {Dapergolas}, {David}, {de
  Laverny}, {De Luise}, {De March}, {de Souza}, {de Torres}, {del Peloso}, {del
  Pozo}, {Delbo}, {Delgado}, {Delisle}, {Demouchy}, {Dharmawardena}, {Diakite},
  {Diener}, {Distefano}, {Dolding}, {Enke}, {Fabre}, {Fabrizio}, {Faigler},
  {Fedorets}, {Fernique}, {Figueras}, {Fournier}, {Fouron}, {Fragkoudi}, {Gai},
  {Garcia-Gutierrez}, {Garcia-Reinaldos}, {Garc{\'\i}a-Torres}, {Garofalo},
  {Gavel}, {Gavras}, {Gerlach}, {Geyer}, {Giacobbe}, {Gilmore}, {Girona},
  {Giuffrida}, {Gomel}, {Gomez}, {Gonz{\'a}lez-N{\'u}{\~n}ez},
  {Gonz{\'a}lez-Santamar{\'\i}a}, {Gonz{\'a}lez-Vidal}, {Granvik}, {Guillout},
  {Guiraud}, {Guti{\'e}rrez-S{\'a}nchez}, {Guy}, {Hatzidimitriou}, {Hauser},
  {Haywood}, {Helmer}, {Helmi}, {Sarmiento}, {Hidalgo}, {H{\l}adczuk}, {Hobbs},
  {Holland}, {Huckle}, {Jardine}, {Jasniewicz}, {Jean-Antoine Piccolo},
  {Jim{\'e}nez-Arranz}, {Juaristi Campillo}, {Julbe}, {Karbevska}, {Kervella},
  {Khanna}, {Kordopatis}, {Korn}, {K{\'o}sp{\'a}l}, {Kostrzewa-Rutkowska},
  {Kruszy{\'n}ska}, {Kun}, {Laizeau}, {Lambert}, {Lanza}, {Lasne}, {Le
  Campion}, {Lebreton}, {Lebzelter}, {Leccia}, {Leclerc}, {Lecoeur-Taibi},
  {Liao}, {Licata}, {Lindstr{\o}m}, {Lister}, {Livanou}, {Lobel}, {Lorca},
  {Loup}, {Madrero Pardo}, {Magdaleno Romeo}, {Managau}, {Mann}, {Manteiga},
  {Marchant}, {Marconi}, {Marcos}, {Marcos Santos}, {Mar{\'\i}n Pina},
  {Marinoni}, {Marocco}, {Marshall}, {Martin Polo}, {Mart{\'\i}n-Fleitas},
  {Marton}, {Mary}, {Masip}, {Massari}, {Mastrobuono-Battisti}, {Mazeh},
  {McMillan}, {Messina}, {Michalik}, {Millar}, {Mints}, {Molina}, {Molinaro},
  {Moln{\'a}r}, {Monari}, {Mongui{\'o}}, {Montegriffo}, {Montero}, {Mor},
  {Mora}, {Morbidelli}, {Morel}, {Morris}, {Muraveva}, {Murphy}, {Musella},
  {Nagy}, {Noval}, {Oca{\~n}a}, {Ogden}, {Ordenovic}, {Osinde}, {Pagani},
  {Pagano}, {Palicio}, {Pallas-Quintela}, {Panahi}, {Payne-Wardenaar},
  {Pe{\~n}alosa Esteller}, {Penttil{\"a}}, {Pichon}, {Piersimoni}, {Pineau},
  {Plachy}, {Plum}, {Poggio}, {Pr{\v{s}}a}, {Pulone}, {Racero}, {Ragaini},
  {Rainer}, {Raiteri}, {Ramos}, {Ramos-Lerate}, {Re Fiorentin}, {Regibo},
  {Richards}, {Rios Diaz}, {Riva}, {Rix}, {Rixon}, {Robichon}, {Robin},
  {Robin}, {Roelens}, {Rogues}, {Rohrbasser}, {Romero-G{\'o}mez}, {Rowell},
  {Royer}, {Ruz Mieres}, {Rybicki}, {Sadowski}, {S{\'a}ez N{\'u}{\~n}ez},
  {Sagrist{\`a} Sell{\'e}s}, {Sahlmann}, {Salguero}, {Samaras}, {Sanchez
  Gimenez}, {Sanna}, {Santove{\~n}a}, {Sarasso}, {Schultheis}, {Sciacca},
  {Segol}, {Segovia}, {S{\'e}gransan}, {Semeux}, {Shahaf}, {Siddiqui},
  {Siebert}, {Siltala}, {Silvelo}, {Slezak}, {Slezak}, {Smart}, {Snaith},
  {Solano}, {Solitro}, {Souami}, {Souchay}, {Spagna}, {Spina}, {Spoto},
  {Steele}, {Steidelm{\"u}ller}, {Stephenson}, {S{\"u}veges}, {Surdej},
  {Szabados}, {Szegedi-Elek}, {Taris}, {Taylor}, {Teixeira}, {Tolomei},
  {Tonello}, {Torra}, {Torra}, {Torralba Elipe}, {Trabucchi}, {Tsounis},
  {Turon}, {Ulla}, {Unger}, {Vaillant}, {vanDillen}, {van Reeven}, {Vanel},
  {Vecchiato}, {Viala}, {Vicente}, {Voutsinas}, {Weiler}, {Wevers},
  {Wyrzykowski}, {Yoldas}, {Yvard}, {Zhao}, {Zorec}, {Zucker}, and
  {Zwitter}}]{Gaia2023e}
{Gaia Collaboration}, {De Ridder} J, {Ripepi} V, et~al (2023) {Gaia Data
  Release 3. Pulsations in main sequence OBAF-type stars}. \aap 674:A36.
  \doi{10.1051/0004-6361/202243767},
  {\href{https://arxiv.org/abs/2206.06075}{{arXiv:2206.06075}}} {[astro-ph.SR]}

\bibitem[{{Garc{\'\i}a} and {Ballot}(2019)}]{Garcia_R_2019}
{Garc{\'\i}a} RA, {Ballot} J (2019) {Asteroseismology of solar-type stars}.
  Living Reviews in Solar Physics 16(1):4. \doi{10.1007/s41116-019-0020-1},
  {\href{https://arxiv.org/abs/1906.12262}{{arXiv:1906.12262}}} {[astro-ph.SR]}

\bibitem[{{Gebruers} et~al(2022){Gebruers}, {Tkachenko}, {Bowman}, {Van Reeth},
  {Burssens}, {IJspeert}, {Mahy}, {Straumit}, {Xiang}, {Rix}, and
  {Aerts}}]{Gebruers2022a}
{Gebruers} S, {Tkachenko} A, {Bowman} DM, et~al (2022) {Analysis of
  high-resolution FEROS spectroscopy for a sample of variable B-type stars
  assembled from TESS photometry}. \aap 665:A36.
  \doi{10.1051/0004-6361/202243839},
  {\href{https://arxiv.org/abs/2206.11280}{{arXiv:2206.11280}}} {[astro-ph.SR]}

\bibitem[{{Georgy} et~al(2013){Georgy}, {Ekstr{\"o}m}, {Eggenberger}, {Meynet},
  {Haemmerl{\'e}}, {Maeder}, {Granada}, {Groh}, {Hirschi}, {Mowlavi}, {Yusof},
  {Charbonnel}, {Decressin}, and {Barblan}}]{Georgy2013c}
{Georgy} C, {Ekstr{\"o}m} S, {Eggenberger} P, et~al (2013) {Grids of stellar
  models with rotation. III. Models from 0.8 to 120 M$_{\odot}$ at a
  metallicity Z = 0.002}. \aap 558:A103. \doi{10.1051/0004-6361/201322178},
  {\href{https://arxiv.org/abs/1308.2914}{{arXiv:1308.2914}}} {[astro-ph.SR]}

\bibitem[{{Goldstein} and {Townsend}(2020)}]{Goldstein_J_2020a}
{Goldstein} J, {Townsend} RHD (2020) {The Contour Method: a New Approach to
  Finding Modes of Nonadiabatic Stellar Pulsations}. \apj 899(2):116.
  \doi{10.3847/1538-4357/aba748},
  {\href{https://arxiv.org/abs/2006.13223}{{arXiv:2006.13223}}} {[astro-ph.SR]}

\bibitem[{{Grassitelli} et~al(2015){Grassitelli}, {Fossati},
  {Sim{\'o}n-Di{\'a}z}, {Langer}, {Castro}, and {Sanyal}}]{Grassitelli2015a}
{Grassitelli} L, {Fossati} L, {Sim{\'o}n-Di{\'a}z} S, et~al (2015)
  {Observational Consequences of Turbulent Pressure in the Envelopes of Massive
  Stars}. \apjl 808:L31. \doi{10.1088/2041-8205/808/1/L31},
  {\href{https://arxiv.org/abs/1507.03988}{{arXiv:1507.03988}}} {[astro-ph.SR]}

\bibitem[{{Grunhut} et~al(2017){Grunhut}, {Wade}, {Neiner}, {Oksala}, {Petit},
  {Alecian}, {Bohlender}, {Bouret}, {Henrichs}, {Hussain}, {Kochukhov}, and
  {MiMeS Collaboration}}]{Grunhut2017}
{Grunhut} JH, {Wade} GA, {Neiner} C, et~al (2017) {The MiMeS survey of
  Magnetism in Massive Stars: magnetic analysis of the O-type stars}. \mnras
  465:2432--2470. \doi{10.1093/mnras/stw2743},
  {\href{https://arxiv.org/abs/1610.07895}{{arXiv:1610.07895}}} {[astro-ph.SR]}

\bibitem[{{Handler} et~al(2004){Handler}, {Shobbrook}, {Jerzykiewicz},
  {Krisciunas}, {Tshenye}, {Rodr{\'{\i}}guez}, {Costa}, {Zhou}, {Medupe},
  {Phorah}, {Garrido}, {Amado}, {Papar{\'o}}, {Zsuffa}, {Ramokgali}, {Crowe},
  {Purves}, {Avila}, {Knight}, {Brassfield}, {Kilmartin}, and
  {Cottrell}}]{Handler2004b}
{Handler} G, {Shobbrook} RR, {Jerzykiewicz} M, et~al (2004) {Asteroseismology
  of the {$\beta$} Cephei star {$\nu$} Eridani - I. Photometric observations
  and pulsational frequency analysis}. \mnras 347:454--462.
  \doi{10.1111/j.1365-2966.2004.07214.x},
  {\href{https://arxiv.org/abs/astro-ph/0501263}{{astro-ph/0501263}}}

\bibitem[{{Handler} et~al(2012){Handler}, {Shobbrook}, {Uytterhoeven},
  {Briquet}, {Neiner}, {Tshenye}, {Ngwato}, {van Winckel}, {Guggenberger},
  {Raskin}, {Rodr{\'{\i}}guez}, {Mazumdar}, {Barban}, {Lorenz},
  {Vandenbussche}, {{\c S}ahin}, {Medupe}, and {Aerts}}]{Handler2012a}
{Handler} G, {Shobbrook} RR, {Uytterhoeven} K, et~al (2012) {A multisite
  photometric study of two unusual {$\beta$} Cep stars: the magnetic V2052 Oph
  and the massive rapid rotator V986 Oph}. \mnras 424:2380--2391.
  \doi{10.1111/j.1365-2966.2012.21414.x},
  {\href{https://arxiv.org/abs/1205.6401}{{arXiv:1205.6401}}} {[astro-ph.SR]}

\bibitem[{{Handler} et~al(2019){Handler}, {Pigulski},
  {Daszy{\'n}ska-Daszkiewicz}, {Irrgang}, {Kilkenny}, {Guo}, {Przybilla},
  {Kahraman Ali{\c c}avu{\c s}}, {Kallinger}, {Pascual-Granado}, {Niemczura},
  {R{\'o}{\.z}a{\'n}ski}, {Chowdhury}, {Buzasi}, {Mirouh}, {Bowman},
  {Johnston}, {Pedersen}, {Sim{\'o}n-D{\'{\i}}az}, {Moravveji}, {Gazeas}, {De
  Cat}, {Vanderspek}, and {Ricker}}]{Handler2019a}
{Handler} G, {Pigulski} A, {Daszy{\'n}ska-Daszkiewicz} J, et~al (2019)
  {Asteroseismology of Massive Stars with the TESS Mission: The Runaway
  {$\beta$} Cep Pulsator PHL 346 = HN Aqr}. \apjl 873:L4.
  \doi{10.3847/2041-8213/ab095f},
  {\href{https://arxiv.org/abs/1902.08312}{{arXiv:1902.08312}}} {[astro-ph.SR]}

\bibitem[{Harris et~al(2020)Harris, Millman, van~der Walt, Gommers, Virtanen,
  Cournapeau, Wieser, Taylor, Berg, Smith, Kern, Picus, Hoyer, van Kerkwijk,
  Brett, Haldane, del R{\'{i}}o, Wiebe, Peterson, G{\'{e}}rard-Marchant,
  Sheppard, Reddy, Weckesser, Abbasi, Gohlke, and Oliphant}]{Numpy_2020}
Harris CR, Millman KJ, van~der Walt SJ, et~al (2020) Array programming with
  {NumPy}. Nature 585(7825):357--362. \doi{10.1038/s41586-020-2649-2},
  \urlprefix\url{https://doi.org/10.1038/s41586-020-2649-2}

\bibitem[{{Hekker} and {Christensen-Dalsgaard}(2017)}]{Hekker2017a}
{Hekker} S, {Christensen-Dalsgaard} J (2017) {Giant star seismology}. \aapr
  25:1. \doi{10.1007/s00159-017-0101-x},
  {\href{https://arxiv.org/abs/1609.07487}{{arXiv:1609.07487}}} {[astro-ph.SR]}

\bibitem[{{Herwig} et~al(2023){Herwig}, {Woodward}, {Mao}, {Thompson},
  {Denissenkov}, {Lau}, {Blouin}, {Andrassy}, and {Paul}}]{Herwig2023a}
{Herwig} F, {Woodward} PR, {Mao} H, et~al (2023) {3D hydrodynamic simulations
  of massive main-sequence stars - I. Dynamics and mixing of convection and
  internal gravity waves}. \mnras 525(2):1601--1629.
  \doi{10.1093/mnras/stad2157},
  {\href{https://arxiv.org/abs/2303.05495}{{arXiv:2303.05495}}} {[astro-ph.SR]}

\bibitem[{{Hirschi} et~al(2005){Hirschi}, {Meynet}, and
  {Maeder}}]{Hirschi2005c}
{Hirschi} R, {Meynet} G, {Maeder} A (2005) {Yields of rotating stars at solar
  metallicity}. \aap 433(3):1013--1022. \doi{10.1051/0004-6361:20041554},
  {\href{https://arxiv.org/abs/astro-ph/0412454}{{arXiv:astro-ph/0412454}}}
  {[astro-ph]}

\bibitem[{{Horst} et~al(2020){Horst}, {Edelmann}, {Andr{\'a}ssy}, {R{\"o}pke},
  {Bowman}, {Aerts}, and {Ratnasingam}}]{Horst2020a}
{Horst} L, {Edelmann} PVF, {Andr{\'a}ssy} R, et~al (2020) {Fully compressible
  simulations of waves and core convection in main-sequence stars}. \aap
  641:A18. \doi{10.1051/0004-6361/202037531},
  {\href{https://arxiv.org/abs/2006.03011}{{arXiv:2006.03011}}} {[astro-ph.SR]}

\bibitem[{{Howarth} et~al(1997){Howarth}, {Siebert}, {Hussain}, and
  {Prinja}}]{Howarth1997}
{Howarth} ID, {Siebert} KW, {Hussain} GAJ, et~al (1997) {Cross-correlation
  characteristics of OB stars from IUE spectroscopy}. \mnras 284:265--285.
  \doi{10.1093/mnras/284.2.265}

\bibitem[{{Howell} et~al(2014){Howell}, {Sobeck}, {Haas}, {Still}, {Barclay},
  {Mullally}, {Troeltzsch}, {Aigrain}, {Bryson}, {Caldwell}, {Chaplin},
  {Cochran}, {Huber}, {Marcy}, {Miglio}, {Najita}, {Smith}, {Twicken}, and
  {Fortney}}]{Howell2014}
{Howell} SB, {Sobeck} C, {Haas} M, et~al (2014) {The K2 Mission:
  Characterization and Early Results}. \pasp 126:398--408.
  \doi{10.1086/676406},
  {\href{https://arxiv.org/abs/1402.5163}{{arXiv:1402.5163}}} {[astro-ph.IM]}

\bibitem[{{Huat} et~al(2009){Huat}, {Hubert}, {Baudin}, {Floquet}, {Neiner},
  {Fr{\'e}mat}, {Guti{\'e}rrez-Soto}, {Andrade}, {de Batz}, {Diago}, {Emilio},
  {Espinosa Lara}, {Fabregat}, {Janot-Pacheco}, {Leroy}, {Martayan}, {Semaan},
  {Suso}, {Auvergne}, {Catala}, {Michel}, and {Samadi}}]{Huat2009c}
{Huat} AL, {Hubert} AM, {Baudin} F, et~al (2009) {The B0.5IVe CoRoT target HD
  49330. I. Photometric analysis from CoRoT data}. \aap 506(1):95--101.
  \doi{10.1051/0004-6361/200911928}

\bibitem[{{Hunter}(2007)}]{Matplotlib_2007}
{Hunter} JD (2007) {Matplotlib: A 2D Graphics Environment}. Computing in
  Science and Engineering 9:90--95. \doi{10.1109/MCSE.2007.55}

\bibitem[{{Iglesias} and {Rogers}(1996)}]{Iglesias1996}
{Iglesias} CA, {Rogers} FJ (1996) {Updated Opal Opacities}. \apj 464:943.
  \doi{10.1086/177381}

\bibitem[{{Jenkins} et~al(2016){Jenkins}, {Twicken}, {McCauliff}, {Campbell},
  {Sanderfer}, {Lung}, {Mansouri-Samani}, {Girouard}, {Tenenbaum}, {Klaus},
  {Smith}, {Caldwell}, {Chacon}, {Henze}, {Heiges}, {Latham}, {Morgan},
  {Swade}, {Rinehart}, and {Vanderspek}}]{Jenkins2016b}
{Jenkins} JM, {Twicken} JD, {McCauliff} S, et~al (2016) {The TESS science
  processing operations center}. In: Software and Cyberinfrastructure for
  Astronomy IV, p 99133E, \doi{10.1117/12.2233418}

\bibitem[{{Jermyn} et~al(2022){Jermyn}, {Anders}, and
  {Cantiello}}]{Jermyn2022a}
{Jermyn} AS, {Anders} EH, {Cantiello} M (2022) {A Transparent Window into
  Early-type Stellar Variability}. \apj 926(2):221.
  \doi{10.3847/1538-4357/ac4e89},
  {\href{https://arxiv.org/abs/2201.10567}{{arXiv:2201.10567}}} {[astro-ph.SR]}

\bibitem[{{Jermyn} et~al(2023){Jermyn}, {Bauer}, {Schwab}, {Farmer}, {Ball},
  {Bellinger}, {Dotter}, {Joyce}, {Marchant}, {Mombarg}, {Wolf}, {Sunny Wong},
  {Cinquegrana}, {Farrell}, {Smolec}, {Thoul}, {Cantiello}, {Herwig}, {Toloza},
  {Bildsten}, {Townsend}, and {Timmes}}]{Jermyn2023a}
{Jermyn} AS, {Bauer} EB, {Schwab} J, et~al (2023) {Modules for Experiments in
  Stellar Astrophysics (MESA): Time-dependent Convection, Energy Conservation,
  Automatic Differentiation, and Infrastructure}. \apjs 265(1):15.
  \doi{10.3847/1538-4365/acae8d},
  {\href{https://arxiv.org/abs/2208.03651}{{arXiv:2208.03651}}} {[astro-ph.SR]}

\bibitem[{{Jiang}(2023)}]{Jiang_Y_2023a}
{Jiang} YF (2023) {Three Dimensional Natures of Massive Star Envelopes}.
  Galaxies 11(5):105. \doi{10.3390/galaxies11050105},
  {\href{https://arxiv.org/abs/2310.07829}{{arXiv:2310.07829}}} {[astro-ph.SR]}

\bibitem[{{Johnston}(2021)}]{Johnston2021b}
{Johnston} C (2021) {One size does not fit all: Evidence for a range of mixing
  efficiencies in stellar evolution calculations}. \aap 655:A29.
  \doi{10.1051/0004-6361/202141080},
  {\href{https://arxiv.org/abs/2107.09075}{{arXiv:2107.09075}}} {[astro-ph.SR]}

\bibitem[{{Kaiser} et~al(2020){Kaiser}, {Hirschi}, {Arnett}, {Georgy}, {Scott},
  and {Cristini}}]{Kaiser2020a}
{Kaiser} EA, {Hirschi} R, {Arnett} WD, et~al (2020) {Relative importance of
  convective uncertainties in massive stars}. \mnras 496(2):1967--1989.
  \doi{10.1093/mnras/staa1595},
  {\href{https://arxiv.org/abs/2006.01877}{{arXiv:2006.01877}}} {[astro-ph.SR]}

\bibitem[{{Keszthelyi} et~al(2019){Keszthelyi}, {Meynet}, {Georgy}, {Wade},
  {Petit}, and {David-Uraz}}]{Keszthelyi2019}
{Keszthelyi} Z, {Meynet} G, {Georgy} C, et~al (2019) {The effects of surface
  fossil magnetic fields on massive star evolution: I. Magnetic field
  evolution, mass-loss quenching, and magnetic braking}. \mnras
  485(4):5843--5860. \doi{10.1093/mnras/stz772},
  {\href{https://arxiv.org/abs/1902.09333}{{arXiv:1902.09333}}} {[astro-ph.SR]}

\bibitem[{{Keszthelyi} et~al(2020){Keszthelyi}, {Meynet}, {Shultz},
  {David-Uraz}, {ud-Doula}, {Townsend}, {Wade}, {Georgy}, {Petit}, and
  {Owocki}}]{Keszthelyi2020a}
{Keszthelyi} Z, {Meynet} G, {Shultz} ME, et~al (2020) {The effects of surface
  fossil magnetic fields on massive star evolution - II. Implementation of
  magnetic braking in MESA and implications for the evolution of surface
  rotation in OB stars}. \mnras 493(1):518--535. \doi{10.1093/mnras/staa237},
  {\href{https://arxiv.org/abs/2001.06239}{{arXiv:2001.06239}}} {[astro-ph.SR]}

\bibitem[{{Keszthelyi} et~al(2021){Keszthelyi}, {Meynet}, {Martins}, {de
  Koter}, and {David-Uraz}}]{Keszthelyi2021a}
{Keszthelyi} Z, {Meynet} G, {Martins} F, et~al (2021) {The effects of surface
  fossil magnetic fields on massive star evolution - III. The case of
  {\ensuremath{\tau}} Sco}. \mnras 504(2):2474--2492.
  \doi{10.1093/mnras/stab893},
  {\href{https://arxiv.org/abs/2103.13465}{{arXiv:2103.13465}}} {[astro-ph.SR]}

\bibitem[{{Koch} et~al(2010){Koch}, {Borucki}, {Basri}, {Batalha}, {Brown},
  {Caldwell}, {Christensen-Dalsgaard}, {Cochran}, {DeVore}, {Dunham},
  {Gautier}, {Geary}, {Gilliland}, {Gould}, {Jenkins}, {Kondo}, {Latham},
  {Lissauer}, {Marcy}, {Monet}, {Sasselov}, {Boss}, {Brownlee}, {Caldwell},
  {Dupree}, {Howell}, {Kjeldsen}, {Meibom}, {Morrison}, {Owen}, {Reitsema},
  {Tarter}, {Bryson}, {Dotson}, {Gazis}, {Haas}, {Kolodziejczak}, {Rowe}, {Van
  Cleve}, {Allen}, {Chandrasekaran}, {Clarke}, {Li}, {Quintana}, {Tenenbaum},
  {Twicken}, and {Wu}}]{Koch2010}
{Koch} DG, {Borucki} WJ, {Basri} G, et~al (2010) {Kepler Mission Design,
  Realized Photometric Performance, and Early Science}. \apjl 713:L79.
  \doi{10.1088/2041-8205/713/2/L79},
  {\href{https://arxiv.org/abs/1001.0268}{{arXiv:1001.0268}}} {[astro-ph.EP]}

\bibitem[{{Krti{\v c}ka} and {Feldmeier}(2018)}]{Krticka2018e}
{Krti{\v c}ka} J, {Feldmeier} A (2018) {Light variations due to the line-driven
  wind instability and wind blanketing in O stars}. \aap 617:A121.
  \doi{10.1051/0004-6361/201731614},
  {\href{https://arxiv.org/abs/1807.09407}{{arXiv:1807.09407}}} {[astro-ph.SR]}

\bibitem[{{Krti{\v{c}}ka} and {Feldmeier}(2021)}]{Krticka2021b}
{Krti{\v{c}}ka} J, {Feldmeier} A (2021) {Stochastic light variations in hot
  stars from wind instability: finding photometric signatures and testing
  against the TESS data}. \aap 648:A79. \doi{10.1051/0004-6361/202040148},
  {\href{https://arxiv.org/abs/2103.08755}{{arXiv:2103.08755}}} {[astro-ph.SR]}

\bibitem[{{Kurtz}(2022)}]{Kurtz2022a}
{Kurtz} DW (2022) {Asteroseismology Across the Hertzsprung-Russell Diagram}.
  \araa 60:31--71. \doi{10.1146/annurev-astro-052920-094232}

\bibitem[{{Kurtz} et~al(2014){Kurtz}, {Saio}, {Takata}, {Shibahashi}, {Murphy},
  and {Sekii}}]{Kurtz2014}
{Kurtz} DW, {Saio} H, {Takata} M, et~al (2014) {Asteroseismic measurement of
  surface-to-core rotation in a main-sequence A star, KIC 11145123}. \mnras
  444:102--116. \doi{10.1093/mnras/stu1329},
  {\href{https://arxiv.org/abs/1405.0155}{{arXiv:1405.0155}}} {[astro-ph.SR]}

\bibitem[{{Kurtz} et~al(2015){Kurtz}, {Shibahashi}, {Murphy}, {Bedding}, and
  {Bowman}}]{Kurtz2015b}
{Kurtz} DW, {Shibahashi} H, {Murphy} SJ, et~al (2015) {A unifying explanation
  of complex frequency spectra of {$\gamma$} Dor, SPB and Be stars: combination
  frequencies and highly non-sinusoidal light curves}. \mnras 450:3015--3029.
  \doi{10.1093/mnras/stv868},
  {\href{https://arxiv.org/abs/1504.04245}{{arXiv:1504.04245}}} {[astro-ph.SR]}

\bibitem[{{Labadie-Bartz} et~al(2021){Labadie-Bartz}, {Baade}, {Carciofi},
  {Rubio}, {Rivinius}, {Borre}, {Martayan}, and {Siverd}}]{Labadie-Bartz2021a}
{Labadie-Bartz} J, {Baade} D, {Carciofi} AC, et~al (2021) {Short-term
  variability and mass loss in Be stars - VI. Frequency groups in
  {\ensuremath{\gamma}} Cas detected by TESS}. \mnras 502(1):242--259.
  \doi{10.1093/mnras/staa3913},
  {\href{https://arxiv.org/abs/2012.06454}{{arXiv:2012.06454}}} {[astro-ph.SR]}

\bibitem[{{Labadie-Bartz} et~al(2022){Labadie-Bartz}, {Carciofi}, {Henrique de
  Amorim}, {Rubio}, {Luiz Figueiredo}, {Ticiani dos Santos}, and
  {Thomson-Paressant}}]{Labadie-Bartz2022a}
{Labadie-Bartz} J, {Carciofi} AC, {Henrique de Amorim} T, et~al (2022)
  {Classifying Be Star Variability With TESS. I. The Southern Ecliptic}. \aj
  163(5):226. \doi{10.3847/1538-3881/ac5abd}

\bibitem[{{Langer}(2012)}]{Langer2012}
{Langer} N (2012) {Presupernova Evolution of Massive Single and Binary Stars}.
  \araa 50:107--164. \doi{10.1146/annurev-astro-081811-125534},
  {\href{https://arxiv.org/abs/1206.5443}{{arXiv:1206.5443}}} {[astro-ph.SR]}

\bibitem[{{Langer} and {Kudritzki}(2014)}]{Langer2014a}
{Langer} N, {Kudritzki} RP (2014) {The spectroscopic Hertzsprung-Russell
  diagram}. \aap 564:A52. \doi{10.1051/0004-6361/201423374},
  {\href{https://arxiv.org/abs/1403.2212}{{arXiv:1403.2212}}} {[astro-ph.SR]}

\bibitem[{{Le Saux} et~al(2023){Le Saux}, {Baraffe}, {Guillet}, {Vlaykov},
  {Morison}, {Pratt}, {Constantino}, and {Goffrey}}]{LeSaux2023a}
{Le Saux} A, {Baraffe} I, {Guillet} T, et~al (2023) {Two-dimensional
  simulations of internal gravity waves in a 5 M$_{{\ensuremath{\odot}}}$
  zero-age-main-sequence model}. \mnras 522(2):2835--2849.
  \doi{10.1093/mnras/stad1067},
  {\href{https://arxiv.org/abs/2304.02508}{{arXiv:2304.02508}}} {[astro-ph.SR]}

\bibitem[{{Lecoanet} and {Edelmann}(2023)}]{Lecoanet2023a}
{Lecoanet} D, {Edelmann} PVF (2023) {Multidimensional Simulations of Core
  Convection}. Galaxies 11(4):89. \doi{10.3390/galaxies11040089},
  {\href{https://arxiv.org/abs/2307.15730}{{arXiv:2307.15730}}} {[astro-ph.SR]}

\bibitem[{{Lecoanet} et~al(2019){Lecoanet}, {Cantiello}, {Quataert}, {Couston},
  {Burns}, {Pope}, {Jermyn}, {Favier}, and {Le Bars}}]{Lecoanet2019a}
{Lecoanet} D, {Cantiello} M, {Quataert} E, et~al (2019) {Low-frequency
  Variability in Massive Stars: Core Generation or Surface Phenomenon?} \apjl
  886(1):L15. \doi{10.3847/2041-8213/ab5446},
  {\href{https://arxiv.org/abs/1910.01643}{{arXiv:1910.01643}}} {[astro-ph.SR]}

\bibitem[{{Lecoanet} et~al(2021){Lecoanet}, {Cantiello}, {Anders}, {Quataert},
  {Couston}, {Bouffard}, {Favier}, and {Le Bars}}]{Lecoanet2021a}
{Lecoanet} D, {Cantiello} M, {Anders} EH, et~al (2021) {Surface manifestation
  of stochastically excited internal gravity waves}. \mnras 508(1):132--143.
  \doi{10.1093/mnras/stab2524},
  {\href{https://arxiv.org/abs/2105.04558}{{arXiv:2105.04558}}} {[astro-ph.SR]}

\bibitem[{{Lecoanet} et~al(2022){Lecoanet}, {Bowman}, and {Van
  Reeth}}]{Lecoanet2022a}
{Lecoanet} D, {Bowman} DM, {Van Reeth} T (2022) {Asteroseismic inference of the
  near-core magnetic field strength in the main-sequence B star HD 43317}.
  \mnras 512(1):L16--L20. \doi{10.1093/mnrasl/slac013},
  {\href{https://arxiv.org/abs/2202.03440}{{arXiv:2202.03440}}} {[astro-ph.SR]}

\bibitem[{{Li} et~al(2020){Li}, {Van Reeth}, {Bedding}, {Murphy}, {Antoci},
  {Ouazzani}, and {Barbara}}]{Li_G_2020a}
{Li} G, {Van Reeth} T, {Bedding} TR, et~al (2020) {Gravity-mode period spacings
  and near-core rotation rates of 611 {\ensuremath{\gamma}} Doradus stars with
  Kepler}. \mnras 491(3):3586--3605. \doi{10.1093/mnras/stz2906},
  {\href{https://arxiv.org/abs/1910.06634}{{arXiv:1910.06634}}} {[astro-ph.SR]}

\bibitem[{{Lucy}(1976)}]{Lucy1976e}
{Lucy} LB (1976) {An analysis of the variable radial velocity of Alpha Cygni.}
  \apj 206:499--508. \doi{10.1086/154405}

\bibitem[{{Maeder}(2009)}]{Maeder_rotation_BOOK}
{Maeder} A (2009) {Physics, Formation and Evolution of Rotating Stars}.
  Springer, \doi{10.1007/978-3-540-76949-1}

\bibitem[{{Maeder} and {Meynet}(2005)}]{Maeder2005b}
{Maeder} A, {Meynet} G (2005) {Stellar evolution with rotation and magnetic
  fields. III. The interplay of circulation and dynamo}. \aap
  440(3):1041--1049. \doi{10.1051/0004-6361:20053261},
  {\href{https://arxiv.org/abs/astro-ph/0506347}{{arXiv:astro-ph/0506347}}}
  {[astro-ph]}

\bibitem[{{Marchant} and {Bodensteiner}(2023)}]{Marchant2023a*}
{Marchant} P, {Bodensteiner} J (2023) {The Evolution of Massive Binary Stars}.
  arXiv e-prints arXiv:2311.01865. \doi{10.48550/arXiv.2311.01865},
  {\href{https://arxiv.org/abs/2311.01865}{{arXiv:2311.01865}}} {[astro-ph.SR]}

\bibitem[{{Martins} and {Palacios}(2013)}]{Martins2013c}
{Martins} F, {Palacios} A (2013) {A comparison of evolutionary tracks for
  single Galactic massive stars}. \aap 560:A16.
  \doi{10.1051/0004-6361/201322480},
  {\href{https://arxiv.org/abs/1310.7218}{{arXiv:1310.7218}}} {[astro-ph.SR]}

\bibitem[{{Mestel}(1999)}]{Mestel1999a}
{Mestel} L (1999) {Stellar magnetism}, International series of monographs on
  physics, vol~99. Oxford

\bibitem[{{Michielsen} et~al(2021){Michielsen}, {Aerts}, and
  {Bowman}}]{Michielsen2021a}
{Michielsen} M, {Aerts} C, {Bowman} DM (2021) {Probing the temperature gradient
  in the core boundary layer of stars with gravito-inertial modes. The case of
  KIC 7760680}. \aap 650:A175. \doi{10.1051/0004-6361/202039926},
  {\href{https://arxiv.org/abs/2104.04531}{{arXiv:2104.04531}}} {[astro-ph.SR]}

\bibitem[{{Mombarg} et~al(2023){Mombarg}, {Rieutord}, and {Espinosa
  Lara}}]{Mombarg2023b}
{Mombarg} JSG, {Rieutord} M, {Espinosa Lara} F (2023) {The first
  two-dimensional stellar structure and evolution models of rotating stars.
  Calibration to {\ensuremath{\beta}} Cephei pulsator HD 192575}. \aap 677:L5.
  \doi{10.1051/0004-6361/202347454},
  {\href{https://arxiv.org/abs/2308.07362}{{arXiv:2308.07362}}} {[astro-ph.SR]}

\bibitem[{{Moravveji}(2016)}]{Moravveji2016a}
{Moravveji} E (2016) {The impact of enhanced iron opacity on massive star
  pulsations: updated instability strips}. \mnras 455:L67--L71.
  \doi{10.1093/mnrasl/slv142},
  {\href{https://arxiv.org/abs/1509.08652}{{arXiv:1509.08652}}} {[astro-ph.SR]}

\bibitem[{{Moravveji} et~al(2015){Moravveji}, {Aerts}, {P{\'a}pics}, {Triana},
  and {Vandoren}}]{Moravveji2015b}
{Moravveji} E, {Aerts} C, {P{\'a}pics} PI, et~al (2015) {Tight asteroseismic
  constraints on core overshooting and diffusive mixing in the slowly rotating
  pulsating B8.3V star KIC 10526294}. \aap 580:A27.
  \doi{10.1051/0004-6361/201425290},
  {\href{https://arxiv.org/abs/1505.06902}{{arXiv:1505.06902}}} {[astro-ph.SR]}

\bibitem[{{Moravveji} et~al(2016){Moravveji}, {Townsend}, {Aerts}, and
  {Mathis}}]{Moravveji2016b}
{Moravveji} E, {Townsend} RHD, {Aerts} C, et~al (2016) {Sub-inertial Gravity
  Modes in the B8V Star KIC 7760680 Reveal Moderate Core Overshooting and Low
  Vertical Diffusive Mixing}. \apj 823:130. \doi{10.3847/0004-637X/823/2/130},
  {\href{https://arxiv.org/abs/1604.02680}{{arXiv:1604.02680}}} {[astro-ph.SR]}

\bibitem[{{Naz{\'e}} et~al(2020){Naz{\'e}}, {Rauw}, and {Pigulski}}]{Naze2020c}
{Naz{\'e}} Y, {Rauw} G, {Pigulski} A (2020) {TESS light curves of
  {\ensuremath{\gamma}} Cas stars}. \mnras 498(3):3171--3183.
  \doi{10.1093/mnras/staa2553},
  {\href{https://arxiv.org/abs/2008.08334}{{arXiv:2008.08334}}} {[astro-ph.SR]}

\bibitem[{{Neiner} et~al(2015){Neiner}, {Mathis}, {Alecian}, {Emeriau},
  {Grunhut}, {BinaMIcS}, and {MiMeS Collaborations}}]{Neiner2015d}
{Neiner} C, {Mathis} S, {Alecian} E, et~al (2015) {The origin of magnetic
  fields in hot stars}. In: {Nagendra} KN, {Bagnulo} S, {Centeno} R, et~al
  (eds) Polarimetry, pp 61--66, \doi{10.1017/S1743921315004524},
  \eprint{1502.00226}

\bibitem[{{Neiner} et~al(2020){Neiner}, {Lee}, {Mathis}, {Saio}, {Lovekin}, and
  {Augustson}}]{Neiner2020b}
{Neiner} C, {Lee} U, {Mathis} S, et~al (2020) {Transport of angular momentum by
  stochastically excited waves as an explanation for the outburst of the
  rapidly rotating Be star HD49330}. \aap 644:A9.
  \doi{10.1051/0004-6361/201935858},
  {\href{https://arxiv.org/abs/2007.08977}{{arXiv:2007.08977}}} {[astro-ph.SR]}

\bibitem[{{P{\'a}pics} et~al(2012){P{\'a}pics}, {Briquet}, {Baglin}, {Poretti},
  {Aerts}, {Degroote}, {Tkachenko}, {Morel}, {Zima}, {Niemczura}, {Rainer},
  {Hareter}, {Baudin}, {Catala}, {Michel}, {Samadi}, and
  {Auvergne}}]{Papics2012a}
{P{\'a}pics} PI, {Briquet} M, {Baglin} A, et~al (2012) {Gravito-inertial and
  pressure modes detected in the B3 IV CoRoT target HD 43317}. \aap 542:A55.
  \doi{10.1051/0004-6361/201218809},
  {\href{https://arxiv.org/abs/1203.5231}{{arXiv:1203.5231}}} {[astro-ph.SR]}

\bibitem[{{Paxton} et~al(2011){Paxton}, {Bildsten}, {Dotter}, {Herwig},
  {Lesaffre}, and {Timmes}}]{Paxton2011}
{Paxton} B, {Bildsten} L, {Dotter} A, et~al (2011) {Modules for Experiments in
  Stellar Astrophysics (MESA)}. \apjs 192:3. \doi{10.1088/0067-0049/192/1/3},
  {\href{https://arxiv.org/abs/1009.1622}{{arXiv:1009.1622}}} {[astro-ph.SR]}

\bibitem[{{Paxton} et~al(2013){Paxton}, {Cantiello}, {Arras}, {Bildsten},
  {Brown}, {Dotter}, {Mankovich}, {Montgomery}, {Stello}, {Timmes}, and
  {Townsend}}]{Paxton2013}
{Paxton} B, {Cantiello} M, {Arras} P, et~al (2013) {Modules for Experiments in
  Stellar Astrophysics (MESA): Planets, Oscillations, Rotation, and Massive
  Stars}. \apjs 208:4. \doi{10.1088/0067-0049/208/1/4},
  {\href{https://arxiv.org/abs/1301.0319}{{arXiv:1301.0319}}} {[astro-ph.SR]}

\bibitem[{{Paxton} et~al(2015){Paxton}, {Marchant}, {Schwab}, {Bauer},
  {Bildsten}, {Cantiello}, {Dessart}, {Farmer}, {Hu}, {Langer}, {Townsend},
  {Townsley}, and {Timmes}}]{Paxton2015}
{Paxton} B, {Marchant} P, {Schwab} J, et~al (2015) {Modules for Experiments in
  Stellar Astrophysics (MESA): Binaries, Pulsations, and Explosions}. \apjs
  220:15. \doi{10.1088/0067-0049/220/1/15},
  {\href{https://arxiv.org/abs/1506.03146}{{arXiv:1506.03146}}} {[astro-ph.SR]}

\bibitem[{{Paxton} et~al(2018){Paxton}, {Schwab}, {Bauer}, {Bildsten},
  {Blinnikov}, {Duffell}, {Farmer}, {Goldberg}, {Marchant}, {Sorokina},
  {Thoul}, {Townsend}, and {Timmes}}]{Paxton2018}
{Paxton} B, {Schwab} J, {Bauer} EB, et~al (2018) {Modules for Experiments in
  Stellar Astrophysics (MESA): Convective Boundaries, Element Diffusion, and
  Massive Star Explosions}. \apjs 234:34. \doi{10.3847/1538-4365/aaa5a8},
  {\href{https://arxiv.org/abs/1710.08424}{{arXiv:1710.08424}}} {[astro-ph.SR]}

\bibitem[{{Paxton} et~al(2019){Paxton}, {Smolec}, {Schwab}, {Gautschy},
  {Bildsten}, {Cantiello}, {Dotter}, {Farmer}, {Goldberg}, {Jermyn}, {Kanbur},
  {Marchant}, {Thoul}, {Townsend}, {Wolf}, {Zhang}, and {Timmes}}]{Paxton2019}
{Paxton} B, {Smolec} R, {Schwab} J, et~al (2019) {Modules for Experiments in
  Stellar Astrophysics (MESA): Pulsating Variable Stars, Rotation, Convective
  Boundaries, and Energy Conservation}. \apjs 243(1):10.
  \doi{10.3847/1538-4365/ab2241},
  {\href{https://arxiv.org/abs/1903.01426}{{arXiv:1903.01426}}} {[astro-ph.SR]}

\bibitem[{{Pedersen} et~al(2019){Pedersen}, {Chowdhury}, {Johnston}, {Bowman},
  {Aerts}, {Handler}, {De Cat}, {Neiner}, {David-Uraz}, {Buzasi}, {Tkachenko},
  {Sim{\'o}n-D{\'{\i}}az}, {Moravveji}, {Sikora}, {Mirouh}, {Lovekin},
  {Cantiello}, {Daszy{\'n}ska-Daszkiewicz}, {Pigulski}, {Vanderspek}, and
  {Ricker}}]{Pedersen2019a}
{Pedersen} MG, {Chowdhury} S, {Johnston} C, et~al (2019) {Diverse Variability
  of O and B Stars Revealed from 2-minute Cadence Light Curves in Sectors 1 and
  2 of the TESS Mission: Selection of an Asteroseismic Sample}. \apjl 872:L9.
  \doi{10.3847/2041-8213/ab01e1},
  {\href{https://arxiv.org/abs/1901.07576}{{arXiv:1901.07576}}} {[astro-ph.SR]}

\bibitem[{{Pedersen} et~al(2021){Pedersen}, {Aerts}, {P{\'a}pics},
  {Michielsen}, {Gebruers}, {Rogers}, {Molenberghs}, {Burssens}, {Garcia}, and
  {Bowman}}]{Pedersen2021a}
{Pedersen} MG, {Aerts} C, {P{\'a}pics} PI, et~al (2021) {Internal mixing of
  rotating stars inferred from dipole gravity modes}. Nature Astronomy
  5:715--722. \doi{10.1038/s41550-021-01351-x},
  {\href{https://arxiv.org/abs/2105.04533}{{arXiv:2105.04533}}} {[astro-ph.SR]}

\bibitem[{{Pope} et~al(2019){Pope}, {Davies}, {Hawkins}, {White}, {Stokholm},
  {Bieryla}, {Latham}, {Lucey}, {Aerts}, {Aigrain}, {Antoci}, {Bedding},
  {Bowman}, {Caldwell}, {Chontos}, {Esquerdo}, {Huber}, {Jofr{\'e}}, {Murphy},
  {van Reeth}, {Silva Aguirre}, and {Yu}}]{Pope2019b}
{Pope} BJS, {Davies} GR, {Hawkins} K, et~al (2019) {The Kepler Smear Campaign:
  Light Curves for 102 Very Bright Stars}. \apjs 244(1):18.
  \doi{10.3847/1538-4365/ab2c04},
  {\href{https://arxiv.org/abs/1905.09831}{{arXiv:1905.09831}}} {[astro-ph.SR]}

\bibitem[{{Prat} et~al(2019){Prat}, {Mathis}, {Buysschaert}, {Van Beeck},
  {Bowman}, {Aerts}, and {Neiner}}]{Prat2019a}
{Prat} V, {Mathis} S, {Buysschaert} B, et~al (2019) {Period spacings of gravity
  modes in rapidly rotating magnetic stars. I. Axisymmetric fossil field with
  poloidal and toroidal components}. \aap 627:A64.
  \doi{10.1051/0004-6361/201935462},
  {\href{https://arxiv.org/abs/1903.05620}{{arXiv:1903.05620}}} {[astro-ph.SR]}

\bibitem[{{Prat} et~al(2020){Prat}, {Mathis}, {Neiner}, {Van Beeck}, {Bowman},
  and {Aerts}}]{Prat2020a}
{Prat} V, {Mathis} S, {Neiner} C, et~al (2020) {Period spacings of gravity
  modes in rapidly rotating magnetic stars. II. The case of an oblique dipolar
  fossil magnetic field}. \aap 636:A100. \doi{10.1051/0004-6361/201937398},
  {\href{https://arxiv.org/abs/2003.08218}{{arXiv:2003.08218}}} {[astro-ph.SR]}

\bibitem[{{Puls} et~al(2020){Puls}, {Najarro}, {Sundqvist}, and
  {Sen}}]{Puls2020a}
{Puls} J, {Najarro} F, {Sundqvist} JO, et~al (2020) {Atmospheric NLTE models
  for the spectroscopic analysis of blue stars with winds. V. Complete comoving
  frame transfer, and updated modeling of X-ray emission}. \aap 642:A172.
  \doi{10.1051/0004-6361/202038464},
  {\href{https://arxiv.org/abs/2011.02310}{{arXiv:2011.02310}}} {[astro-ph.SR]}

\bibitem[{{Ramiaramanantsoa} et~al(2014){Ramiaramanantsoa}, {Moffat},
  {Chen{\'e}}, {Richardson}, {Henrichs}, {Desforges}, {Antoci}, {Rowe},
  {Matthews}, {Kuschnig}, {Weiss}, {Sasselov}, {Rucinski}, and
  {Guenther}}]{Rami2014}
{Ramiaramanantsoa} T, {Moffat} AFJ, {Chen{\'e}} AN, et~al (2014) {MOST detects
  corotating bright spots on the mid-O-type giant {$\xi$} Persei}. \mnras
  441:910--917. \doi{10.1093/mnras/stu619},
  {\href{https://arxiv.org/abs/1403.7843}{{arXiv:1403.7843}}} {[astro-ph.SR]}

\bibitem[{{Raskin} et~al(2011){Raskin}, {van Winckel}, {Hensberge}, {Jorissen},
  {Lehmann}, {Waelkens}, {Avila}, {de Cuyper}, {Degroote}, {Dubosson},
  {Dumortier}, {Fr{\'e}mat}, {Laux}, {Michaud}, {Morren}, {Perez Padilla},
  {Pessemier}, {Prins}, {Smolders}, {van Eck}, and {Winkler}}]{Raskin2011}
{Raskin} G, {van Winckel} H, {Hensberge} H, et~al (2011) {HERMES: a
  high-resolution fibre-fed spectrograph for the Mercator telescope}. \aap
  526:A69. \doi{10.1051/0004-6361/201015435},
  {\href{https://arxiv.org/abs/1011.0258}{{arXiv:1011.0258}}} {[astro-ph.IM]}

\bibitem[{{Raskin} et~al(2018){Raskin}, {Delabie}, {De Munter}, {Sana},
  {Vandenbussche}, {Vandoren}, {Antoci}, {Kjeldsen}, {Karoff}, {de Koter},
  {D{\'e}sert}, {Mladenov}, and {Vandepitte}}]{Raskin2018}
{Raskin} G, {Delabie} T, {De Munter} W, et~al (2018) {CUBESPEC: low-cost
  space-based astronomical spectroscopy}. In: {Lystrup} M, {MacEwen} HA,
  {Fazio} GG, et~al (eds) Space Telescopes and Instrumentation 2018: Optical,
  Infrared, and Millimeter Wave, p 106985R, \doi{10.1117/12.2314074},
  \eprint{1805.11848}

\bibitem[{{Rasmussen} and {Williams}(2006)}]{Rasmussen2006}
{Rasmussen} CE, {Williams} CKI (2006) {Gaussian Processes for Machine
  Learning}. Massachusetts Institute of Technology

\bibitem[{{Ratnasingam} et~al(2019){Ratnasingam}, {Edelmann}, and
  {Rogers}}]{Ratnasingam2019a}
{Ratnasingam} RP, {Edelmann} PVF, {Rogers} TM (2019) {Onset of non-linear
  internal gravity waves in intermediate-mass stars}. \mnras 482:5500--5512.
  \doi{10.1093/mnras/sty3086},
  {\href{https://arxiv.org/abs/1812.01046}{{arXiv:1812.01046}}} {[astro-ph.SR]}

\bibitem[{{Ratnasingam} et~al(2020){Ratnasingam}, {Edelmann}, and
  {Rogers}}]{Ratnasingam2020a}
{Ratnasingam} RP, {Edelmann} PVF, {Rogers} TM (2020) {Two-dimensional
  simulations of internal gravity waves in the radiation zones of
  intermediate-mass stars}. \mnras 497(4):4231--4245.
  \doi{10.1093/mnras/staa2296},
  {\href{https://arxiv.org/abs/2008.03306}{{arXiv:2008.03306}}} {[astro-ph.SR]}

\bibitem[{{Ratnasingam} et~al(2023){Ratnasingam}, {Rogers}, {Chowdhury},
  {Handler}, {Vanon}, {Varghese}, and {Edelmann}}]{Ratnasingam2023a}
{Ratnasingam} RP, {Rogers} TM, {Chowdhury} S, et~al (2023) {Internal gravity
  waves in massive stars. II. Frequency analysis across stellar mass}. \aap
  674:A134. \doi{10.1051/0004-6361/202245727},
  {\href{https://arxiv.org/abs/2305.06379}{{arXiv:2305.06379}}} {[astro-ph.SR]}

\bibitem[{{Rauer} et~al(2014){Rauer}, {Catala}, {Aerts}, {Appourchaux}, {Benz},
  {Brandeker}, {Christensen-Dalsgaard}, {Deleuil}, {Gizon}, {Goupil},
  {G{\"u}del}, {Janot-Pacheco}, {Mas-Hesse}, {Pagano}, {Piotto}, {Pollacco},
  {Santos}, {Smith}, {Su{\'a}rez}, {Szab{\'o}}, {Udry}, {Adibekyan}, {Alibert},
  {Almenara}, {Amaro-Seoane}, {Eiff}, {Asplund}, {Antonello}, {Barnes},
  {Baudin}, {Belkacem}, {Bergemann}, {Bihain}, {Birch}, {Bonfils}, {Boisse},
  {Bonomo}, {Borsa}, {Brand{\~a}o}, {Brocato}, {Brun}, {Burleigh}, {Burston},
  {Cabrera}, {Cassisi}, {Chaplin}, {Charpinet}, {Chiappini}, {Church},
  {Csizmadia}, {Cunha}, {Damasso}, {Davies}, {Deeg}, {D{\'{\i}}az}, {Dreizler},
  {Dreyer}, {Eggenberger}, {Ehrenreich}, {Eigm{\"u}ller}, {Erikson}, {Farmer},
  {Feltzing}, {de Oliveira Fialho}, {Figueira}, {Forveille}, {Fridlund},
  {Garc{\'{\i}}a}, {Giommi}, {Giuffrida}, {Godolt}, {Gomes da Silva},
  {Granzer}, {Grenfell}, {Grotsch-Noels}, {G{\"u}nther}, {Haswell}, {Hatzes},
  {H{\'e}brard}, {Hekker}, {Helled}, {Heng}, {Jenkins}, {Johansen},
  {Khodachenko}, {Kislyakova}, {Kley}, {Kolb}, {Krivova}, {Kupka}, {Lammer},
  {Lanza}, {Lebreton}, {Magrin}, {Marcos-Arenal}, {Marrese}, {Marques},
  {Martins}, {Mathis}, {Mathur}, {Messina}, {Miglio}, {Montalban}, {Montalto},
  {Monteiro}, {Moradi}, {Moravveji}, {Mordasini}, {Morel}, {Mortier},
  {Nascimbeni}, {Nelson}, {Nielsen}, {Noack}, {Norton}, {Ofir}, {Oshagh},
  {Ouazzani}, {P{\'a}pics}, {Parro}, {Petit}, {Plez}, {Poretti}, {Quirrenbach},
  {Ragazzoni}, {Raimondo}, {Rainer}, {Reese}, {Redmer}, {Reffert},
  {Rojas-Ayala}, {Roxburgh}, {Salmon}, {Santerne}, {Schneider}, {Schou},
  {Schuh}, {Schunker}, {Silva-Valio}, {Silvotti}, {Skillen}, {Snellen}, {Sohl},
  {Sousa}, {Sozzetti}, {Stello}, {Strassmeier}, {{\v S}vanda}, {Szab{\'o}},
  {Tkachenko}, {Valencia}, {Van Grootel}, {Vauclair}, {Ventura}, {Wagner},
  {Walton}, {Weingrill}, {Werner}, {Wheatley}, and {Zwintz}}]{Rauer2014}
{Rauer} H, {Catala} C, {Aerts} C, et~al (2014) {The PLATO 2.0 mission}.
  Experimental Astronomy 38:249--330. \doi{10.1007/s10686-014-9383-4},
  {\href{https://arxiv.org/abs/1310.0696}{{arXiv:1310.0696}}} {[astro-ph.EP]}

\bibitem[{{Ricker} et~al(2015){Ricker}, {Winn}, {Vanderspek}, {Latham},
  {Bakos}, {Bean}, {Berta-Thompson}, {Brown}, {Buchhave}, {Butler}, {Butler},
  {Chaplin}, {Charbonneau}, {Christensen-Dalsgaard}, {Clampin}, {Deming},
  {Doty}, {De Lee}, {Dressing}, {Dunham}, {Endl}, {Fressin}, {Ge}, {Henning},
  {Holman}, {Howard}, {Ida}, {Jenkins}, {Jernigan}, {Johnson}, {Kaltenegger},
  {Kawai}, {Kjeldsen}, {Laughlin}, {Levine}, {Lin}, {Lissauer}, {MacQueen},
  {Marcy}, {McCullough}, {Morton}, {Narita}, {Paegert}, {Palle}, {Pepe},
  {Pepper}, {Quirrenbach}, {Rinehart}, {Sasselov}, {Sato}, {Seager},
  {Sozzetti}, {Stassun}, {Sullivan}, {Szentgyorgyi}, {Torres}, {Udry}, and
  {Villasenor}}]{Ricker2015}
{Ricker} GR, {Winn} JN, {Vanderspek} R, et~al (2015) {Transiting Exoplanet
  Survey Satellite (TESS)}. Journal of Astronomical Telescopes, Instruments,
  and Systems 1(1):014003. \doi{10.1117/1.JATIS.1.1.014003}

\bibitem[{{Rieutord} et~al(2016){Rieutord}, {Espinosa Lara}, and
  {Putigny}}]{Rieutord2016c}
{Rieutord} M, {Espinosa Lara} F, {Putigny} B (2016) {An algorithm for computing
  the 2D structure of fast rotating stars}. Journal of Computational Physics
  318:277--304. \doi{10.1016/j.jcp.2016.05.011},
  {\href{https://arxiv.org/abs/1605.02359}{{arXiv:1605.02359}}} {[astro-ph.SR]}

\bibitem[{{Rivinius} et~al(2013){Rivinius}, {Carciofi}, and
  {Martayan}}]{Rivinius2013c}
{Rivinius} T, {Carciofi} AC, {Martayan} C (2013) {Classical Be stars. Rapidly
  rotating B stars with viscous Keplerian decretion disks}. \aapr 21:69.
  \doi{10.1007/s00159-013-0069-0},
  {\href{https://arxiv.org/abs/1310.3962}{{arXiv:1310.3962}}} {[astro-ph.SR]}

\bibitem[{{Rogers}(2015)}]{Rogers2015}
{Rogers} TM (2015) {On the Differential Rotation of Massive Main-sequence
  Stars}. \apjl 815:L30. \doi{10.1088/2041-8205/815/2/L30},
  {\href{https://arxiv.org/abs/1511.03809}{{arXiv:1511.03809}}} {[astro-ph.SR]}

\bibitem[{{Rogers} and {McElwaine}(2017)}]{Rogers2017c}
{Rogers} TM, {McElwaine} JN (2017) {On the Chemical Mixing Induced by Internal
  Gravity Waves}. \apjl 848:L1. \doi{10.3847/2041-8213/aa8d13},
  {\href{https://arxiv.org/abs/1709.04920}{{arXiv:1709.04920}}} {[astro-ph.SR]}

\bibitem[{{Rogers} et~al(2013){Rogers}, {Lin}, {McElwaine}, and
  {Lau}}]{Rogers2013b}
{Rogers} TM, {Lin} DNC, {McElwaine} JN, et~al (2013) {Internal Gravity Waves in
  Massive Stars: Angular Momentum Transport}. \apj 772:21.
  \doi{10.1088/0004-637X/772/1/21},
  {\href{https://arxiv.org/abs/1306.3262}{{arXiv:1306.3262}}} {[astro-ph.SR]}

\bibitem[{{Saio} et~al(2006){Saio}, {Kuschnig}, {Gautschy}, {Cameron},
  {Walker}, {Matthews}, {Guenther}, {Moffat}, {Rucinski}, {Sasselov}, and
  {Weiss}}]{Saio2006b}
{Saio} H, {Kuschnig} R, {Gautschy} A, et~al (2006) {MOST Detects g- and p-Modes
  in the B Supergiant HD 163899 (B2 Ib/II)}. \apj 650:1111--1118.
  \doi{10.1086/507409},
  {\href{https://arxiv.org/abs/astro-ph/0606712}{{astro-ph/0606712}}}

\bibitem[{{Saio} et~al(2015){Saio}, {Kurtz}, {Takata}, {Shibahashi}, {Murphy},
  {Sekii}, and {Bedding}}]{Saio2015b}
{Saio} H, {Kurtz} DW, {Takata} M, et~al (2015) {Asteroseismic measurement of
  slow, nearly uniform surface-to-core rotation in the main-sequence F star KIC
  9244992}. \mnras 447:3264--3277. \doi{10.1093/mnras/stu2696},
  {\href{https://arxiv.org/abs/1412.5362}{{arXiv:1412.5362}}} {[astro-ph.SR]}

\bibitem[{{Salmon} et~al(2012){Salmon}, {Montalb{\'a}n}, {Morel}, {Miglio},
  {Dupret}, and {Noels}}]{Salmon2012}
{Salmon} S, {Montalb{\'a}n} J, {Morel} T, et~al (2012) {Testing the effects of
  opacity and the chemical mixture on the excitation of pulsations in B stars
  of the Magellanic Clouds}. \mnras 422:3460--3474.
  \doi{10.1111/j.1365-2966.2012.20857.x},
  {\href{https://arxiv.org/abs/1203.0527}{{arXiv:1203.0527}}} {[astro-ph.SR]}

\bibitem[{{Salmon} et~al(2022{\natexlab{a}}){Salmon}, {Eggenberger},
  {Montalb{\'a}n}, {Miglio}, {Noels}, {Buldgen}, {Moyano}, and
  {Meynet}}]{Salmon2022a}
{Salmon} SJAJ, {Eggenberger} P, {Montalb{\'a}n} J, et~al (2022{\natexlab{a}})
  {Asteroseismology of {\ensuremath{\beta}} Cephei stars: The stellar
  inferences tested in hare and hound exercises}. \aap 659:A142.
  \doi{10.1051/0004-6361/202142483},
  {\href{https://arxiv.org/abs/2112.04064}{{arXiv:2112.04064}}} {[astro-ph.SR]}

\bibitem[{{Salmon} et~al(2022{\natexlab{b}}){Salmon}, {Moyano}, {Eggenberger},
  {Haemmerl{\'e}}, and {Buldgen}}]{Salmon2022b}
{Salmon} SJAJ, {Moyano} FD, {Eggenberger} P, et~al (2022{\natexlab{b}})
  {Backtracing the internal rotation history of the {\ensuremath{\beta}} Cep
  star HD 129929}. \aap 664:L1. \doi{10.1051/0004-6361/202243961},
  {\href{https://arxiv.org/abs/2207.11051}{{arXiv:2207.11051}}} {[astro-ph.SR]}

\bibitem[{{Sana} et~al(2012){Sana}, {de Mink}, {de Koter}, {Langer}, {Evans},
  {Gieles}, {Gosset}, {Izzard}, {Le Bouquin}, and {Schneider}}]{Sana2012b}
{Sana} H, {de Mink} SE, {de Koter} A, et~al (2012) {Binary Interaction
  Dominates the Evolution of Massive Stars}. Science 337:444.
  \doi{10.1126/science.1223344},
  {\href{https://arxiv.org/abs/1207.6397}{{arXiv:1207.6397}}} {[astro-ph.SR]}

\bibitem[{{Schneider} et~al(2015){Schneider}, {Izzard}, {Langer}, and {de
  Mink}}]{Schneider_F_2015a}
{Schneider} FRN, {Izzard} RG, {Langer} N, et~al (2015) {Evolution of Mass
  Functions of Coeval Stars through Wind Mass Loss and Binary Interactions}.
  \apj 805(1):20. \doi{10.1088/0004-637X/805/1/20},
  {\href{https://arxiv.org/abs/1504.01735}{{arXiv:1504.01735}}} {[astro-ph.SR]}

\bibitem[{{Schneider} et~al(2019){Schneider}, {Ohlmann}, {Podsiadlowski},
  {R{\"o}pke}, {Balbus}, {Pakmor}, and {Springel}}]{Schneider_F_2019a}
{Schneider} FRN, {Ohlmann} ST, {Podsiadlowski} P, et~al (2019) {Stellar mergers
  as the origin of magnetic massive stars}. \nat 574(7777):211--214.
  \doi{10.1038/s41586-019-1621-5},
  {\href{https://arxiv.org/abs/1910.14058}{{arXiv:1910.14058}}} {[astro-ph.SR]}

\bibitem[{{Schneider} et~al(2020){Schneider}, {Ohlmann}, {Podsiadlowski},
  {R{\"o}pke}, {Balbus}, and {Pakmor}}]{Schneider_F_2020a}
{Schneider} FRN, {Ohlmann} ST, {Podsiadlowski} P, et~al (2020) {Long-term
  evolution of a magnetic massive merger product}. \mnras 495(3):2796--2812.
  \doi{10.1093/mnras/staa1326},
  {\href{https://arxiv.org/abs/2005.05335}{{arXiv:2005.05335}}} {[astro-ph.SR]}

\bibitem[{{Schultz} et~al(2022){Schultz}, {Bildsten}, and
  {Jiang}}]{Schultz2022a}
{Schultz} WC, {Bildsten} L, {Jiang} YF (2022) {Stochastic Low-frequency
  Variability in Three-dimensional Radiation Hydrodynamical Models of Massive
  Star Envelopes}. \apjl 924(1):L11. \doi{10.3847/2041-8213/ac441f},
  {\href{https://arxiv.org/abs/2110.13944}{{arXiv:2110.13944}}} {[astro-ph.SR]}

\bibitem[{{Scott} et~al(2021){Scott}, {Hirschi}, {Georgy}, {Arnett}, {Meakin},
  {Kaiser}, {Ekstr{\"o}m}, and {Yusof}}]{Scott2021a}
{Scott} LJA, {Hirschi} R, {Georgy} C, et~al (2021) {Convective core entrainment
  in 1D main-sequence stellar models}. \mnras 503(3):4208--4220.
  \doi{10.1093/mnras/stab752},
  {\href{https://arxiv.org/abs/2103.06196}{{arXiv:2103.06196}}} {[astro-ph.SR]}

\bibitem[{{Seaton} et~al(1994){Seaton}, {Yan}, {Mihalas}, and
  {Pradhan}}]{Seaton1994}
{Seaton} MJ, {Yan} Y, {Mihalas} D, et~al (1994) {Opacities for Stellar
  Envelopes}. \mnras 266:805. \doi{10.1093/mnras/266.4.805}

\bibitem[{{Serebriakova} et~al(2023){Serebriakova}, {Tkachenko}, {Gebruers},
  {Bowman}, {Van Reeth}, {Mahy}, {Burssens}, {IJspeert}, {Sana}, and
  {Aerts}}]{Serebriakova2023a}
{Serebriakova} N, {Tkachenko} A, {Gebruers} S, et~al (2023) {The ESO UVES/FEROS
  Large Programs of TESS OB pulsators. I. Global stellar parameters from
  high-resolution spectroscopy}. \aap 676:A85.
  \doi{10.1051/0004-6361/202346108},
  {\href{https://arxiv.org/abs/2305.19948}{{arXiv:2305.19948}}} {[astro-ph.SR]}

\bibitem[{{Shenar} et~al(2020){Shenar}, {Bodensteiner}, {Abdul-Masih}, {Fabry},
  {Mahy}, {Marchant}, {Banyard}, {Bowman}, {Dsilva}, {Hawcroft}, {Reggiani},
  and {Sana}}]{Shenar2020b}
{Shenar} T, {Bodensteiner} J, {Abdul-Masih} M, et~al (2020) {The ``hidden''
  companion in LB-1 unveiled by spectral disentangling}. \aap 639:L6.
  \doi{10.1051/0004-6361/202038275},
  {\href{https://arxiv.org/abs/2004.12882}{{arXiv:2004.12882}}} {[astro-ph.SR]}

\bibitem[{{Shibahashi} and {Aerts}(2000)}]{Shibahashi2000b}
{Shibahashi} H, {Aerts} C (2000) {Asteroseismology and Oblique Pulsator Model
  of {$\beta$} Cephei}. \apjl 531:L143--L146. \doi{10.1086/312533},
  {\href{https://arxiv.org/abs/astro-ph/0001372}{{astro-ph/0001372}}}

\bibitem[{{Shultz} et~al(2019){Shultz}, {Wade}, {Rivinius}, {Alecian},
  {Neiner}, {Petit}, {Owocki}, {ud-Doula}, {Kochukhov}, {Bohlender},
  {Keszthelyi}, {MiMeS Collaboration}, and {BinaMIcS
  Collaboration}}]{Shultz2019d}
{Shultz} ME, {Wade} GA, {Rivinius} T, et~al (2019) {The magnetic early B-type
  stars - III. A main-sequence magnetic, rotational, and magnetospheric
  biography}. \mnras 490(1):274--295. \doi{10.1093/mnras/stz2551},
  {\href{https://arxiv.org/abs/1909.02530}{{arXiv:1909.02530}}} {[astro-ph.SR]}

\bibitem[{{Sim{\'o}n-D{\'{\i}}az} and {Herrero}(2014)}]{Simon-Diaz2014a}
{Sim{\'o}n-D{\'{\i}}az} S, {Herrero} A (2014) {The IACOB project. I. Rotational
  velocities in northern Galactic O- and early B-type stars revisited. The
  impact of other sources of line-broadening}. \aap 562:A135.
  \doi{10.1051/0004-6361/201322758},
  {\href{https://arxiv.org/abs/1311.3360}{{arXiv:1311.3360}}} {[astro-ph.SR]}

\bibitem[{{Sim{\'o}n-D{\'{\i}}az} et~al(2017){Sim{\'o}n-D{\'{\i}}az}, {Godart},
  {Castro}, {Herrero}, {Aerts}, {Puls}, {Telting}, and
  {Grassitelli}}]{Simon-Diaz2017a}
{Sim{\'o}n-D{\'{\i}}az} S, {Godart} M, {Castro} N, et~al (2017) {The IACOB
  project . III. New observational clues to understand macroturbulent
  broadening in massive O- and B-type stars}. \aap 597:A22.
  \doi{10.1051/0004-6361/201628541},
  {\href{https://arxiv.org/abs/1608.05508}{{arXiv:1608.05508}}} {[astro-ph.SR]}

\bibitem[{{Southworth} and {Bowman}(2022)}]{Southworth2022a}
{Southworth} J, {Bowman} DM (2022) {High-mass pulsators in eclipsing binaries
  observed using TESS}. \mnras 513(3):3191--3209. \doi{10.1093/mnras/stac875},
  {\href{https://arxiv.org/abs/2203.15365}{{arXiv:2203.15365}}} {[astro-ph.SR]}

\bibitem[{{Stankov} and {Handler}(2005)}]{Stankov2005}
{Stankov} A, {Handler} G (2005) {Catalog of Galactic {$\beta$} Cephei Stars}.
  \apjs 158:193--216. \doi{10.1086/429408},
  {\href{https://arxiv.org/abs/astro-ph/0506495}{{astro-ph/0506495}}}

\bibitem[{{Sun} et~al(2023){Sun}, {Townsend}, and {Guo}}]{Sun_M_2023b}
{Sun} M, {Townsend} RHD, {Guo} Z (2023) {gyre\_tides: Modeling Binary Tides
  within the GYRE Stellar Oscillation Code}. \apj 945(1):43.
  \doi{10.3847/1538-4357/acb33a},
  {\href{https://arxiv.org/abs/2301.06599}{{arXiv:2301.06599}}} {[astro-ph.SR]}

\bibitem[{{Sundqvist} et~al(2019){Sundqvist}, {Bj{\"o}rklund}, {Puls}, and
  {Najarro}}]{Sundqvist2019}
{Sundqvist} JO, {Bj{\"o}rklund} R, {Puls} J, et~al (2019) {New predictions for
  radiation-driven, steady-state mass-loss and wind-momentum from hot, massive
  stars. I. Method and first results}. \aap 632:A126.
  \doi{10.1051/0004-6361/201936580},
  {\href{https://arxiv.org/abs/1910.06586}{{arXiv:1910.06586}}} {[astro-ph.SR]}

\bibitem[{{Szewczuk} and {Daszy{\'n}ska-Daszkiewicz}(2017)}]{Szewczuk2017a}
{Szewczuk} W, {Daszy{\'n}ska-Daszkiewicz} J (2017) {Domains of pulsational
  instability of low-frequency modes in rotating upper main sequence stars}.
  \mnras 469:13--46. \doi{10.1093/mnras/stx738},
  {\href{https://arxiv.org/abs/1703.08075}{{arXiv:1703.08075}}} {[astro-ph.SR]}

\bibitem[{{Szewczuk} and {Daszy{\'n}ska-Daszkiewicz}(2018)}]{Szewczuk2018a}
{Szewczuk} W, {Daszy{\'n}ska-Daszkiewicz} J (2018) {KIC 3240411 - the hottest
  known SPB star with the asymptotic g-mode period spacing}. \mnras
  478:2243--2256. \doi{10.1093/mnras/sty1126},
  {\href{https://arxiv.org/abs/1805.07100}{{arXiv:1805.07100}}} {[astro-ph.SR]}

\bibitem[{{Talon} et~al(1997){Talon}, {Zahn}, {Maeder}, and
  {Meynet}}]{Talon1997b}
{Talon} S, {Zahn} JP, {Maeder} A, et~al (1997) {Rotational mixing in early-type
  stars: the main-sequence evolution of a 9M$_{sun}$\_ star.} \aap
  322:209--217.
  {\href{https://arxiv.org/abs/astro-ph/9611131}{{astro-ph/9611131}}}

\bibitem[{{Telting} and {Schrijvers}(1998)}]{Telting1998c}
{Telting} JH, {Schrijvers} C (1998) {A new bright beta Cephei star:
  line-profile variability in omega\^1 SCO}. \aap 339:150--158

\bibitem[{{Thompson} et~al(2023){Thompson}, {Herwig}, {Woodward}, {Mao},
  {Denissenkov}, {Bowman}, and {Blouin}}]{Thompson_W_2023a*}
{Thompson} W, {Herwig} F, {Woodward} PR, et~al (2023) {3D hydrodynamic
  simulations of massive main-sequence stars II. Convective excitation and
  spectra of internal gravity waves}. arXiv e-prints arXiv:2303.06125.
  \doi{10.48550/arXiv.2303.06125},
  {\href{https://arxiv.org/abs/2303.06125}{{arXiv:2303.06125}}} {[astro-ph.SR]}

\bibitem[{{Tkachenko} et~al(2012){Tkachenko}, {Aerts}, {Pavlovski},
  {Southworth}, {Degroote}, {Debosscher}, {Still}, {Bryson}, {Molenberghs},
  {Bloemen}, {de Vries}, {Hrudkova}, {Lombaert}, {Neyskens}, {P{\'a}pics},
  {Raskin}, {Van Winckel}, {Morris}, {Sanderfer}, and
  {Seader}}]{Tkachenko2012b}
{Tkachenko} A, {Aerts} C, {Pavlovski} K, et~al (2012) {Detection of gravity
  modes in the massive binary V380 Cyg from Kepler space-based photometry and
  high-resolution spectroscopy}. \mnras 424:L21--L25.
  \doi{10.1111/j.1745-3933.2012.01277.x},
  {\href{https://arxiv.org/abs/1205.0554}{{arXiv:1205.0554}}} {[astro-ph.SR]}

\bibitem[{{Tkachenko} et~al(2016){Tkachenko}, {Matthews}, {Aerts}, {Pavlovski},
  {P{\'a}pics}, {Zwintz}, {Cameron}, {Walker}, {Kuschnig}, {Degroote},
  {Debosscher}, {Moravveji}, {Kolbas}, {Guenther}, {Moffat}, {Rowe},
  {Rucinski}, {Sasselov}, and {Weiss}}]{Tkachenko2016}
{Tkachenko} A, {Matthews} JM, {Aerts} C, et~al (2016) {Stellar modelling of
  Spica, a high-mass spectroscopic binary with a {$\beta$} Cep variable primary
  component}. \mnras 458:1964--1976. \doi{10.1093/mnras/stw255},
  {\href{https://arxiv.org/abs/1601.08069}{{arXiv:1601.08069}}} {[astro-ph.SR]}

\bibitem[{{Townsend}(2005{\natexlab{a}})}]{Townsend2005b}
{Townsend} RHD (2005{\natexlab{a}}) {Influence of the Coriolis force on the
  instability of slowly pulsating B stars}. \mnras 360:465--476.
  \doi{10.1111/j.1365-2966.2005.09002.x},
  {\href{https://arxiv.org/abs/astro-ph/0503192}{{astro-ph/0503192}}}

\bibitem[{{Townsend}(2005{\natexlab{b}})}]{Townsend2005e}
{Townsend} RHD (2005{\natexlab{b}}) {Kappa-mechanism excitation of retrograde
  mixed modes in rotating B-type stars}. \mnras 364:573--582.
  \doi{10.1111/j.1365-2966.2005.09585.x},
  {\href{https://arxiv.org/abs/astro-ph/0506580}{{astro-ph/0506580}}}

\bibitem[{{Townsend}(2020)}]{Townsend2020a}
{Townsend} RHD (2020) {Improved asymptotic expressions for the eigenvalues of
  Laplace's tidal equations}. \mnras 497(3):2670--2679.
  \doi{10.1093/mnras/staa2159},
  {\href{https://arxiv.org/abs/2006.12596}{{arXiv:2006.12596}}} {[astro-ph.SR]}

\bibitem[{{Townsend} and {Teitler}(2013)}]{Townsend2013b}
{Townsend} RHD, {Teitler} SA (2013) {GYRE: an open-source stellar oscillation
  code based on a new Magnus Multiple Shooting scheme}. \mnras 435:3406--3418.
  \doi{10.1093/mnras/stt1533},
  {\href{https://arxiv.org/abs/1308.2965}{{arXiv:1308.2965}}} {[astro-ph.SR]}

\bibitem[{{Townsend} et~al(2013){Townsend}, {Rivinius}, {Rowe}, {Moffat},
  {Matthews}, {Bohlender}, {Neiner}, {Telting}, {Guenther}, {Kallinger},
  {Kuschnig}, {Rucinski}, {Sasselov}, and {Weiss}}]{Townsend2013a}
{Townsend} RHD, {Rivinius} T, {Rowe} JF, et~al (2013) {MOST Observations of
  {$\sigma$} Ori E: Challenging the Centrifugal Breakout Narrative}. \apj
  769:33. \doi{10.1088/0004-637X/769/1/33},
  {\href{https://arxiv.org/abs/1304.2392}{{arXiv:1304.2392}}} {[astro-ph.SR]}

\bibitem[{{Townsend} et~al(2018){Townsend}, {Goldstein}, and
  {Zweibel}}]{Townsend2018a}
{Townsend} RHD, {Goldstein} J, {Zweibel} EG (2018) {Angular momentum transport
  by heat-driven g-modes in slowly pulsating B stars}. \mnras 475:879--893.
  \doi{10.1093/mnras/stx3142},
  {\href{https://arxiv.org/abs/1712.02420}{{arXiv:1712.02420}}} {[astro-ph.SR]}

\bibitem[{{Uytterhoeven} et~al(2004){Uytterhoeven}, {Telting}, {Aerts}, and
  {Willems}}]{Uytterhoeven2004b}
{Uytterhoeven} K, {Telting} JH, {Aerts} C, et~al (2004) {Interpretation of the
  variability of the {\ensuremath{\beta}} Cephei star {\ensuremath{\lambda}}
  Scorpii. II. The line-profile diagnostics}. \aap 427:593--605.
  \doi{10.1051/0004-6361:20041224}

\bibitem[{{Uytterhoeven} et~al(2005){Uytterhoeven}, {Briquet}, {Aerts},
  {Telting}, {Harmanec}, {Lefever}, and {Cuypers}}]{Uytterhoeven2005a}
{Uytterhoeven} K, {Briquet} M, {Aerts} C, et~al (2005) {Disentangling component
  spectra of {\ensuremath{\kappa}} Scorpii, a spectroscopic binary with a
  pulsating primary. II. Interpretation of the line-profile variability}. \aap
  432(3):955--967. \doi{10.1051/0004-6361:20041444}

\bibitem[{{Van Beeck} et~al(2020){Van Beeck}, {Prat}, {Van Reeth}, {Mathis},
  {Bowman}, {Neiner}, and {Aerts}}]{VanBeeck2020a}
{Van Beeck} J, {Prat} V, {Van Reeth} T, et~al (2020) {Detecting axisymmetric
  magnetic fields using gravity modes in intermediate-mass stars}. \aap
  638:A149. \doi{10.1051/0004-6361/201937363},
  {\href{https://arxiv.org/abs/2005.02411}{{arXiv:2005.02411}}} {[astro-ph.SR]}

\bibitem[{{Vanbeveren} et~al(1998){Vanbeveren}, {De Loore}, and {Van
  Rensbergen}}]{Vanbeveren1998a}
{Vanbeveren} D, {De Loore} C, {Van Rensbergen} W (1998) {Massive stars}. \aapr
  9(1-2):63--152. \doi{10.1007/s001590050015}

\bibitem[{{Vanlaer} et~al(2024){Vanlaer}, {Aerts}, {Bugnet}, {Burssens}, and
  {Bowman}}]{Vanlaer2024a**}
{Vanlaer} V, {Aerts} C, {Bugnet} L, et~al (2024) {Rotation and magnetic
  inversion of HD~192575}. \aap, submitted

\bibitem[{{Vanon} et~al(2023){Vanon}, {Edelmann}, {Ratnasingam}, {Varghese},
  and {Rogers}}]{Vanon2023a}
{Vanon} R, {Edelmann} PVF, {Ratnasingam} RP, et~al (2023) {Three-dimensional
  Simulations of Massive Stars. II. Age Dependence}. \apj 954(2):171.
  \doi{10.3847/1538-4357/ace9db},
  {\href{https://arxiv.org/abs/2307.15109}{{arXiv:2307.15109}}} {[astro-ph.SR]}

\bibitem[{{Varghese} et~al(2023){Varghese}, {Ratnasingam}, {Vanon}, {Edelmann},
  and {Rogers}}]{Varghese2023a}
{Varghese} A, {Ratnasingam} RP, {Vanon} R, et~al (2023) {Chemical Mixing
  Induced by Internal Gravity Waves in Intermediate-mass Stars}. \apj
  942(1):53. \doi{10.3847/1538-4357/aca092},
  {\href{https://arxiv.org/abs/2211.06432}{{arXiv:2211.06432}}} {[astro-ph.SR]}

\bibitem[{{Vink} et~al(2023){Vink}, {Mehner}, {Crowther}, {Fullerton},
  {Garcia}, {Martins}, {Morrell}, {Oskinova}, {St-Louis}, {ud-Doula}, {Sander},
  {Sana}, {Bouret}, {Kub{\'a}tov{\'a}}, {Marchant}, {Martins}, {Wofford}, {van
  Loon}, {Grace Telford}, {G{\"o}tberg}, {Bowman}, {Erba}, {Kalari},
  {Abdul-Masih}, {Alkousa}, {Backs}, {Barbosa}, {Berlanas}, {Bernini-Peron},
  {Bestenlehner}, {Blomme}, {Bodensteiner}, {Brands}, {Evans}, {David-Uraz},
  {Driessen}, {Dsilva}, {Geen}, {G{\'o}mez-Gonz{\'a}lez}, {Grassitelli},
  {Hamann}, {Hawcroft}, {Herrero}, {Higgins}, {John Hillier}, {Ignace},
  {Istrate}, {Kaper}, {Kee}, {Kehrig}, {Keszthelyi}, {Klencki}, {de Koter},
  {Kuiper}, {Laplace}, {Larkin}, {Lefever}, {Leitherer}, {Lennon}, {Mahy},
  {Ma{\'\i}z Apell{\'a}niz}, {Maravelias}, {Marcolino}, {McLeod}, {de Mink},
  {Najarro}, {Oey}, {Parsons}, {Pauli}, {Pedersen}, {Prinja}, {Ramachandran},
  {Ram{\'\i}rez-Tannus}, {Sabhahit}, {Schootemeijer}, {Reyero Serantes},
  {Shenar}, {Stringfellow}, {Sudnik}, {Tramper}, and {Wang}}]{Vink2023a}
{Vink} JS, {Mehner} A, {Crowther} PA, et~al (2023) {X-Shooting ULLYSES: Massive
  stars at low metallicity. I. Project description}. \aap 675:A154.
  \doi{10.1051/0004-6361/202245650},
  {\href{https://arxiv.org/abs/2305.06376}{{arXiv:2305.06376}}} {[astro-ph.SR]}

\bibitem[{{Wade} et~al(2016){Wade}, {Neiner}, {Alecian}, {Grunhut}, {Petit},
  {Batz}, {Bohlender}, {Cohen}, {Henrichs}, {Kochukhov}, {Landstreet},
  {Manset}, {Martins}, {Mathis}, {Oksala}, {Owocki}, {Rivinius}, {Shultz},
  {Sundqvist}, {Townsend}, {ud-Doula}, {Bouret}, {Braithwaite}, {Briquet},
  {Carciofi}, {David-Uraz}, {Folsom}, {Fullerton}, {Leroy}, {Marcolino},
  {Moffat}, {Naz{\'e}}, {Louis}, {Auri{\`e}re}, {Bagnulo}, {Bailey},
  {Barb{\'a}}, {Blaz{\`e}re}, {B{\"o}hm}, {Catala}, {Donati}, {Ferrario},
  {Harrington}, {Howarth}, {Ignace}, {Kaper}, {L{\"u}ftinger}, {Prinja},
  {Vink}, {Weiss}, and {Yakunin}}]{Wade2016a}
{Wade} GA, {Neiner} C, {Alecian} E, et~al (2016) {The MiMeS survey of magnetism
  in massive stars: introduction and overview}. \mnras 456:2--22.
  \doi{10.1093/mnras/stv2568},
  {\href{https://arxiv.org/abs/1511.08425}{{arXiv:1511.08425}}} {[astro-ph.SR]}

\bibitem[{{Wade} et~al(2020){Wade}, {Pigulski}, {Begy}, {Shultz}, {Hand ler},
  {Sikora}, {Neilson}, {Cugier}, {Erba}, {Moffat}, {Pablo}, {Popowicz},
  {Weiss}, and {Zwintz}}]{Wade2020a}
{Wade} GA, {Pigulski} A, {Begy} S, et~al (2020) {Evolving pulsation of the
  slowly rotating magnetic {\ensuremath{\beta}} Cep star
  {\ensuremath{\xi}}$^{1}$ CMa}. \mnras 492(2):2762--2774.
  \doi{10.1093/mnras/staa025},
  {\href{https://arxiv.org/abs/1912.08347}{{arXiv:1912.08347}}} {[astro-ph.SR]}

\bibitem[{{Walczak} et~al(2015){Walczak}, {Fontes}, {Colgan}, {Kilcrease}, and
  {Guzik}}]{Walczak2015}
{Walczak} P, {Fontes} CJ, {Colgan} J, et~al (2015) {Wider pulsation instability
  regions for {\ensuremath{\beta}} Cephei and SPB stars calculated using new
  Los Alamos opacities}. \aap 580:L9. \doi{10.1051/0004-6361/201526824}

\bibitem[{{Walczak} et~al(2019){Walczak}, {Daszy{\'n}ska-Daszkiewicz},
  {Pigulski}, {Pamyatnykh}, {Moffat}, {Handler}, {Pablo}, {Popowicz}, {Wade},
  {Weiss}, and {Zwintz}}]{Walczak2019a}
{Walczak} P, {Daszy{\'n}ska-Daszkiewicz} J, {Pigulski} A, et~al (2019) {Seismic
  modelling of early B-type pulsators observed by BRITE - I.
  {\ensuremath{\theta}} Ophiuchi}. \mnras 485(3):3544--3557.
  \doi{10.1093/mnras/stz639},
  {\href{https://arxiv.org/abs/1903.04224}{{arXiv:1903.04224}}} {[astro-ph.SR]}

\bibitem[{{Walker} et~al(2003){Walker}, {Matthews}, {Kuschnig}, {Johnson},
  {Rucinski}, {Pazder}, {Burley}, {Walker}, {Skaret}, {Zee}, {Grocott},
  {Carroll}, {Sinclair}, {Sturgeon}, and {Harron}}]{Walker2003}
{Walker} G, {Matthews} J, {Kuschnig} R, et~al (2003) {The MOST Asteroseismology
  Mission: Ultraprecise Photometry from Space}. \pasp 115:1023--1035.
  \doi{10.1086/377358}

\bibitem[{{Walker} et~al(2005{\natexlab{a}}){Walker}, {Kuschnig}, {Matthews},
  {Cameron}, {Saio}, {Lee}, {Kambe}, {Masuda}, {Guenther}, {Moffat},
  {Rucinski}, {Sasselov}, and {Weiss}}]{Walker2005b}
{Walker} GAH, {Kuschnig} R, {Matthews} JM, et~al (2005{\natexlab{a}}) {MOST
  Detects g-Modes in the Be Star HD 163868}. \apjl 635:L77--L80.
  \doi{10.1086/499362}

\bibitem[{{Walker} et~al(2005{\natexlab{b}}){Walker}, {Kuschnig}, {Matthews},
  {Reegen}, {Kallinger}, {Kambe}, {Saio}, {Harmanec}, {Guenther}, {Moffat},
  {Rucinski}, {Sasselov}, {Weiss}, {Bohlender}, {Bo{\v z}i{\'c}}, {Hashimoto},
  {Koubsk{\'y}}, {Mann}, {Ru{\v z}djak}, {{\v S}koda}, {{\v S}lechta}, {Sudar},
  {Wolf}, and {Yang}}]{Walker2005a}
{Walker} GAH, {Kuschnig} R, {Matthews} JM, et~al (2005{\natexlab{b}})
  {Pulsations of the Oe Star {$\zeta$} Ophiuchi from MOST Satellite Photometry
  and Ground-based Spectroscopy}. \apjl 623:L145--L148. \doi{10.1086/430254}

\bibitem[{Waskom(2021)}]{Seaborn_2021}
Waskom ML (2021) seaborn: statistical data visualization. Journal of Open
  Source Software 6(60):3021. \doi{10.21105/joss.03021},
  \urlprefix\url{https://doi.org/10.21105/joss.03021}

\bibitem[{{Weiss} et~al(2014){Weiss}, {Rucinski}, {Moffat},
  {Schwarzenberg-Czerny}, {Koudelka}, {Grant}, {Zee}, {Kuschnig}, {Mochnacki},
  {Matthews}, {Orleanski}, {Pamyatnykh}, {Pigulski}, {Alves}, {Guedel},
  {Handler}, {Wade}, and {Zwintz}}]{Weiss2014}
{Weiss} WW, {Rucinski} SM, {Moffat} AFJ, et~al (2014) {BRITE-Constellation:
  Nanosatellites for Precision Photometry of Bright Stars}. \pasp 126(940):573.
  \doi{10.1086/677236},
  {\href{https://arxiv.org/abs/1406.3778}{{arXiv:1406.3778}}} {[astro-ph.IM]}

\bibitem[{{Weiss} et~al(2021){Weiss}, {Zwintz}, {Kuschnig}, {Handler},
  {Moffat}, {Baade}, {Bowman}, {Granzer}, {Kallinger}, {Koudelka}, {Lovekin},
  {Neiner}, {Pablo}, {Pigulski}, {Popowicz}, {Ramiaramanantsoa}, {Rucinski},
  {Strassmeier}, and {Wade}}]{Weiss2021a}
{Weiss} WW, {Zwintz} K, {Kuschnig} R, et~al (2021) {Space Photometry with
  Brite-Constellation}. Universe 7(6):199. \doi{10.3390/universe7060199},
  {\href{https://arxiv.org/abs/2106.12952}{{arXiv:2106.12952}}} {[astro-ph.SR]}

\bibitem[{{Zahn}(1991)}]{Zahn1991}
{Zahn} JP (1991) {Convective penetration in stellar interiors}. \aap
  252:179--188

\bibitem[{{Zwintz} et~al(2023){Zwintz}, {Pigulski}, {Kuschnig}, {Wade},
  {Doherty}, {Earl}, {Lovekin}, {Muellner}, {Pich{\'e}-Perrier}, {Steindl},
  {Beck}, {Bicz}, {Bowman}, {Handler}, {Pablo}, {Popowicz}, {Rozanski},
  {Miko{\l}ajczyk}, {Baade}, {Koudelka}, {Moffat}, {Neiner}, {Orleanski},
  {Smolec}, {St. Louis}, {Weiss}, {Wenger}, and {Zoclonska}}]{Zwintz2023a*}
{Zwintz} K, {Pigulski} A, {Kuschnig} R, et~al (2023) {Catalogue of
  BRITE-Constellation targets I. Fields 1 to 14 (November 2013 - April 2016)}.
  arXiv e-prints arXiv:2311.18382. \doi{10.48550/arXiv.2311.18382},
  {\href{https://arxiv.org/abs/2311.18382}{{arXiv:2311.18382}}} {[astro-ph.SR]}

\end{thebibliography}


\section*{Statements and Declarations}

\subsection*{Funding}
This work was supported by a senior postdoctoral fellowship of the Research Foundation Flanders (FWO; grant number [1286521N]), a UK Research and Innovation (UKRI) Frontier Research grant under the UK government's Horizon Europe funding guarantee (SYMPHONY; grant number [EP/Y031059/1]), and a Royal Society University Research Fellowship (grant number: URF{\textbackslash}R1{\textbackslash}231631).

\subsection*{Competing Interests}
The author DMB has no relevant financial or non-financial interests to disclose.

\subsection*{Ethics approval}
Not applicable.

\subsection*{Consent to participate}
Not applicable.

\subsection*{Consent for publication}
Not applicable.

\subsection*{Availability of data and materials}
Not applicable.

\subsection*{Code availability}
Not applicable.

\subsection*{Data availability}
TESS data are freely available from the Mikulski Archive for Space Telescopes (MAST) archive: \url {https://archive.stsci.edu/}. CoRoT data are freely available from the mission's dedicated archive: \url{http://idoc-corot.ias.u-psud.fr/sitools/client-user/COROT_N2_PUBLIC_DATA/project-index.html}

\subsection*{Author Contributions}
The author DMB is the sole contributor to the preparation and writing of this work.


\end{document}